\documentclass[journal]{IEEEtran}
\usepackage{graphicx}
\usepackage{epsfig}
\usepackage{latexsym}
\usepackage{amsfonts}
\usepackage{here}
\usepackage{rawfonts}
\usepackage[latin1]{inputenc}
\usepackage[T1]{fontenc}
\usepackage{calc}
\usepackage{url}
\usepackage{enumerate}
\usepackage{color}
\usepackage[tbtags]{amsmath}
\usepackage{amssymb}
\usepackage{upref}
\usepackage{epic,eepic}
\usepackage{times}
\usepackage{dsfont}
\usepackage{comment}
\usepackage{cite}

\usepackage{graphicx}
\usepackage[font={small}]{caption}
\usepackage[font={small}]{subcaption}

\usepackage{stfloats}
\usepackage{mathtools}
\usepackage{amsmath}

\usepackage{algorithm}
\usepackage{algpseudocode}

\newtheorem{theorem}{\bf{Theorem}}

\newtheorem{corollary}{\bf{Corollary}}

\newtheorem{remark}{Remark}

\ifCLASSINFOpdf
\else
\fi
\begin{document}

\title{Scalable Hierarchical Over-the-Air Federated Learning}

\author{{Seyed Mohammad Azimi-Abarghouyi and Viktoria Fodor}

	\thanks{The authors are with the School of Electrical Engineering and Computer Science, KTH Royal Institute of Technology and Digital Futures, Stockholm, Sweden (Emails: $\bigl\{$seyaa,vjfodor$\bigr\}$@kth.se). A part of this work has been accepted for presentation at IEEE WCNC 2024 \cite{conf}.}
}

%%%%%%%%%%%%%%%%%%%%%%%%%%%%%%%%%%%%%
%\author{
%\authorblockN{Behrouz Maham$^\dag$$^\ddag$ and Are Hj{\o}rungnes$^\dag$}\vspace*{0.5em}
%\authorblockA{$^\dag$UNIK -- University Graduate Center, University of Oslo, Norway\\
%$^\ddag$Department of Electrical
%    Engineering, Stanford University, USA\\
%Email: \protect\url{bmaham@stanford.edu,
%arehj@unik.no}}\vspace*{-2.1em}
%    \thanks{This work was supported by the Research Council of Norway
%    through the project 176773/S10 entitled "Optimized Heterogeneous Multiuser MIMO Networks -- OptiMO".}%
%  }
%%%%%%%%%%%%%%%%%%%%%%%%%%%%%%%%%%%%%
% make the title area
\maketitle

\vspace{-15pt}
\begin{abstract}
When implementing hierarchical federated learning over wireless networks, scalability assurance and the ability to handle both interference and device data heterogeneity are crucial. This work introduces a new two-level learning method designed to address these challenges, along with
a scalable over-the-air aggregation scheme for the uplink and a bandwidth-limited broadcast scheme for the downlink that efficiently use a single wireless resource. To provide resistance against data heterogeneity, we employ gradient aggregations. Meanwhile, the impact of uplink and downlink interference is minimized through optimized receiver normalizing factors. We present a comprehensive mathematical approach to derive the convergence bound for the proposed algorithm, applicable to a multi-cluster wireless network encompassing any count of collaborating clusters, and provide special cases and design remarks. As a key step to enable a tractable analysis, we develop a spatial model for the setup
by modeling devices as a Poisson cluster process over the edge
servers and rigorously quantify
uplink and downlink error terms due to the interference. Finally,
we show that despite the interference and data heterogeneity, the proposed algorithm
not only achieves high learning accuracy for a variety of parameters but also significantly outperforms the conventional hierarchical learning algorithm.

\end{abstract}
\vspace{0pt}
\begin{IEEEkeywords}
Federated learning, machine learning, hierarchical systems, over-the-air computation.
\end{IEEEkeywords}
\vspace{-5pt}
\section{Introduction}
With the growing pervasiveness and computational power of wireless edge devices, i.e., phones, smart watches, sensors, and autonomous vehicles, there is an increasing demand for enabling machine learning to train a global model from the diverse distributed data over the edge devices \cite{viktoria}. However, loading such enormous amounts of data from the devices to a central server is not often feasible due to strict constraints on latency, power, and bandwidth, or concerns on data privacy. A promising and practically feasible distributed approach is federated learning (FL), which is able to implement machine learning directly at the wireless edge while under no circumstances the data leaves the devices \cite{mcmahan}. In this approach, the model training is performed locally at each single device with the help of a parameter server, such that synchronous
model update at all devices and model aggregation at the server are repeated until convergence. Most studies on FL consider a single server. However, to support more devices that can collaborate and to speed up the learning process, recent studies have proposed hierarchical architectures for FL that incorporate a core server and multiple edge servers \cite{peter, letaief, bennis, tony, yang1, niyato, xuli, gunduz_ota}. The hierarchical FL has two levels of aggregation of model parameters: the intra-cluster aggregation at the edge servers (edge aggregation) and then an inter-cluster aggregation at the core server (core aggregation). The goal of this paper is to propose a hierarchical learning scheme that is both resource efficient and resilient to data heterogeneity and wireless interference from parallel learning processes, and to provide a modeling methodology that allows to express and evaluate the interference and its effect.
\vspace{-5pt}
\subsection{Prior Art}
The convergence properties of hierarchical FL, without considering the limitations of a wireless environment is evaluated in \cite{peter,letaief,niyato}, proposing server selection, node scheduling and data compression solutions. The limited resources and wireless channel impairments under orthogonal transmissions and fixed network configurations are taken into account for example in \cite{bennis, tony, yang1, xuli}. It is recognized however that the orthogonal transmissions lead to performance bottlenecks when a high number of devices participate in the learning. 
%In practical scenarios, the FL faces the challenge of transmitting data over unreliable wireless networks with limited resources. In such networks, numerous devices and edge servers communicate through a shared wireless propagation medium, making communication efficiency one of the most critical obstacles to overcome in the FL. As the conventional approach, orthogonal multiple access techniques
%have been utilized in FL via individual transmissions of different devices to the servers \cite{peter, letaief, bennis, tony, yang1, yang2, niyato, xuli}. 
An effective and desirable approach for these scenarios is known as over-the-air FL, which utilizes the over-the-air computation scheme \cite{nazer, azimi_cmp}. This approach leverages interference caused by simultaneous multi-access transmissions from edge devices to perform aggregation. Through the integration of communication and computation, over-the-air FL can operate with significantly fewer resources and lower latencies in both communication and computation compared to FL using orthogonal transmissions \cite{viktoria}. An extensive survey of opportunities and challenges of FL in wireless networks is presented in \cite{isl}.

The majority of research in the field of over-the-air FL focuses on single-cell learning scenarios, with an emphasis on the uplink transmissions \cite{ding, huang_analog, gunduz3,cao}. Recently, several works include specific aspects of uplink interference. Interferers distributed according to Poisson point processes (PPPs) are considered in \cite{yang_sg, huang_sg, salehi_sg}.
	An abstract interference model with heavy tail is considered in \cite{yang3}, and it is shown that while heavy tail slows down the learning process, it may improve the generalization capability. In \cite{huang_turning}, transmission power is controlled to escape from saddle points. Studies have been also conducted to examine bandwidth-limited downlink in both single-cell \cite{downlink} and multi-cell \cite{multicell} settings. 

Hierarchical FL using over-the-air computation is studied in \cite{gunduz_ota}, and that is the work most closely related to our paper. However, that study assumes the presence of multiple antennas at the edge servers, an ideal downlink transmission, and most importantly, does not account for inter-cell interference.

To understand the effect of interference in hierarchical FL and provide a comprehensive network analysis, we turn to stochastic geometry. Stochastic geometry has been developed as a tool to characterize interference in networks, taking the location of the nodes into account \cite{haenggi_book, halim}. Within the field, there are two \emph{canonical} approaches to characterize wireless networks. One of them is PPPs, which assumes uniform device and server placements, and has applications in cellular networks \cite{andrews, huang_sg, salehi_sg}. The other approach is based on Poisson cluster processes (PCPs), which takes into account non-uniformity and allows to model the correlation between the locations of the devices and servers within and among clusters. PCPs are justified by Third Generation Partnership Project (3GPP) \cite{3gpp, dhillon} and widely-adopted in deployments where devices are frequently grouped together,
meaning that they are commonly found in specific areas, known as ``hotspots'' \cite{qzhu, dhillon, azimi, azimi_cloud, hole}. The PCP fits to the application area of distributed learning on hierarchical networks, which are composed of explicitly defined clusters, such as transportation and other smart city sensing applications, environmental monitoring or industrial IoT. Moreover, it is designed to allow performance models where the effects of the network parameters are pronounced. Therefore, this is the approach we follow in this paper.

\vspace{-5pt}
\subsection{Key Contributions}
This paper develops a learning and transmission scheme for hierarchical FL utilizing over-the-air computation, and provides modeling solution that can capture the effect of interference. The key contributions are as follows:

\textit{Learning Method:} We propose a new iterative learning method with
	two-level aggregation named {\fontfamily{lmtt}\selectfont MultiAirFed}, that combines intra-cluster gradient and inter-cluster model parameter based aggregations, and includes also multi-step local training. The method is well suited for the hierarchical network structure with unreliable wireless and reliable backhaul links, while also being resilient to non-i.i.d. data.

%\textit{Learning Framework:} \textcolor{blue}{For more robustness to channel noise, we propose a new iterative learning method with
%a two-level aggregation named {\fontfamily{lmtt}\selectfont
%	MultiAirFed}. 
%Different from \cite{peter, letaief, bennis, tony, yang1, yang2, niyato, xuli, gunduz_ota}, the gradients of the model updates
%are combined in the intra-cluster aggregation, while model parameters are 
%aggregated during the inter-cluster aggregation. Also, an approach is proposed which involves multiple local updates and their aggregation, in order to further leverage the benefits of local training. The proposed method
%is well suited for the unreliable wireless and for the reliable backhaul 
%transmission, respectively.}

\textit{Transmission Scheme:} We propose scalable clustered over-the-air aggregation scheme for the uplink and bandwidth-limited analog broadcast scheme for the downlink transmission. The schemes are independent of the number of clusters and devices, and each iteration all over the network is performed only in one single resource block, for both the uplink and the downlink. %In particular, we consider the realities of imperfect downlink, an unlimited number of single-antenna nodes, and diverse tasks that are prevalent in scalable wireless networks \cite{3gpp, qzhu, hole, andrews, azimi, dhillon, haenggi_book,  azimi_cloud, guard, fan}. 
We design uplink and downlink power control schemes, and propose receiver normalizing factors that minimize the distortion of the recovered gradients or model vectors, taking task diversity, data heterogeneity, wireless channel impairments and interference into account. While the proposed transmission scheme is general, we apply it for {\fontfamily{lmtt}\selectfont MultiAirFed}, and express the intra- and inter-cluster aggregation errors caused by the uplink and downlink interference.

%\textit{Transmission Scheme:} Under limited transmit powers at the devices and the servers, we propose a scalable clustered over-the-air aggregation scheme for the uplink and a bandwidth-limited broadcast scheme for the downlink. Using the schemes, we implement {\fontfamily{lmtt}\selectfont
%MultiAirFed} over a large-scale wireless network where independent of the number of clusters and devices, each iteration all over the network is performed only in one single resource block, for both the uplink and the downlink. \textcolor{blue}{The proposed schemes can be also utilized for other  hierarchical FL algorithms, which further reinforces the generality of this work. In contrast to the literature that typically assumes ideal orthogonal downlink, a fixed number of nodes, and a single learning task, as demonstrated by \cite{letaief, bennis, tony, gunduz_ota, huang_analog, huang_sg, niyato, xuli, gunduz3, yang1, peter, yang2, yang3, cao}, we take into account the realities of imperfect bandwidth-limited downlink, an unlimited number of nodes, and various tasks, which are common in scalable wireless networks \cite{3gpp, andrews, azimi, dhillon, haenggi_book, azimi_cloud, guard, fan}. Furthermore, the proposed schemes are designed to operate effectively under the conditions of single-antenna servers and inter-cluster interference.} 

\textit{Tractable Modeling and Convergence Analysis:} We utilize the tools of stochastic geometry, and specifically PCPs, and quantify the intra- and inter-cluster aggregation error terms. Then, we characterize the learning convergence in terms of the optimality gap for the setup, as a function of the network parameters, and present design remarks and special cases. The optimiality gap is tractable and well structured, where it is easy to identify the effects of the learning parameters, the transmission scheme and the network topology.

\textit{System Design Insights:} Our analysis reveals that due to interference, there is a non-zero optimality gap after convergence. Increasing the number of devices in each cluster or the number of collaborating clusters increases the learning accuracy, however, the improvement is limited by increasing the interference. A higher density of cluster centers has also a degrading effect on the learning performance. Our numerical results show that  {\fontfamily{lmtt}\selectfont MultiAirFed}, based on the combination of gradient and model parameter aggregations, can provide significantly better learning accuracy compared to the other hierarchical FL algorithm in the proposed wireless setup.

\vspace{-4pt}
\section{System Model}

\begin{figure*}[tb!]
	\centering
	\vspace{-22pt}
	\includegraphics[width =.9\linewidth]{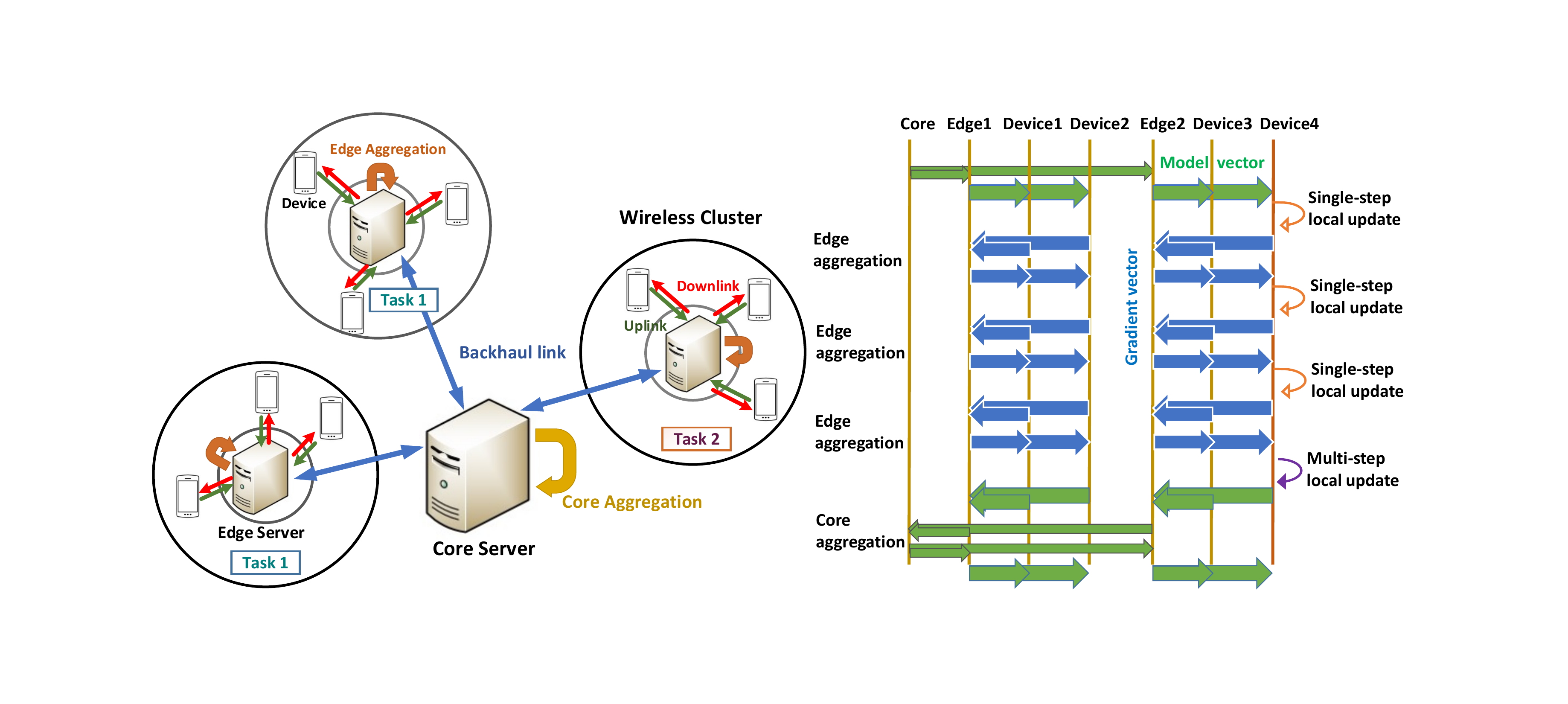}
	\vspace{-35pt}
	\caption{Representation of the FL system with three clusters, and the message exchange in the {\fontfamily{lmtt}\selectfont MultiAirFed} method.}
	\vspace{-15pt}
\end{figure*}
\vspace{-3pt}
\subsection{Network Topology}
We consider groups of devices clustered around edge servers, performing FL, as shown on Fig.~1. The clusters may have different or similar learning tasks. For example, a university campus is mostly interested in science learning tasks in contrast with a farm where the type of sensors and data are completely different. One or more core servers support the clusters in the learning process. The clusters with the same task connect to the same core server and collaborate to allow hierarchical learning. The clusters themselves are wireless cells, with the edge servers located at the base stations. The edge servers are connected to the core server by a wired backhaul link.

We model the emerging network topology with the help of PCP \cite{haenggi_book}.
A PCP $\Phi$ is formally defined as a union of offspring points in $\mathbb{R}^2$ that are located around parent points. In our case, the parent points are the edge servers, while the offspring points are the devices.
The parent point process is a homogeneous Poisson point process (PPP) $\Phi_\text{p}$ with density $\lambda_\text{p}$. Also, the offspring point processes are conditionally independent. The set of offspring points of $\mathbf{x} \in \Phi_\text{p}$ is denoted by cluster ${\cal N}^{{\mathbf{x}}}$, such that $\Phi = \cup_{\mathbf{x}\in \Phi_{\rm p}}{\cal N}^{{\mathbf{x}}}$. The PDF of each element of ${\cal N}^{{\mathbf{x}}}$ being at a location $\mathbf{y}+\mathbf{x} \in \mathbb{R}^2$ inside the cluster is shown by $f_{\|\mathbf{y}\|}(y)$. 

In wireless networks, clusters are mostly modeled as disk-shaped regions where nodes are distributed uniformly. When these clusters are employed in a PCP, the resulting point process is
referred to as Mat{\'e}rn cluster process (MCP) \cite{haenggi_book}. Research has demonstrated that MCP can be utilized to accurately simulate actual setups utilized by 3GPP \cite{dhillon, 3gpp}. Also, since the edge servers are usually located
inside base station buildings or integrated with antenna towers,
there are protective zones over them where devices and other edge servers cannot
be located. These zones are additionally motivated to suppress interference by inhibiting nearby devices \cite{azimi_cloud, guard}. Therefore, we
further consider a modified type of MCP, named MCP with holes at the cluster centers (MCP-H) \cite{hole, azimi_cloud}, where the points are distributed
around cluster centers with uniform distribution inside rings with inner radius $r_0$ and outer radius $R$ as
\begin{align}
f_{\|\mathbf{y}\|}(y) = \frac{2y}{R^2-r_0^2}, \ r_0\leq y \leq R.
\end{align}
%From the set of nano-machines located in the cluster centered at $\mathbf{x} \in \Phi_\text{p}$, we assume a subset ${\cal B}^{{\mathbf{x}}}\subseteq {\cal N}^{{\mathbf{x}}}$ has information and is simultaneously active. Furthermore, the number of active nano-machines $|{\cal B}^{{\mathbf{x}}}|$ is assumed to have a Poisson distribution with mean $n_a$.

The number of devices in each cluster is assumed to be $M$, i.e., $|{\cal N}^{{\mathbf{x}}}| = M$. Also, among all the devices of a cluster $\mathbf{x}$, the set of active devices in a time slot is denoted by $\cal A^{\mathbf{x}} \subseteq {\cal N}^{{\mathbf{x}}}$. The term "active device" denotes a device that participates in the aggregation phase of the FL by its uplink transmission. 
\vspace{-5pt}
\subsection{Channel Model}
All the nodes are single-antenna units. For the wireless links between devices and edge servers, we assume single-slope power-law path loss and small-scale i.i.d.  Rayleigh fading. The pathloss exponent is denoted by parameter $\alpha$. The uplink fading between a device $\mathbf{y} \in {\cal N}^{\mathbf{x}}$ and a server at $\mathbf{z}$ is modeled by $f_{\mathbf{y}\mathbf{z}}^\mathbf{x} \in \mathbb{C}$, such that the channel gain $|f_{\mathbf{y}\mathbf{z}}^\mathbf{x}|^2 \sim \exp(1)$. The downlink fading between a server $\mathbf{x}$ and a reference device is $f^\mathbf{x}$ with the gain $|f^\mathbf{x}|^2 \sim \exp(\sigma_\text{d}^2)$. Also,
all channel gains are assumed invariant during one time slot required for
an uplink or downlink transmission, while they change independently from one time slot to another. The communication between the edge servers and the respective core server is performed over high-capacity backhaul links \cite{fan}. This communication is considered error free, and its optimization is not part of this work.
\vspace{-5pt}
\section{Proposed Learning Method}
Assume that there are $C$ collaborating clusters including a reference cluster with its center at the origin $\mathbf{o}$ that have a same learning task. The centers of these clusters are denoted by a set $\mathcal{C}$. Also, a device at a location $\mathbf{y}$ in a cluster $\mathbf{x}$ has its private dataset ${\cal D}_{\mathbf{y}}^{\mathbf{x}}$. The $l$-th sample in ${\cal D}_{\mathbf{y}}^{\mathbf{x}}$ is $\xi_l$. The learning model is parametrized by the parameter vector $\mathbf{w} \in \mathbb{R}^d$, where $d$ denotes the learning model size. Then, the local loss function of the model parameter vector $\mathbf{w}$ over ${\cal D}_{\mathbf{y}}^{\mathbf{x}}$ is
\begin{align}
\label{localloss}
F_{\mathbf{y}}^{\mathbf{x}}(\mathbf{w}) =  \frac{1}{D_\mathbf{y}^\mathbf{x}}\sum_{\xi_l\in {\cal D}_{\mathbf{y}}^{\mathbf{x}}}^{}\ell(\mathbf{w},\xi_l),
\end{align}
where $D_\mathbf{y}^\mathbf{x} = |{\cal D}_{\mathbf{y}}^{\mathbf{x}}|$ is the dataset size and $\ell(\mathbf{w},\xi_l)$ is the sample-wise loss function that quantifies the
prediction error of the model vector $\mathbf{w}$ on $\xi_l$. Then, the global loss function on the distributed datasets $\cup_{{\mathbf{x}} \in {\mathcal{C}}}\cup_{\mathbf{y} \in {\cal N}^\mathbf{x}} {\cal D}_{\mathbf{y}}^{\mathbf{x}}$ over the clusters in $\mathcal{C}$ is computed as
\begin{align}
\label{lossfunction}
F(\mathbf{w}) = \frac{1}{\sum_{{\mathbf{x}} \in {\mathcal{C}}} \sum_{\mathbf{y} \in {\cal N}^\mathbf{x}} D_\mathbf{y}^\mathbf{x}}\sum_{\mathbf{x}\in {\mathcal{C}}}^{} \sum_{\mathbf{y} \in {\cal N}^\mathbf{x}}^{} D_\mathbf{y}^\mathbf{x} F_{\mathbf{y}}^{\mathbf{x}}(\mathbf{w}).
\end{align}
Thus, the learning process has the objective to find the desired model parameter vector $\mathbf{w}$ as
\begin{align}
\label{globalopt}
\mathbf{w}^* = \min_{\mathbf{w}} F(\mathbf{w}).
\end{align}
%In general, FL algorithms solve \eqref{globalopt} in two steps: local update and aggregation. These steps are distributedly repeated over a number of iterations based on gradient descent techniques \cite{mcmahan}. 

To solve \eqref{globalopt} for the hierarchical systems, we propose a new two-level algorithm named {\fontfamily{lmtt}\selectfont
	MultiAirFed}. The {\fontfamily{lmtt}\selectfont
	MultiAirFed} is a combination of intra-cluster gradient and inter-cluster
model parameter aggregation. Gradient aggregation has been shown to be robust to noise and interference in
\cite{huang_sg, yang3}, and to non-i.i.d. data in \cite{fedsgd-is-better}, and therefore is a good candidate for the intra-cluster learning process over the interfering wireless links.
The model-parameter aggregation at the core server at the same time allows multiple inter-cluster iterations.
The resulting hierarchical learning process is shown on Fig. 1 and Algorithm 1. It is as follows: Consider $T$ global inter-cluster iterations. In a particular iteration $t$, consider $\tau$ intra-cluster iterations. In a particular intra-cluster iteration $i$, each device $\mathbf{y}$ in a cluster $\mathbf{x}$ computes the local gradient of the loss function in \eqref{localloss} from its local dataset, indexed by $\left\{i,t\right\}$, as
\begin{align}
\label{localgrad}
\mathbf{g}_{\mathbf{y},i,t}^{\mathbf{x}} = \nabla F_{\mathbf{y}}^{\mathbf{x}}(\mathbf{w}_\mathbf{y}^\mathbf{x},\xi_\mathbf{y}^\mathbf{x}),
\end{align}
where $\mathbf{w}_\mathbf{y}^\mathbf{x}$ is its parameter vector and ${\xi}_{\mathbf{y}}^{\mathbf{x}}$ is the local mini-batch
chosen uniformly at random from ${\cal D}_{\mathbf{y}}^{\mathbf{x}}$.\footnote{The mini-batch size is chosen to ensure that each device can complete its local computation before the subsequent transmission.} Then, devices upload (transmit) their local gradients to their servers for intra-cluster aggregation. The server of cluster $\mathbf{x}$ averages of the local gradients from its active devices and broadcasts the generated intra-cluster gradient
\begin{align}
\label{intragrad}
\mathbf{g}_{i,t}^{{\mathbf{x}}} = \frac{1}{|{\cal A}_{i,t}^\mathbf{x}|}\sum_{\mathbf{y} \in {\cal A}_{i,t}^\mathbf{x}}^{} \mathbf{g}_{\mathbf{y},i,t}^{\mathbf{x}},
\end{align}
where $|{\cal A}_{i,t}^\mathbf{x}|$ is the number of active devices in the cluster
$\mathbf{x}$ for the iteration index $\left\{i,t\right\}$, i.e., $|{\cal A}_{i,t}^\mathbf{x}| = \sum_{\mathbf{y} \in {\cal N}^\mathbf{x}}^{}\mathds{1}_{\mathbf{y},i,t}^\mathbf{x}$, where $\mathds{1}_{\mathbf{y},i,t}^\mathbf{x}$ equals 1 if the device is active and 0 otherwise. Then, the servers broadcast the intra-cluster gradients $\mathbf{g}_{i,t}^{\mathbf{x}}, \forall \mathbf{x}$ to their devices. Utilizing $\mathbf{g}_{i,t}^{\mathbf{x}}$, each device $\mathbf{y}$ in any cluster $\mathbf{x}$ updates its local model following
a one-step gradient descent as
\begin{align}
\label{localcomp}
\mathbf{w}_{\mathbf{y},i+1,t}^{\mathbf{x}} = \mathbf{w}_{\mathbf{y},i,t}^{\mathbf{x}} -\mu_{t} \mathbf{g}_{i,t}^{\mathbf{x}},
\end{align}
where $\mu_{t}$ is the learning rate at the global iteration $t$. After completing $\tau$ intra-cluster iterations, each device performs a $\gamma$-step gradient descent locally as
\begin{align}
\label{final_local00}
\mathbf{w}_{\mathbf{y},\tau,0,t}^{\mathbf{x}} = \mathbf{w}_{\mathbf{y},\tau,t}^{\mathbf{x}},
\end{align}
\vspace{-20pt}
\begin{align}
\label{final_local01}
\mathbf{w}_{\mathbf{y},\tau,j,t}^{\mathbf{x}} = \mathbf{w}_{\mathbf{y},\tau,j-1,t}^{\mathbf{x}} - \mu_{t}\nabla F_{\mathbf{y}}^{\mathbf{x}}(\mathbf{w}_{\mathbf{y},\tau,j-1,t}^{\mathbf{x}},\xi_{\mathbf{y},\tau,j-1,t}^\mathbf{x}), \nonumber\\ j = \left\{1,\cdots,\gamma\right\}.
\end{align} 
To start the inter-cluster iteration, the devices upload their model parameters, i.e., $\mathbf{w}_{\mathbf{y},\tau,\gamma,t}^\mathbf{x}, \forall \mathbf{y}, \mathbf{x}$, to their servers. Accordingly, each server $\mathbf{x}$ computes an intra-cluster model parameter vector with the following average
\begin{align}
\label{intergrad}
\mathbf{w}_{t+1}^{{\mathbf{x}}} = \frac{1}{|{\cal A}_{\tau,t}^\mathbf{x}|}\sum_{\mathbf{y} \in {\cal A}_{\tau,t}^\mathbf{x}}^{} \mathbf{w}_{\mathbf{y},\tau,\gamma,t}^{\mathbf{x}}.
\end{align}
Then, the collaborating servers upload their intra-cluster model parameter vectors to the core server for a global inter-cluster model parameter aggregation as

\begin{align}
\label{globalgrad}
\mathbf{w}_{t+1}^{\text{G}} = \frac{1}{|{\cal A}_{\tau,t}|} \sum_{\mathbf{x} \in {\mathcal{C}}}^{} |{\cal A}_{\tau,t}^{\mathbf{x}}| \mathbf{w}_{t+1}^{\mathbf{x}},
\end{align}
which is the average of all model parameter vectors from the active
devices in the clusters of ${\mathcal{C}}$. Also, $|{\cal A}_{\tau,t}| = \sum_{\mathbf{x}\in{\mathcal{C}}}^{}|{\cal A}_{\tau,t}^\mathbf{x}|$
denotes the number of the active devices for the iteration index
$\left\{\tau,t\right\}$. Then, the servers broadcast $\mathbf{w}_{t+1}^{\text{G}}$ to the devices to update their initial model parameter vector for the next global iteration $t+1$ as $\mathbf{w}_{\mathbf{y},0,t+1}^{\mathbf{x}} =  \mathbf{w}_{t+1}^{\text{G}}, \forall \mathbf{y}, \mathbf{x}$. This global update synchronizes all the devices in the collaborating clusters and prevents a high deviation of the local training processes. In line with \cite{letaief, bennis, ding, huang_analog, gunduz3, yang3}, we have not included the dataset sizes across devices into the average terms in \eqref{intragrad} and \eqref{intergrad}. However, for a weighted average extension, one can adjust the local gradient or model vector for the intra- or inter-cluster aggregation uploads by replacing $\mathbf{g}_{\mathbf{y},i,t}^{\mathbf{x}}$ with $D_\mathbf{y}^\mathbf{x}\mathbf{g}_{\mathbf{y},i,t}^{\mathbf{x}}$ and $\mathbf{w}_{\mathbf{y},\tau,\gamma,t}^{\mathbf{x}}$ with $D_\mathbf{y}^\mathbf{x} \mathbf{w}_{\mathbf{y},\tau,\gamma,t}^{\mathbf{x}}$. To calculate the number of active devices in each cluster, the cardinality $\sum_{\mathbf{y} \in {\cal N}^\mathbf{x}}^{}D_\mathbf{y}^\mathbf{x}\mathds{1}_{\mathbf{y},i,t}^\mathbf{x}$ can be used instead of $|{\cal A}_{i,t}^\mathbf{x}|$. Then, the proposed algorithm can be followed. The proposed transmission scheme in Section IV can also be readily modified using these changes.

Compared to other hierarchical methods in \cite{peter, letaief, bennis, tony, xuli, niyato, gunduz_ota} which are based on model parameter transmissions, gradient transmission in {\fontfamily{lmtt}\selectfont
	MultiAirFed} over wireless links is expected to be more robust to channel noise and interference. This is because for each device local update, a noisy model parameter aggregation leads to imperfections both on the initial model parameter vector update and the local gradient function evaluation in \eqref{localgrad}. Moreover, the resulting errors propagate and reinforce through multiple local steps. As learning convergence demands high accuracy in the gradient direction, particularly in the vicinity of the optimal solution, noisy model parameter aggregation may hinder the model to converge. However, when gradients are transmitted, devices can download the aggregated gradient \eqref{intragrad} as a same gradient term for their update without the need for local computations, and the initial model parameter vector remains unaffected by any noise. This further guarantees that the local steps taken during an inter-cluster iteration will continuously approach convergence. Moreover, in general, gradient based aggregation is shown to be resilient against heterogeneity and non-i.i.d. data when compared to approaches that transmit model parameters \cite{fedsgd-is-better}. These will be further justified through experimental results in Section VI. Also, even though gradient transmission does not permit multiple local iterations at each intra-cluster iteration, the proposed final gradient descent \eqref{final_local00}-\eqref{final_local01} serve as a reinforcement for integrating local training into the learning process. 

Before we conclude this discussion, please note that the proposed learning method is not limited to the choice of the spatial model in Subsection II.A and the sequel transmission scheme in Section IV. It can be easily applied to different scenarios of multi-server FL.
\vspace{-4pt}
\begin{algorithm}
	\small
	\caption{{\fontfamily{lmtt}\selectfont
				MultiAirFed} algorithm}
	\begin{algorithmic}
			\vspace{-1pt}
			\State Initialize the global model $\mathbf{w}_0^\text{G}$\
			\vspace{0pt}
			\State \textbf{for} inter-cluster iteration $t=1,...,T$ \textbf{do}
			\vspace{0pt}
			\State \hspace{10pt}Each device updates its model by $\mathbf{w}_{t}^\text{G}$ 
			\vspace{0pt}
			\State \hspace{10pt}\textbf{for} intra-cluster iteration $i=1,...,\tau$ \textbf{do}
			\vspace{0pt}
			\State \hspace{20pt}Each device obtains its local gradient from\\ \hspace{30pt} $\mathbf{g}_{\mathbf{y},i,t}^{\mathbf{x}} = \nabla F_{\mathbf{y}}^{\mathbf{x}}(\mathbf{w}_{\mathbf{y},i,t}^\mathbf{x},\xi_{\mathbf{y},i,t}^\mathbf{x})$
			\vspace{0pt}
			\State \hspace{20pt}Each edge server obtains its intra-cluster gradient from\\ \hspace{30pt} $\mathbf{g}_{i,t}^{{\mathbf{x}}} = \frac{1}{|{\cal A}_{i,t}^\mathbf{x}|}\sum_{\mathbf{y} \in {\cal A}_{i,t}^\mathbf{x}}^{} \mathbf{g}_{\mathbf{y},i,t}^{\mathbf{x}}$
			\vspace{0pt}
			\State \hspace{20pt}Each device updates its local model as \\ \hspace{30pt}$\mathbf{w}_{\mathbf{y},i+1,t}^{\mathbf{x}} = \mathbf{w}_{\mathbf{y},i,t}^{\mathbf{x}} -\mu_{t} \mathbf{g}_{i,t}^{\mathbf{x}}$
			\vspace{0pt}
			\State \hspace{20pt}\textbf{if} $i = \tau$ \textbf{do}
			\vspace{0pt}
			\State \hspace{30pt}Each device updates its local model as\\ \hspace{40pt}$\small{\mathbf{w}_{\mathbf{y},\tau,0,t}^{\mathbf{x}} = \mathbf{w}_{\mathbf{y},\tau,t}^{\mathbf{x}},\
				\mathbf{w}_{\mathbf{y},\tau,j,t}^{\mathbf{x}} = \mathbf{w}_{\mathbf{y},\tau,j-1,t}^{\mathbf{x}} - \mu_{t}}\times$ \\\hspace{56pt}$\small{\nabla F_{\mathbf{y}}^{\mathbf{x}}(\mathbf{w}_{\mathbf{y},\tau,j-1,t}^{\mathbf{x}},\xi_{\mathbf{y},\tau,j-1,t}^\mathbf{x}), \ j = \left\{1,\cdots,\gamma\right\}.}$
			\vspace{0pt}
			\State \hspace{10pt}Each edge server obtains its intra-cluster model from\\ \hspace{20pt}$\mathbf{w}_{t+1}^{{\mathbf{x}}} = \frac{1}{|{\cal A}_{\tau,t}^\mathbf{x}|}\sum_{\mathbf{y} \in {\cal A}_{\tau,t}^\mathbf{x}}^{} \mathbf{w}_{\mathbf{y},\tau,\gamma,t}^{\mathbf{x}}$
			\vspace{0pt}
			\State \hspace{10pt}Core server obtains global model from \\ \hspace{20pt}$\mathbf{w}_{t+1}^{\text{G}} = \frac{1}{|{\cal A}_{\tau,t}|} \sum_{\mathbf{x} \in {\mathcal{C}}}^{} |{\cal A}_{\tau,t}^{\mathbf{x}}| \mathbf{w}_{t+1}^{\mathbf{x}}$
	\end{algorithmic}

\end{algorithm}

\vspace{-12pt}
\section{Clustered Transmission Scheme}
To implement {\fontfamily{lmtt}\selectfont
	MultiAirFed}, we propose a scalable transmission scheme including two types of analog transmissions for uplink and downlink, where each is done simultaneously over the clusters in a single resource block, as shown on the message exchange diagram on Fig. 1. It is inspired from \cite{downlink} which shows that analog downlink approach significantly outperforms the digital one. Synchronization is required within a cluster, this is a common assumption in the literature, see e.g., \cite{multicell, yang_sg, huang_sg}. It is also worth noting that the transmission scheme is not constrained to the spatial model choice presented in Subsection II.A, but the spatial model is needed to derive analytic expressions for the gradient and model aggregations. From here, we ignore the iteration indexes for simplicity of presentation.
\vspace{-5pt}
\subsection{Uplink} 
For the uplink, we propose a clustered over-the-air aggregation scheme. The term "over-the-air" stems from the facts that devices transmit simultaneously and the objective is to construct the aggregation vectors \eqref{intragrad} and \eqref{globalgrad} at the edge servers based on the additive nature of wireless multiple-access channels. The term "clustered" comes from the fact that the power allocation in each cluster is distinct from other clusters.

Depending on an intra- or inter-cluster iteration, the gradient parameters or model parameters at each
device are normalized before transmission to have zero mean and unit variance. There are two advantages to normalizing the parameters. First, when the parameters have zero-mean entries, the estimates obtained in the sequel are unbiased. Second, when the entries have unit variance, the interference and consequently the error terms depend only on the power control, and do not depend on the specific values of the model or gradient parameters. 

For an intra-cluster iteration, the local gradient vector at a device $\mathbf{y} \in {\cal N}^{\mathbf{x}}$, i.e., ${{\mathbf{g}}_{\mathbf{y}}^\mathbf{x}}$, is normalized as ${\bar{\mathbf{g}}_{\mathbf{y}}^\mathbf{x}} = \frac{{{\mathbf{g}}_{\mathbf{y}}^\mathbf{x}}-\mu_{\text{g},\mathbf{y}}^\mathbf{x}\mathbf{1}}{\sigma_{\text{g},\mathbf{y}}^\mathbf{x}}$, where $\mathbf{1}$ is the all one vector, and $\mu_{\text{g},\mathbf{y}}^\mathbf{x}$ and ${\sigma_{\text{g},\mathbf{y}}^\mathbf{x}}$ denote the mean and standard deviation of the $d$ entries of the gradient given by
\begin{align}
\mu_{\text{g},\mathbf{y}}^\mathbf{x} = \frac{1}{d}\sum_{i=1}^{d} {\mathbf{g}}_\mathbf{y}^\mathbf{x}(i),\
\sigma_{\text{g},\mathbf{y}}^{\mathbf{x}\ 2} = \frac{1}{d}\sum_{i=1}^{d}({\mathbf{g}}_\mathbf{y}^\mathbf{x}(i)-\mu_{\text{g},\mathbf{y}}^\mathbf{x})^2,
\end{align}
where $\mathbf{g}_\mathbf{y}^\mathbf{x}(i)$ is the $i$-th entry of the vector.
 Also, for an inter-cluster iteration, the normalized local model parameter vector is ${\bar{\mathbf{w}}_{\mathbf{y}}^\mathbf{x}} = \frac{{{\mathbf{w}}_{\mathbf{y}}^\mathbf{x}}-\mu_{\text{w},\mathbf{y}}^\mathbf{x}\mathbf{1}}{\sigma_{\text{w},\mathbf{y}}^\mathbf{x}}$, where the mean and variance are
 \begin{align}
 \mu_{\text{w},\mathbf{y}}^\mathbf{x} = \frac{1}{d}\sum_{i=1}^{d} {\mathbf{w}}_\mathbf{y}^\mathbf{x}(i),\
 \sigma_{\text{w},\mathbf{y}}^{\mathbf{x}\ 2} = \frac{1}{d}\sum_{i=1}^{d}({\mathbf{w}}_\mathbf{y}^\mathbf{x}(i)-\mu_{\text{w},\mathbf{y}}^\mathbf{x})^2.
 \end{align}
Then, at each device $\mathbf{y}$ in the cluster $\mathbf{x}$, the normalized vector ${\bar{\mathbf{g}}_{\mathbf{y}}^\mathbf{x}}$ or ${\bar{\mathbf{w}}_{\mathbf{y}}^\mathbf{x}}$ is analog modulated and
transmitted as ${p_{\mathbf{y}}^{\mathbf{x}}} {\bar{\mathbf{g}}_{\mathbf{y}}^\mathbf{x}}$ or ${p_{\mathbf{y}}^{\mathbf{x}}} {\bar{\mathbf{w}}_{\mathbf{y}}^\mathbf{x}}$ simultaneously with other devices in all the clusters, where $|{p_{\mathbf{y}}^{\mathbf{x}}}|^2$ denotes the transmission power. Thus, the received signal at a server located at $\mathbf{z}$ is
\begin{align}
	\label{uplink}
	\mathbf{v}_\text{u}^{\mathbf{z}} &= \sum_{\mathbf{y} \in {\cal N}^{\mathbf{z}}}^{}{p_{\mathbf{y}}^{\mathbf{z}}} \|\mathbf{y}\|^{-\frac{\alpha}{2}} f_{\mathbf{y}\mathbf{z}}^\mathbf{z} {\bar{\mathbf{s}}_{\mathbf{y}}^\mathbf{z}}+\sum_{\mathbf{x} \in \Phi_\text{p}\backslash \left\{\mathbf{z}\right\}}^{}\sum_{\mathbf{y} \in {\cal N}^{\mathbf{x}}}^{} {p_{\mathbf{y}}^{\mathbf{x}}}\times\nonumber\\& \|\mathbf{x}+\mathbf{y}-\mathbf{z}\|^{-\frac{\alpha}{2}} f_{\mathbf{y}\mathbf{z}}^\mathbf{x} {\bar{\mathbf{s}}_{\mathbf{y}}^\mathbf{x}} + \mathbf{n}_\text{u}^\mathbf{z}, \ \bar{\mathbf{s}}_{\mathbf{y}}^\mathbf{x} = \left\{{\bar{\mathbf{g}}_{\mathbf{y}}^\mathbf{x}},{\bar{\mathbf{w}}_{\mathbf{y}}^\mathbf{x}}\right\},
\end{align}
where the first term is the useful signal, the second is the inter-cluster interference and $\mathbf{n}_\text{u}^\mathbf{z} \in \mathbb{C}^{d \times 1}$ is the additive white Gaussian noise (AWGN) at the receiver of the server with zero mean and variance $\sigma_\text{en}^2$ for each entry.

Each device $\mathbf{y} \in {\cal N}^\mathbf{x}$ of cluster $\mathbf{x}$ follows a truncated power allocation \cite{huang_analog} as
\begin{align}
\label{poweralloc}
p_\mathbf{y}^{\mathbf{x}} = \left\{\begin{matrix}
\frac{\sqrt{\rho}}{\|\mathbf{y}\|^{-\frac{\alpha}{2}}f_{\mathbf{y}\mathbf{x}}^{\mathbf{x}}}\ &|f_{\mathbf{y}\mathbf{x}}^{\mathbf{x}}|^2 \geq \text{th}_1,\\
0\ &|f_{\mathbf{y}\mathbf{x}}^{\mathbf{x}}|^2 < \text{th}_1,
\end{matrix}\right.
\end{align}
where $\rho$ is the power allocation parameter and $\text{th}_1$ is a threshold. We assume that the device knows this channel, the uplink channel to its server. In \eqref{poweralloc}, devices with deep fades do not transmit but the channel pathloss is not included in the conditions. By enabling the inclusion of devices with high pathloss, the learning process can ensure fair device deployment and leverage data diversity from all devices \cite{huang_analog, huang_sg}. In \eqref{poweralloc}, to meet a maximum average power $P_\text{u}$ in each device, we have
\begin{align}
&\mathbb{E}\left\{|p_\mathbf{y}^\mathbf{x}|^2\right\} = \mathbb{E}\left\{\frac{\rho}{\|\mathbf{y}\|^{-\alpha}|f_{\mathbf{y}\mathbf{x}}^{\mathbf{x}}|^2}\right\} \nonumber\\
&= \rho \mathbb{E}\left\{\left.\frac{1}{|f_{\mathbf{y}\mathbf{x}}^{\mathbf{x}}|^2}\right \vert |f_{\mathbf{y}\mathbf{x}}^{\mathbf{x}}|^2 > \text{th}_1 \right\} \mathbb{E}\left\{\|\mathbf{y}\|^\alpha\right\} = \rho \text{Ei}(\text{th}_1)\times \nonumber
\end{align}
\begin{align}
&\int_{r_0}^{R} \frac{2y^{1+\alpha}}{R^2-r_0^2} \mathrm{d}y = \frac{2\rho}{2+\alpha} \text{Ei}(\text{th}_1) \frac{R^{\alpha+2}-r_0^{\alpha+2}}{R^2-r_0^2} \leq P_\text{u},
\end{align}
where $\text{Ei}(x) =\int_{x}^{\infty}\frac{e^{-t}}{t}\mathrm{d}t$ is the exponential integral function. Thus, $\rho$ for all the devices can be selected as
\begin{align}
\rho = \frac{(2+\alpha)(R^2-r_0^2)}{2 \text{Ei}(\text{th}_1)(R^{\alpha+2}-r_0^{\alpha+2})}P_\text{u}.
\end{align}
In each time slot, the activity set in the cluster $\mathbf{x}$ is defined as ${\cal A}^{\mathbf{x}} = \left\{\mathbf{y} \in {{\cal N}^\mathbf{x}}: |f_{\mathbf{y}\mathbf{x}}^{\mathbf{x}}|^2 \geq \text{th}_1\right\}$, whose cardinality $|{\cal A}^\mathbf{x}|$ has the binomial distribution with probability $\mathbb{P}\left\{|f_{\mathbf{y}\mathbf{x}}^{\mathbf{x}}|^2\geq \text{th}_1\right\} = e^{-\text{th}_1}$ and $|{\cal A}^\mathbf{x}|, \forall \mathbf{x}$ are independent. If ${\cal A}^{\mathbf{x}}$ is found to be empty, the device $\mathbf{y} \in {\cal N}^\mathbf{x}$ with the highest $|f_{\mathbf{y}\mathbf{x}}^{\mathbf{x}}|$ is selected to transmit at $P_\text{u}$, and we consider $|{\cal A}^\mathbf{x}| = 1$. However, for the sake of simplicity and to gain more insight into the analytical results, we make the assumption that the probability of the emptiness event, which is equal to $(1-e^{-\text{th}_1})^M$, is sufficiently small that it can be disregarded and not included as a condition in the sequel analytical derivations for \eqref{upper_error_derive} and \eqref{intraactive}-\eqref{udb}.

\vspace{-5pt}
\subsection{Downlink for intra-cluster iteration} 
The downlink transmission happens parallel at each edge server, via bandwidth limited analog broadcast. As $\mathbb{E}\left\{\mathbf{v}_\text{u}^{\mathbf{x}}\right\} = \mathbf{0}, \forall \mathbf{x}$, each server at a location $\mathbf{x} \in \Phi_\text{p}$ normalizes its received signal $\mathbf{v}_\text{u}^{\mathbf{x}}$ with its variance, which is $\mathbb{E}\left\{\|\mathbf{v}_\text{u}^{\mathbf{x}}\|^2\right\}$, as $\frac{\mathbf{v}_\text{u}^{\mathbf{x}}}{\sqrt{\mathbb{E}\left\{\|\mathbf{v}_\text{u}^{\mathbf{x}}\|^2\right\}}}$. Then, all the servers transmit the normalized
signals simultaneously. Therefore, the received signal at a reference device at $\mathbf{y}_0$ in the reference cluster $\mathbf{o}$ \footnote{The locations of the reference device and cluster can be anywhere in $\mathbb{R}^2$. The origin $\mathbf{o}$ and $\mathbf{y}_0$ are relatively determined.} is
\begin{align}
	\label{intradownlink}
	\mathbf{v}_{\text{d}_{\mathbf{y}_0}}^{\mathbf{o}} = \sum_{\mathbf{x} \in \Phi_\text{p}}^{} \sqrt{\frac{{P_{\text{d}}}}{{\mathbb{E}\left\{\|\mathbf{v}_\text{u}^{\mathbf{x}}\|^2\right\}}}} \|\mathbf{x}+\mathbf{y}_0\|^{-\frac{\alpha}{2}} f^\mathbf{x} {\mathbf{v}_\text{u}^{\mathbf{x}}} + \mathbf{n}_\text{d},
\end{align}
where $P_{\text{d}}$ is the transmission power constraint of the servers, and $\mathbf{n}_\text{d} \in \mathbb{C}^{d \times 1}$ is the AWGN at the device with zero mean and variance $\sigma_\text{dn}^2$ for each entry.\footnote{The error-free downlink is a special case of this work. For that, in the sequel, all downlink error terms should be set to zero.} In general, the server can estimate $\mathbb{E}\left\{\|\mathbf{v}_\text{u}^\mathbf{x}\|^2\right\}$ by taking measurements of the received signal over time and its entries and calculating the average power of those samples. However, the MCP-H modeling allows us to express $\mathbb{E}\left\{\|\mathbf{v}_\text{u}^\mathbf{x}\|^2\right\}$  as a function of the network parameters and the power control. Specifically, from \eqref{uplink} and \eqref{poweralloc}
\begin{align}
\mathbb{E}\left\{\|\mathbf{v}_\text{u}^\mathbf{x}\|^2\right\} = \rho|{\cal A}^\mathbf{x}|+\Psi,
\end{align}
where $\Psi = $
\begin{align}
\label{upper_error_derive}
&\rho \mathbb{E}\left\{\sum_{\mathbf{x}\in \Phi_\text{p}\backslash \left\{\mathbf{o}\right\}}\sum_{\mathbf{y} \in {\cal N}^{\mathbf{x}}} \mathds{1}\left(|f_{\mathbf{y}\mathbf{x}}^{\mathbf{x}}|^2>\text{th}_1\right) \frac{\|\mathbf{x}+\mathbf{y}\|^{-{\alpha}}}{\|\mathbf{y}\|^{-{\alpha}}} \left|\frac{f_{{\mathbf{y}\mathbf{o}}}^\mathbf{x}}{f_{\mathbf{y}\mathbf{x}}^{\mathbf{x}}}\right|^2\right\}\nonumber\\
&+\sigma_\text{en}^2= \rho\mathbb{E}\Biggl\{\sum_{\mathbf{x}\in \Phi_\text{p}}\sum_{\mathbf{y} \in {\cal N}^{\mathbf{x}}} \mathds{1}\left(|{f}_{\mathbf{y}\mathbf{x}}^{\mathbf{x}}|^2>\text{th}_1\right)\mathbb{E}\biggl\{\biggl.\frac{|f_{\mathbf{y}\mathbf{o}}^\mathbf{x}|^2}{|f_{\mathbf{y}\mathbf{x}}^{\mathbf{x}}|^2}\biggr\vert \nonumber\\
& |{f}_{\mathbf{y}\mathbf{x}}^{\mathbf{x}}|^2>\text{th}_1\biggr\} \frac{\left(\|\mathbf{x}\|^2+\|\mathbf{y}\|^2-2\|\mathbf{x}\|\|\mathbf{y}\|\cos(\theta_\mathbf{xy})\right)^{-\frac{\alpha}{2}}}{\|\mathbf{y}\|^{-{\alpha}}} \Biggr\}\nonumber\\
&+\sigma_\text{en}^2 =\rho \text{Ei}(\text{th}_1) \mathbb{E}\Biggl\{\sum_{\mathbf{x}\in \Phi_\text{p}}\sum_{m=1}^{M} {M \choose m} e^{-\text{th}_1  m} \left(1-e^{-\text{th}_1}\right)^{M-m}\nonumber
\end{align}
\begin{align}
&m\int_{0}^{\infty}  \left(1+\frac{\|\mathbf{x}\|^2}{y^2}-2\frac{\|\mathbf{x}\|}{y}\cos(\theta_{\mathbf{xy}})\right)^{-\frac{\alpha}{2}} \hspace{-6pt}f_{\|\mathbf{y}\|}(y)\mathrm{d}y\Biggr\}+\sigma_\text{en}^2 \nonumber\\ &\stackrel{(a)}{=}\rho \text{Ei}(\text{th}_1) Me^{-\text{th}_1}  \lambda_\text{p} \int_{0}^{\infty}\int_{0}^{2\pi}\int_{0}^{\infty}  \biggl(1+\frac{x^2}{y^2}-\frac{2x}{y}\cos(\theta)\nonumber\\&\biggr)^{-\frac{\alpha}{2}} f_{\|\mathbf{y}\|}(y)x \mathrm{d}y \mathrm{d}\theta \mathrm{d}x+\sigma_\text{en}^2 \stackrel{(b)}{=} \frac{2M \rho \lambda_\text{p} \text{Ei}(\text{th}_1)e^{-\text{th}_1}}{R^2-r_0^2}\times\nonumber\\&\int_{2r_0}^{\infty}\int_{0}^{2\pi}\int_{r_0}^{R}  y\left(1+\frac{x^2}{y^2}-2\frac{x}{y}\cos(\theta)\right)^{-\frac{\alpha}{2}} x \mathrm{d}y \mathrm{d}\theta \mathrm{d}x+\sigma_\text{en}^2,
\end{align}
where $\mathbb{E}\biggl\{\left.\frac{|f_{{\mathbf{y}\mathbf{o}}}^\mathbf{x}|^2}{|f_{\mathbf{y}\mathbf{x}}^{\mathbf{x}}|^2}\right\vert |{f}_{\mathbf{y}\mathbf{x}}^{\mathbf{x}}|^2>\text{th}_1\biggr\} = \mathbb{E}\left\{|f_{\mathbf{y}\mathbf{o}}^\mathbf{x}|^2\right\}\mathbb{E}\bigl\{\left.\frac{1}{|f_{\mathbf{y}\mathbf{x}}^{\mathbf{x}}|^2}\right\vert |{f}_{\mathbf{y}\mathbf{x}}^{\mathbf{x}}|^2>\text{th}_1\bigr\} = \text{Ei}(\text{th}_1)$. Also, $(a)$ comes from the Campbell's theorem \cite{haenggi_book} and $(b)$ is due to the fact that edge servers
have at least $2r_0$ distance from each other.

Denormalizing received signal, the reference device estimates the intra-cluster gradient \eqref{intragrad} as
\begin{align}
\label{intraestimate1}
{\mathbf{g}}_{\mathbf{y}_0}^\mathbf{o} &= \frac{ \vartheta_{\text{d}_{\mathbf{y}_0}}^{\mathbf{o}} \mathbf{v}_{\text{d}_{\mathbf{y}_0}}^{\mathbf{o}}}{\sqrt{\rho} \sqrt{\frac{{P_{\text{d}}}}{{\mathbb{E}\left\{\|\mathbf{v}_\text{u}^{\mathbf{o}}\|^2\right\}}}}|{\cal A}^\mathbf{o}|f^{\mathbf{o}}\|\mathbf{y}_0\|^{-\frac{\alpha}{2}}} +\frac{1}{|{\cal A}^\mathbf{o}|}\sum_{\mathbf{y} \in {\cal A}^\mathbf{o}} \mu_{\text{g},\mathbf{y}}^\mathbf{o}\mathbf{1},
\end{align}
where $\vartheta_{\text{d}_{\mathbf{y}_0}}^{\mathbf{o}}$ is the intra-cluster receive normalizing factor at the device. For this operation, it is assumed that each device knows its downlink channel from its server \footnote{It can be estimated by an initial training phase before the transmission.} and the reference server shares the scalars $({\mu_{\text{g},\mathbf{y}}^\mathbf{o}}, {\sigma_{\text{g},\mathbf{y}}^\mathbf{o}}), \forall \mathbf{y} \in {\cal A}^\mathbf{o}$ with its devices in an error-free manner. This is needed to support data heterogeneity. This information is however small compared to the gradients, and needs to be shared within a single cluster only. If the downlink channel $|f^\mathbf{o}|$ is lower than a threshold $\text{th}_0$, the device does not update its local model, and will not contribute to the present inter-cluster iteration \footnote{Due to the high elevation, powerful transmitter, and advanced signal processing capability, each server typically provide strong downlink channels, specially within its cluster. It is rare for the event $|f^\mathbf{o}| < \text{th}_0$ to occur, assuming that $\text{th}_0$ is sufficiently small.}. However, retransmission strategies can be utilized in such case. By replacing \eqref{uplink} in \eqref{intradownlink} and expanding the result, \eqref{intraestimate1} can be rewritten as 
\begin{align}
\label{intraestimate}
&{\mathbf{g}}_{\mathbf{y}_0}^\mathbf{o} = \frac{1}{|{\cal A}^\mathbf{o}|} \sum_{\mathbf{y}\in {\cal A}^\mathbf{o}}^{} {\mathbf{g}}_{\mathbf{y}}^\mathbf{o}+\frac{ \vartheta_{\text{d}_{\mathbf{y}_0}}^{\mathbf{o}} \mathbf{v}_{\text{d}_{\mathbf{y}_0}}^{\mathbf{o}}}{\sqrt{\rho} \sqrt{\frac{{P_{\text{d}}}}{{\mathbb{E}\left\{\|\mathbf{v}_\text{u}^{\mathbf{o}}\|^2\right\}}}}|{\cal A}^\mathbf{o}|f^{\mathbf{o}}\|\mathbf{y}_0\|^{-\frac{\alpha}{2}}} -\nonumber\\&\frac{1}{|{\cal A}^\mathbf{o}|}\sum_{\mathbf{y} \in {\cal A}^\mathbf{o}} \left({\mathbf{g}}_{\mathbf{y}}^\mathbf{o}-\mu_{\text{g},\mathbf{y}}^\mathbf{o}\mathbf{1}\right)=\frac{1}{|{\cal A}^\mathbf{o}|} \sum_{\mathbf{y}\in {\cal A}^\mathbf{o}}^{} {\mathbf{g}}_{\mathbf{y}}^\mathbf{o}+\frac{\vartheta_{\text{d}_{\mathbf{y}_0}}^{\mathbf{o}}}{|{\cal A}^\mathbf{o}|}\sum_{\mathbf{y}\in {\cal A}^\mathbf{o}}^{} \bar{\mathbf{g}}_{\mathbf{y}}^\mathbf{o}\nonumber\\
&+ \frac{ \vartheta_{\text{d}_{\mathbf{y}_0}}^{\mathbf{o}}}{ |{\cal A}^{\mathbf{o}}|}\sum_{\mathbf{x}\in \Phi_\text{p}}\sum_{\mathbf{y} \in {\cal N}^{\mathbf{x}}} \mathds{1}\left(|f_{\mathbf{y}\mathbf{x}}^{\mathbf{x}}|^2>\text{th}_1\right) \frac{\|\mathbf{x}+\mathbf{y}\|^{-\frac{\alpha}{2}}}{\|\mathbf{y}\|^{-\frac{\alpha}{2}}} \frac{f_{\mathbf{y}\mathbf{o}}^\mathbf{x}}{f_{\mathbf{y}\mathbf{x}}^{\mathbf{x}}}\bar{\mathbf{g}}_{\mathbf{y}}^\mathbf{x} +\nonumber\\
&\frac{\vartheta_{\text{d}_{\mathbf{y}_0}}^{\mathbf{o}}}{\sqrt{\rho}|{\cal A}^\mathbf{o}|} \mathbf{n}_\text{u}^\mathbf{o} +\frac{\vartheta_{\text{d}_{\mathbf{y}_0}}^{\mathbf{o}}}{\sqrt{\rho}|{\cal A}^\mathbf{o}| \sqrt{\frac{{P_{\text{d}}}}{{\mathbb{E}\left\{\|\mathbf{v}_\text{u}^{\mathbf{o}}\|^2\right\}}}}f^{\mathbf{o}}\|\mathbf{y}_0\|^{-\frac{\alpha}{2}}} \times\nonumber\\
&\sum_{\mathbf{x} \in \Phi_\text{p}}^{} \sqrt{\frac{{P_{\text{d}}}}{{\mathbb{E}\left\{\|\mathbf{v}_\text{u}^{\mathbf{x}}\|^2\right\}}}} \|\mathbf{x}+\mathbf{y}_0\|^{-\frac{\alpha}{2}} f^\mathbf{x} {\mathbf{v}_\text{u}^{\mathbf{x}}}+\nonumber
\end{align}
\begin{align}
&\frac{\vartheta_{\text{d}_{\mathbf{y}_0}}^{\mathbf{o}} \mathbf{n}_\text{d}}{\sqrt{\rho}|{\cal A}^\mathbf{o}| \sqrt{\frac{{P_{\text{d}}}}{{\mathbb{E}\left\{\|\mathbf{v}_\text{u}^{\mathbf{o}}\|^2\right\}}}}|f^{\mathbf{o}}\|\mathbf{y}_0\|^{-\frac{\alpha}{2}}}-\frac{1}{|{\cal A}^\mathbf{o}|}\sum_{\mathbf{y} \in {\cal A}^\mathbf{o}}\sigma_{\text{g},\mathbf{y}}^\mathbf{o} \bar{\mathbf{g}}_{\mathbf{y}}^\mathbf{o} \nonumber\\
&=\frac{1}{|{\cal A}^\mathbf{o}|} \sum_{\mathbf{y}\in {\cal A}^\mathbf{o}}^{} {\mathbf{g}}_{\mathbf{y}}^\mathbf{o}+ \frac{ \boldsymbol{\epsilon}_\text{u}^\mathbf{o}}{ |{\cal A}^{\mathbf{o}}|} +\frac{\boldsymbol{\epsilon}_{\text{d}_{\mathbf{y}_0}}^{\mathbf{o}}}{|{\cal A}^\mathbf{o}|},
\end{align}
where $\boldsymbol{\epsilon}_\text{u}^\mathbf{o}$ is the intra-cluster uplink error given by  
\begin{align}
\label{intrauplinkerror}
&\boldsymbol{\epsilon}_\text{u}^\mathbf{o} =\sum_{\mathbf{y} \in {\cal A}^\mathbf{o}}\left(\vartheta_{\text{d}_{\mathbf{y}_0}}^{\mathbf{o}}-\sigma_{\text{g},\mathbf{y}}^\mathbf{o}\right)\bar{\mathbf{g}}_{\mathbf{y}}^\mathbf{o}+\vartheta_{\text{d}_{\mathbf{y}_0}}^{\mathbf{o}}\sum_{\mathbf{x}\in \Phi_\text{p}}\sum_{\mathbf{y} \in {\cal N}^{\mathbf{x}}}\nonumber\\& \mathds{1}\left(|f_{\mathbf{y}\mathbf{x}}^{\mathbf{x}}|^2>\text{th}_1\right) \frac{\|\mathbf{x}+\mathbf{y}\|^{-\frac{\alpha}{2}}}{\|\mathbf{y}\|^{-\frac{\alpha}{2}}} \frac{f_{\mathbf{y}\mathbf{o}}^\mathbf{x}}{f_{\mathbf{y}\mathbf{x}}^{\mathbf{x}}}\bar{\mathbf{g}}_{\mathbf{y}}^\mathbf{x} +\frac{\vartheta_{\text{d}_{\mathbf{y}_0}}^{\mathbf{o}}}{\sqrt{\rho}} \mathbf{n}_\text{u}^\mathbf{o},
\end{align}
and the intra-cluster downlink error $\boldsymbol{\epsilon}_{\text{d}_{\mathbf{y}_0}}^{\mathbf{o}}$ is
\begin{align}
\label{intradownlinkerror}
&\boldsymbol{\epsilon}_{\text{d}_{\mathbf{y}_0}}^{\mathbf{o}} = \frac{\vartheta_{\text{d}_{\mathbf{y}_0}}^{\mathbf{o}}\sum_{\mathbf{x} \in \Phi_\text{p}}^{} \sqrt{\frac{{P_{\text{d}}}}{{\mathbb{E}\left\{\|\mathbf{v}_\text{u}^{\mathbf{x}}\|^2\right\}}}} \|\mathbf{x}+\mathbf{y}_0\|^{-\frac{\alpha}{2}} f^\mathbf{x} {\mathbf{v}_\text{u}^{\mathbf{x}}}}{\sqrt{\rho} \sqrt{\frac{{P_{\text{d}}}}{{\mathbb{E}\left\{\|\mathbf{v}_\text{u}^{\mathbf{o}}\|^2\right\}}}}f^{\mathbf{o}}\|\mathbf{y}_0\|^{-\frac{\alpha}{2}}} +\nonumber\\
&\frac{\vartheta_{\text{d}_{\mathbf{y}_0}}^{\mathbf{o}} \mathbf{n}_\text{d}}{\sqrt{\rho} \sqrt{\frac{{P_{\text{d}}}}{{\mathbb{E}\left\{\|\mathbf{v}_\text{u}^{\mathbf{o}}\|^2\right\}}}}|f^{\mathbf{o}}\|\mathbf{y}_0\|^{-\frac{\alpha}{2}}}.
\end{align}
These results hold for the learning process under general network topology. For the specific case of the MCP-H, we can select the normalizing factor $\vartheta_{\text{d}_{\mathbf{y}_0}}^{\mathbf{o}}$ in \eqref{intraestimate1} to minimize the distortion of the recovered gradient ${\mathbf{g}}_{\mathbf{y}_0}^\mathbf{o}$ with respect to the ground true gradient $\frac{1}{|{\cal A}^\mathbf{o}|} \sum_{\mathbf{y}\in {\cal A}^\mathbf{o}}^{} {\mathbf{g}}_{\mathbf{y}}^\mathbf{o}$ from \eqref{intraestimate}, which can be measured by the mean squared error (MSE)\cite{mse, cao}, as
\begin{align}
\label{factor_opt}
&\min_{\vartheta_{\text{d}_{\mathbf{y}_0}}^{\mathbf{o}}} \mathbb{E}\left\{\left\Vert\frac{ \boldsymbol{\epsilon}_\text{u}^\mathbf{o}}{|{\cal A}^\mathbf{o}|} +\frac{\boldsymbol{\epsilon}_{\text{d}_{\mathbf{y}_0}}^{\mathbf{o}}}{|{\cal A}^\mathbf{o}|}\right\Vert^2\right\} \nonumber\\&=\frac{1}{|{\cal A}^\mathbf{o}|^2}\left(\mathbb{E}\left\{\|\boldsymbol{\epsilon}_\text{u}^\mathbf{o}\|^2\right\}+\mathbb{E}\left\{\|\boldsymbol{\epsilon}_{\text{d}_{\mathbf{y}_0}}^\mathbf{o}\|^2\right\}\right),
\end{align}
where the equality holds due to the independent error terms and $\mathbb{E}\left\{\|\boldsymbol{\epsilon}_\text{u}^\mathbf{o}\|^2\right\} = \sum_{\mathbf{y} \in {\cal A}^\mathbf{o}}\bigl(\vartheta_{\text{d}_{\mathbf{y}_0}}^{\mathbf{o}}-\sigma_{\text{g},\mathbf{y}}^\mathbf{o}\bigr)^2+\frac{{\vartheta_{\text{d}_{\mathbf{y}_0}}^{\mathbf{o}\ \hspace{-2pt}2}}}{\rho} \Psi$ from \eqref{uplink} and \eqref{intrauplinkerror}, where $\Psi$ is calculated in \eqref{upper_error_derive}. Then, for the expected term on the intra-cluster downlink error in \eqref{intradownlinkerror}, due to the MCP-H network topology, we have
\begin{align}
\label{downlink_error_derive}
&\mathbb{E}\left\{\|\boldsymbol{\epsilon}_{\text{d}_{\mathbf{y}_0}}^\mathbf{o}\|^2\right\}  = \frac{{\vartheta_{\text{d}_{\mathbf{y}_0}}^{\mathbf{o}\ \hspace{-2pt}2}}\mathbb{E}\left\{\sum_{\mathbf{x} \in \Phi_\text{p}}^{} {{{P_{\text{d}}}}} \|\mathbf{x}+\mathbf{y}_0\|^{-{\alpha}} |f^\mathbf{x}|^2\right\}}{{\rho} {\frac{{P_{\text{d}}}}{{\mathbb{E}\left\{\|\mathbf{v}_\text{u}^{\mathbf{o}}\|^2\right\}}}}|f^{\mathbf{o}}|^2\|\mathbf{y}_0\|^{-{\alpha}}}  +\nonumber\\&\frac{ {\vartheta_{\text{d}_{\mathbf{y}_0}}^{\mathbf{o}\ \hspace{-2pt}2}}\sigma_\text{dn}^2}{{\rho} {\frac{{P_{\text{d}}}}{{\mathbb{E}\left\{\|\mathbf{v}_\text{u}^{\mathbf{o}}\|^2\right\}}}}|f^{\mathbf{o}}|^2\|\mathbf{y}_0\|^{-{\alpha}}} \stackrel{(c)}{=}\frac{{\vartheta_{\text{d}_{\mathbf{y}_0}}^{\mathbf{o}\ \hspace{-2pt}2}}\mathbb{E}\left\{\|\mathbf{v}_\text{u}^\mathbf{o}\|^2\right\}}{\rho|f^\mathbf{o}|^2\|\mathbf{y}_0\|^{-\alpha}}  \biggl(\sigma_\text{d}^2\times 2\pi \lambda_\text{p} \int_{r_0}^{\infty} \nonumber\\&x^{1-\alpha}\mathrm{d}x + \frac{\sigma_\text{dn}^2}{P_\text{d}}\biggr)=  \frac{{\vartheta_{\text{d}_{\mathbf{y}_0}}^{\mathbf{o}\ \hspace{-2pt}2}}}{\rho|f^\mathbf{o}|^2\|\mathbf{y}_0\|^{-\alpha}} \left(\frac{2\pi \lambda_\text{p}\sigma_\text{d}^2}{(\alpha-2)r_0^{\alpha-2}}+\frac{\sigma_\text{dn}^2}{P_\text{d}}\right) \times \nonumber\\&(\rho|{\cal A}^\mathbf{o}|+\Psi),
\end{align}
where $(c)$ is due to the Campbell's theorem. To solve \eqref{factor_opt}, we take derivative from the objective and set the result to zero, which leads to
%\begin{align}
%\label{approach}
%&\sum_{\mathbf{y} \in {\cal %A}^\mathbf{o}}\left(\vartheta_{\text{d}_{\mathbf{y}_%0}}^{\mathbf{o}}-\sigma_{\text{g},\mathbf{y}}^\mathb%f{o}\right) + %\frac{{\vartheta_{\text{d}_{\mathbf{y}_0}}^{\mathbf{%o}}}}{\rho} \Psi + %\frac{2{\vartheta_{\text{d}_{\mathbf{y}_0}}^{\mathbf%{o}}}(R^{\alpha+2}-r_0^{\alpha+2})\text{Ei}(\text{th%}_0)}{\rho(\alpha+2)(R^2-r_0^2)}  \left(\frac{2\pi %\lambda_\text{p}}{(\alpha-2)r_0^{\alpha-2}}+\frac{\s%igma_\text{n}^2}{P_\text{d}}\right) (\rho|{\cal %A}^\mathbf{o}|+\Psi)\nonumber\\&=0,
%\end{align}
\begin{align}
\label{intra_dowlik_norm}
\vartheta_{\text{d}_{\mathbf{y}_0}}^{\mathbf{o}} =\frac{\sum_{\mathbf{y} \in {\cal A}^\mathbf{o}}^{} \sigma_{\text{g},\mathbf{y}}^\mathbf{o}}{\left(1+\frac{\beta}{|f^\mathbf{o}|^2\|\mathbf{y}_0\|^{-\alpha}}\right)\left(|{\cal A}^\mathbf{o}|+\frac{\Psi}{\rho}\right)},
\end{align}
where 
\begin{align}
\label{beta_def}
\beta = \frac{2\pi \lambda_\text{p}\sigma_\text{d}^2}{(\alpha-2)r_0^{\alpha-2}}+\frac{\sigma_\text{dn}^2}{P_\text{d}}.
\end{align}

\vspace{-10pt}
\subsection{Downlink for inter-cluster iteration}
The core server sums and redistributes the signals received
from any set of collaborating edge servers without introducing further error. Consider $C_\mathbf{x}$ collaborating clusters having the same learning task with a cluster $\mathbf{x}$, denoted as the set ${\mathcal{C}}_{\mathbf{x}}$. Then, the sum of received signals of the clusters in ${\mathcal{C}}_{\mathbf{x}}$, i.e., $\sum_{\mathbf{z} \in {\mathcal{C}}_\mathbf{x}}^{}\mathbf{v}_\text{u}^{\mathbf{z}}$, is normalized with its variance, which is $\sum_{\mathbf{z} \in {\mathcal{C}}_\mathbf{x}}^{}\mathbb{E}\left\{\|\mathbf{v}_\text{u}^{\mathbf{z}}\|^2\right\}$, as $\frac{\sum_{\mathbf{z} \in {\mathcal{C}}_\mathbf{x}}^{}\mathbf{v}_\text{u}^{\mathbf{z}}}{\sqrt{\sum_{\mathbf{z} \in {\mathcal{C}}_\mathbf{x}}^{}\mathbb{E}\left\{\|\mathbf{v}_\text{u}^{\mathbf{z}}\|^2\right\}}}$. Then, the result is simultaneously transmitted from the servers of the clusters in ${\mathcal{C}}_\mathbf{x}$ to their devices. Therefore, the received signal at the reference device is
\begin{align}
	\label{interdownlink}
	\mathbf{v}_{\text{d}_{\mathbf{y}_0}} &= \sum_{\mathbf{x} \in \Phi_\text{p}}^{} \sqrt{\frac{{P_{\text{d}}}}{\sum_{\mathbf{z} \in {\mathcal{C}}_\mathbf{x}}^{}\mathbb{E}\left\{\|\mathbf{v}_\text{u}^{\mathbf{z}}\|^2\right\}}} \|\mathbf{x}+\mathbf{y}_0\|^{-\frac{\alpha}{2}} f^\mathbf{x} \sum_{\mathbf{z} \in {\mathcal{C}}_\mathbf{x}}^{}{\mathbf{v}_\text{u}^{\mathbf{z}}} \nonumber\\&+ \mathbf{n}_\text{d}.
\end{align}
Then, the reference device can estimate the inter-cluster model parameter vector \eqref{globalgrad} as
\begin{align}
\label{interestimate1}
{\mathbf{w}}_{\mathbf{y}_0}^\mathbf{o} &= \frac{\vartheta_{\text{d}_{\mathbf{y}_0}} \mathbf{v}_{\text{d}_{\mathbf{y}_0}}}{\sqrt{\rho}\sqrt{\frac{{P_{\text{d}}}}{\sum_{\mathbf{x} \in {\mathcal{C}}}^{}\mathbb{E}\left\{\|\mathbf{v}_\text{u}^{\mathbf{x}}\|^2\right\}}} f^{\mathbf{o}}\|\mathbf{y}_0\|^{-\frac{\alpha}{2}}|{\cal A}|} \nonumber\\&+\frac{1}{|{\cal A}|}\sum_{\mathbf{x} \in {\cal C}}\sum_{\mathbf{y} \in {\cal A}^\mathbf{x}} \mu_{\text{w},\mathbf{y}}^\mathbf{x}\mathbf{1},
\end{align}
where due to the symmetry of the network $\sum_{\mathbf{x} \in {\mathcal{C}}}^{}\mathbb{E}\left\{\|\mathbf{v}_\text{u}^{\mathbf{x}}\|^2\right\} = \rho|{\cal A}|+C{\Psi}$ and $|{\cal A}| = \sum_{\mathbf{x}\in {\mathcal{C}}}^{}|{\cal A}^\mathbf{x}|$. Also, $\vartheta_{\text{d}_{\mathbf{y}_0}}$ is the inter-cluster receive normalizing factor. We assume that the reference edge server receives scalars $( {\mu_{\text{w},\mathbf{y}}^\mathbf{x}}, {\sigma_{\text{w},\mathbf{y}}^\mathbf{x}}), \forall \mathbf{x} \in {\cal C}, \mathbf{y}\in {{\cal A}^\mathbf{x}} $ from its core server and shares them among its devices. 

After replacing \eqref{uplink} in \eqref{interdownlink}, \eqref{interestimate1} can be expanded as
\begin{align}
\label{interestimate}
&{\mathbf{w}}_{\mathbf{y}_0}^\mathbf{o} = \frac{1}{|{\cal A}|} \sum_{\mathbf{x}\in {\mathcal{C}}}^{} \hspace{-1pt}\sum_{\mathbf{y}\in {\cal A}^{\mathbf{x}}}^{} {\mathbf{w}}_{\mathbf{y}}^\mathbf{x}+\hspace{-2pt} \frac{\vartheta_{\text{d}_{\mathbf{y}_0}} \mathbf{v}_{\text{d}_{\mathbf{y}_0}}}{\sqrt{\rho}\sqrt{\frac{{P_{\text{d}}}}{\sum_{\mathbf{x} \in {\mathcal{C}}}^{}\mathbb{E}\left\{\|\mathbf{v}_\text{u}^{\mathbf{x}}\|^2\right\}}} f^{\mathbf{o}}\|\mathbf{y}_0\|^{-\frac{\alpha}{2}}|{\cal A}|}\nonumber\\&-\frac{1}{|{\cal A}|} \sum_{\mathbf{x}\in {\mathcal{C}}}^{} \sum_{\mathbf{y}\in {\cal A}^{\mathbf{x}}}^{} \left({\mathbf{w}}_{\mathbf{y}}^\mathbf{x}-\mu_{\text{w},\mathbf{y}}^\mathbf{x}\mathbf{1}\right)=\frac{1}{|{\cal A}|} \sum_{\mathbf{x}\in {\mathcal{C}}}^{} \sum_{\mathbf{y}\in {\cal A}^{\mathbf{x}}}^{} {\mathbf{w}}_{\mathbf{y}}^\mathbf{x}+\frac{\vartheta_{\text{d}_{\mathbf{y}_0}}}{|{\cal A}|}\nonumber\\& \sum_{\mathbf{x}\in {\mathcal{C}}}^{} \sum_{\mathbf{y}\in {\cal A}^{\mathbf{x}}}^{} \bar{\mathbf{w}}_{\mathbf{y}}^\mathbf{x}+ \frac{\vartheta_{\text{d}_{\mathbf{y}_0}}}{|\cal A|} \sum_{\mathbf{z} \in {\mathcal{C}}}^{}\sum_{\mathbf{x}\in \Phi_\text{p}}\sum_{\mathbf{y} \in {\cal N}^{\mathbf{x}}} \mathds{1}\left(|f_{\mathbf{y}\mathbf{x}}^{\mathbf{x}}|^2>\text{th}_1\right)\nonumber\\& \frac{\|\mathbf{x}+\mathbf{y}-\mathbf{z}\|^{-\frac{\alpha}{2}}}{\|\mathbf{y}\|^{-\frac{\alpha}{2}}} \frac{f_{\mathbf{y}\mathbf{z}}^\mathbf{x}}{f_{\mathbf{y}\mathbf{x}}^{\mathbf{x}}}\bar{\mathbf{w}}_{\mathbf{y}}^\mathbf{x} +\frac{\vartheta_{\text{d}_{\mathbf{y}_0}}\sum_{\mathbf{z}\in {\mathcal{C}}}^{}\mathbf{n}_\text{u}^\mathbf{z}}{\sqrt{\rho}|\cal A|} +\nonumber\\&\frac{\vartheta_{\text{d}_{\mathbf{y}_0}}\sum_{\mathbf{x} \in \Phi_\text{p}}^{} \sqrt{\frac{{P_{\text{d}}}}{{\sum_{\mathbf{z} \in {\mathcal{C}}_\mathbf{x}}^{}\mathbb{E}\left\{\|\mathbf{v}_\text{u}^{\mathbf{z}}\|^2\right\}}}} \|\mathbf{x}+\mathbf{y}_0\|^{-\frac{\alpha}{2}} f^\mathbf{x} \sum_{\mathbf{z} \in {\mathcal{C}}_\mathbf{x}}^{}{\mathbf{v}_\text{u}^{\mathbf{z}}}}{\sqrt{\rho}|{\cal A}| \sqrt{\frac{{P_{\text{d}}}}{{\sum_{\mathbf{x} \in {\mathcal{C}}}^{}\mathbb{E}\left\{\|\mathbf{v}_\text{u}^{\mathbf{x}}\|^2\right\}}}}f^{\mathbf{o}}\|\mathbf{y}_0\|^{-\frac{\alpha}{2}}} +\nonumber\\
&\frac{\vartheta_{\text{d}_{\mathbf{y}_0}} \mathbf{n}_\text{d}}{\sqrt{\rho}|{\cal A}| \sqrt{\frac{{P_{\text{d}}}}{{\sum_{\mathbf{x} \in {\mathcal{C}}}^{}\mathbb{E}\left\{\|\mathbf{v}_\text{u}^{\mathbf{x}}\|^2\right\}}}}|f^{\mathbf{o}}\|\mathbf{y}_0\|^{-\frac{\alpha}{2}}}-\frac{1}{|{\cal A}|} \sum_{\mathbf{x}\in {\mathcal{C}}}^{} \sum_{\mathbf{y}\in {\cal A}^{\mathbf{x}}}^{} \sigma_{\text{w},\mathbf{y}}^\mathbf{x} \bar{\mathbf{w}}_\mathbf{y}^\mathbf{x} \nonumber\\&= \frac{1}{|{\cal A}|} \sum_{\mathbf{x}\in {\mathcal{C}}}^{} \sum_{\mathbf{y}\in {\cal A}^{\mathbf{x}}}^{} {\mathbf{w}}_{\mathbf{y}}^\mathbf{x} + \frac{\boldsymbol{\epsilon}_\text{u}}{|\cal A|}+\frac{\boldsymbol{\epsilon}_{\text{d}_{\mathbf{y}_0}}}{|\cal A|},
\end{align}
where $\boldsymbol{\epsilon}_\text{u}$ is the inter-cluster uplink error as  
\begin{align}
\label{interuplinkerror}
&\boldsymbol{\epsilon}_\text{u} = \sum_{\mathbf{x}\in {\cal C}}^{} \sum_{\mathbf{y} \in {\cal A}^\mathbf{x}}^{}\left(\vartheta_{\text{d}_{\mathbf{y}_0}}-\sigma_{\text{w},\mathbf{y}}^\mathbf{x}\right)\bar{\mathbf{w}}_\mathbf{y}^\mathbf{x}+ \vartheta_{\text{d}_{\mathbf{y}_0}} \sum_{\mathbf{z} \in {\mathcal{C}}}^{}\sum_{\mathbf{x}\in \Phi_\text{p}}\sum_{\mathbf{y} \in {\cal N}^{\mathbf{x}}} \mathds{1}\nonumber\\&\left(|f_{\mathbf{y}\mathbf{x}}^{\mathbf{x}}|^2>\text{th}_1\right) \frac{\|\mathbf{x}+\mathbf{y}-\mathbf{z}\|^{-\frac{\alpha}{2}}}{\|\mathbf{y}\|^{-\frac{\alpha}{2}}} \frac{f_{\mathbf{y}\mathbf{z}}^\mathbf{x}}{f_{\mathbf{y}\mathbf{x}}^{\mathbf{x}}}\bar{\mathbf{w}}_{\mathbf{y}}^\mathbf{x} +\frac{\vartheta_{\text{d}_{\mathbf{y}_0}}}{\sqrt{\rho}} \hspace{-2pt}\sum_{\mathbf{z}\in {\mathcal{C}}}^{}\mathbf{n}_\text{u}^\mathbf{z},
\end{align}
and the inter-cluster downlink error $\boldsymbol{\epsilon}_{\text{d}_{\mathbf{y}_0}}$ is
\begin{align}
\label{interdownlinkerror}
&\boldsymbol{\epsilon}_{\text{d}_{\mathbf{y}_0}} =\nonumber\\& \frac{\vartheta_{\text{d}_{\mathbf{y}_0}}\sum_{\mathbf{x} \in \Phi_\text{p}}^{} \sqrt{\frac{{P_{\text{d}}}}{{\sum_{\mathbf{z} \in {\mathcal{C}}_\mathbf{x}}^{}\mathbb{E}\left\{\|\mathbf{v}_\text{u}^{\mathbf{z}}\|^2\right\}}}} \|\mathbf{x}+\mathbf{y}_0\|^{-\frac{\alpha}{2}} f^\mathbf{x} \sum_{\mathbf{z} \in {\mathcal{C}}_\mathbf{x}}^{}{\mathbf{v}_\text{u}^{\mathbf{z}}}}{\sqrt{\rho} \sqrt{\frac{{P_{\text{d}}}}{{\sum_{\mathbf{x} \in {\mathcal{C}}}^{}\mathbb{E}\left\{\|\mathbf{v}_\text{u}^{\mathbf{x}}\|^2\right\}}}}f^{\mathbf{o}}\|\mathbf{y}_0\|^{-\frac{\alpha}{2}}} \nonumber\\&+\frac{\vartheta_{\text{d}_{\mathbf{y}_0}} \mathbf{n}_\text{d}}{\sqrt{\rho} \sqrt{\frac{{P_{\text{d}}}}{{\sum_{\mathbf{x} \in {\mathcal{C}}}^{}\mathbb{E}\left\{\|\mathbf{v}_\text{u}^{\mathbf{x}}\|^2\right\}}}}|f^{\mathbf{o}}\|\mathbf{y}_0\|^{-\frac{\alpha}{2}}}.
\end{align}
Again, the results up to here do not depend on the network topology. For the MCP-H case, we can progress as follows. The normalizing factor $\vartheta_{\text{d}_{\mathbf{y}_0}}$ is selected to minimize the distortion of the recovered
model vector ${\mathbf{w}}_{\mathbf{y}_0}^\mathbf{o}$ with respect to the ground true model vector $\frac{1}{|{\cal A}|} \sum_{\mathbf{x}\in {\mathcal{C}}}^{} \sum_{\mathbf{y}\in {\cal A}^{\mathbf{x}}}^{} {\mathbf{w}}_{\mathbf{y}}^\mathbf{x}$ from \eqref{interestimate} as
\begin{align}
\label{globaldistortion}
&\min_{\vartheta_{\text{d}_{\mathbf{y}_0}}} \mathbb{E}\left\{\left\Vert\frac{ \boldsymbol{\epsilon}_\text{u}}{|{\cal A}|} +\frac{\boldsymbol{\epsilon}_{\text{d}_{\mathbf{y}_0}}}{|{\cal A}|}\right\Vert^2\right\} \nonumber\\&=\frac{1}{{|{\cal A}|^2}}\left(\mathbb{E}\left\{\|\boldsymbol{\epsilon}_\text{u}\|^2\right\}+\mathbb{E}\left\{\|\boldsymbol{\epsilon}_{\text{d}_{\mathbf{y}_0}}\|^2\right\}\right),
\end{align}
where due to the symmetry of the network and the error terms in \eqref{intrauplinkerror}-\eqref{intradownlinkerror} and \eqref{interuplinkerror}-\eqref{interdownlinkerror}, we have $\mathbb{E}\left\{\|\boldsymbol{\epsilon}_\text{u}\|^2\right\} =\sum_{\mathbf{x}\in {\cal C}}^{} \sum_{\mathbf{y} \in {\cal A}^\mathbf{x}}^{}\left(\vartheta_{\text{d}_{\mathbf{y}_0}}-\sigma_{\text{w},\mathbf{y}}^\mathbf{x}\right)^2+ C\frac{\vartheta_{\text{d}_{\mathbf{y}_0}}^2}{\rho}\Psi$ and $\mathbb{E}\left\{\|\boldsymbol{\epsilon}_{\text{d}_{\mathbf{y}_0}}\|^2\right\}  =\frac{{\vartheta_{\text{d}_{\mathbf{y}_0}}^{2}}\beta}{\rho|f^\mathbf{o}|^2\|\mathbf{y}_0\|^{-\alpha}} (\rho|{\cal A}|+C\Psi)$. Therefore, the solution of \eqref{globaldistortion} is 
\begin{align}
\label{inter_error_dowlink}
\vartheta_{\text{d}_{\mathbf{y}_0}} = \frac{\sum_{\mathbf{x} \in {\cal C}}^{}\sum_{\mathbf{y} \in {\cal A}^\mathbf{x}}^{} \sigma_{\text{w},\mathbf{y}}^\mathbf{x}}{\left(1+\frac{\beta}{|f^\mathbf{o}|^2\|\mathbf{y}_0\|^{-\alpha}}\right)\left(|{\cal A}|+\frac{C\Psi}{\rho}\right)}.
\end{align}

Note that in the intra- and inter-cluster iterations, the uplink error is due to the interference of devices in the uplink and the downlink error comes from the interference of edge servers in the downlink. Also, they include the effect of simultaneous transmissions of all the clusters regardless of their learning tasks. Since \eqref{uplink}, \eqref{intradownlink}, and \eqref{interdownlink} utilize only one resource block all over the network containing frequency subchannels equal to the size of the learning model, i.e., $d$, and all nodes have a single antenna, the communication efficiency is validated in terms of bandwidth and antenna resources. According to the uplink and downlink schemes, the expected latency in completing the {\fontfamily{lmtt}\selectfont
	MultiAirFed} algorithm is obtained as
\begin{align}
\label{delay}
T_{\text{L}} = T  \left(t_\text{BH} +2t_\text{BC}+ \gamma t_\text{CM}+\tau \left(t_\text{CM}+2t_\text{BC}\right)\right),
\end{align}
where $t_\text{CM}$ is the local computation latency of each device given by $t_\text{CM} = \frac{cN_\text{b}}{f}$ \cite{compdelay}, where $c$ is the number of CPU cycles required for computing one sample data, $f$ is the CPU cycle frequency, and $N_\text{b}$ is the size of data involved in the local update. In \eqref{delay}, $t_\text{BC}$ is the time needed for uplink or downlink transmission of $d$ model or gradient parameters as $t_\text{BC} = \frac{d}{W}$ \cite{huang_sg}, where a total bandwidth of $W$ is assumed. Also, $t_\text{BH} \gg t_\text{BC}$ denotes the backhaul latency for the inter-cluster process. As observed from \eqref{delay}, the latency is independent of the number of clusters and devices.
\vspace{-5pt}
\section{Convergence Analysis}
The convergence analysis of {\fontfamily{lmtt}\selectfont MultiAirFed} in terms of the optimality gap is presented in the following theorem, based on the estimations in \eqref{intraestimate} and \eqref{interestimate}. The analysis assumes common assumptions found in literature \cite{cao, downlink, gunduz3, letaief, multicell} as

\textbf{Assumption 1 (Lipschitz-Continuous Gradient):} The gradient of loss function $F(\mathbf{w})$ in \eqref{lossfunction} is Lipschitz continuous with a non-negative constant $L > 0$. It means that for any model vectors $\mathbf{w}_1$ and $\mathbf{w}_2$, we have
\begin{align}
&F(\mathbf{w}_2) \leq F(\mathbf{w}_1) + \nabla F(\mathbf{w}_1)^T (\mathbf{w}_2-\mathbf{w}_1) + \frac{L}{2} \|\mathbf{w}_2 - \mathbf{w}_1\|^2,\\
&\|\nabla F(\mathbf{w}_2)-\nabla F(\mathbf{w}_1)\| \leq L \|\mathbf{w}_2 - \mathbf{w}_1\|.
\end{align}

%\textbf{Assumption 2 (Polyak-Lojasiewicz Inequality):} Let ${F^*}$
%denote the optimal loss function value of the problem (4). There exists a constant $\delta \geq 0$ such that 
%\begin{align}
%\|\nabla F(\mathbf{w})\|^2 \geq 2\delta (F(\mathbf{w}) - F^*).
%\end{align}

\textbf{Assumption 2 (Variance Bound):} The local gradient estimate ${\mathbf{g}}$ at a device is an unbiased estimate
of the ground-true gradient $\nabla F(\mathbf{w})$ with bounded variance
\begin{align}
\mathbb{E}\left\{\|\mathbf{g} - \nabla F(\mathbf{w})\|^2\right\} \leq \frac{\sigma^2}{B},
\end{align}
where $B$ is the mini-batch data size.

\textbf{Assumption 3 (Polyak-Lojasiewicz Inequality):} Consider $F^* = F(\mathbf{w}^*)$ from the problem \eqref{globalopt}. There is a constant $\delta \geq 0$ such that the following
condition is satisfied.
\begin{align}
\label{polyak}
\|\nabla F(\mathbf{w})\|^2 \geq 2\delta \left(F(\mathbf{w}) - F^*\right).
\end{align}
The inequality \eqref{polyak} is significantly more general than the assumption of strong convexity \cite{karimi}.

To make the analysis more manageable, we assume that the downlink channel gains from an edge server to the devices in its cluster are greater than $\text{th}_0$. Also, normalizing factors $\vartheta_{\text{d}_{\mathbf{y}_0},i,t}^{\mathbf{o}}$ and $\vartheta_{\text{d}_{\mathbf{y}_0},t}$ in any intra-cluster iteration $i$ and inter-cluster iteration $t$ are lower than constants $\vartheta_{\text{d}_{\mathbf{y}_0}}^{\text{b}\mathbf{o}}$ and $\vartheta_{\text{d}_{\mathbf{y}_0}}^{\text{b}}$, respectively.
\begin{theorem}
Consider a fixed learning rate $\mu_t = \mu$ satisfying
\begin{align}
\label{learningrate_th1}
&\mu(\tau+\gamma)\delta \leq 1,\ 1- \frac{L^2\mu^2 \tau (\tau-1)}{2}-{L\mu \tau} -{L^2 \mu^2 \tau \gamma}\geq 0, \nonumber\\& 1- \frac{L^2\mu^2 \gamma (\gamma-1)}{2}-{L\mu \gamma} \geq 0.
\end{align}
Then, the following optimality gap holds for the local learning model of any reference device.
\begin{align}
\label{convergence}
&\mathbb{E}\left\{F({\mathbf{w}}_{\mathbf{y}_0,0,T}^\mathbf{o})\right\}-F^* \leq {(1-\mu (\tau+\gamma) \delta)}^T\times \nonumber\\& \biggl(F({\mathbf{w}}_{\mathbf{y}_0,0,0}^\mathbf{o})-F^*\biggr)+\frac{1-(1-\mu (\tau+\gamma) \delta)^T}{\mu (\tau+\gamma) \delta}\biggl[\frac{L^2 \mu^3}{2}\frac{\sigma^2}{B}\times\nonumber\\
&\left(\frac{\tau(\tau-1)}{2}+\tau \gamma \right) \mathbb{E}\left\{ \frac{1}{|{\cal A}^\mathbf{o}|}\right\}+\frac{L\mu^2}{2} \frac{\sigma^2}{B} \tau \mathbb{E}\left\{\frac{1}{|{\cal A}^\mathbf{o}|}\right\}\nonumber\\
& \mathbb{E}\left\{\sum_{\mathbf{x} \in {\mathcal{C}}}^{} \frac{|{\cal A}^\mathbf{x}|^2}{{|{\cal A}|}^2}\right\}+\frac{ L^2 \mu^3}{4} \frac{\sigma^2\gamma (\gamma-1)}{B}  +\frac{L\mu^2}{2} \gamma \frac{\sigma^2}{B}\mathbb{E}\left\{\frac{1}{|{\cal A}|}\right\}\nonumber\\&+\mathbb{E}\left\{\frac{1}{{|{\cal A}^\mathbf{o}|^2} }\right\}\biggl(\frac{L^2 \mu^3}{2}\left(\frac{\tau(\tau-1)}{2}+\tau \gamma\right) +\frac{L\mu^2}{2}\tau \nonumber\\&\mathbb{E}\left\{\sum_{\mathbf{x} \in {\mathcal{C}}}^{} \frac{| {\cal A}^\mathbf{x}|^2}{|{\cal A}|^2} \right\}\biggr) \mathbb{E}\left\{\|\boldsymbol\epsilon_\text{u}^{\text{b}\mathbf{o}}\|^2\right\}+\mathbb{E}\left\{\frac{1}{{|{\cal A}^\mathbf{o}|^2} }\right\} \biggl(\frac{L^2 \mu^3}{2}\times \nonumber\\&\left(\frac{\tau(\tau-1)}{2}+\tau \gamma\right) +\frac{L\mu^2}{2} \tau \mathbb{E}\left\{\frac{1}{|{\cal A}|}\right\} \biggr)\mathbb{E}\left\{\|\boldsymbol\epsilon_{\text{d}_{\mathbf{y}_0}}^{\text{b}\mathbf{o}}\|^2\right\}\nonumber
\end{align}
\begin{align}
& +\frac{L}{2}\mathbb{E}\left\{\frac{1}{{|{\cal A}|^2} }\right\}\mathbb{E}\left\{\|\boldsymbol\epsilon_\text{u}^\text{b}\|^2\right\}+\mathbb{E}\left\{\frac{1}{{|{\cal A}|^2}}\right\}\times\nonumber\\&\biggl({L^2 \mu(\tau+{ \gamma})}  +\frac{L}{2} \mathbb{E}\left\{\frac{1}{|{\cal A}|}\right\} +L\biggr)\mathbb{E}\left\{\|\boldsymbol\epsilon_{\text{d}_{\mathbf{y}_0}}^\text{b}\|^2\right\}\biggr].
\end{align}
\end{theorem}
\begin{IEEEproof}
See Appendix A.
\end{IEEEproof}
Considering the error term in the bound, the first four parts reflect the gradient estimation errors. These are followed by the effects of the intra- and inter-cluster uplink and downlink errors. The scaling factors of the error terms depend on the learning parameters and on the device selection, while the error terms depend on the interference, determined by network topology, the wireless environment and the power control. This structure therefore supports a separate design and evaluation of the learning algorithm and the wireless communication.

In order to establish the bound, we need to characterize several expected quantities. These include the expected values of the terms related to the number of active devices, namely $\mathbb{E}\left\{\frac{1}{|{\cal A}^\mathbf{o}|}\right\}$, $\mathbb{E}\left\{\frac{1}{|{\cal A}^\mathbf{o}|^2}\right\}$, $\mathbb{E}\left\{\frac{1}{|{\cal A}|}\right\}$, $\mathbb{E}\left\{\frac{1}{|{\cal A}|^2}\right\}$, and $\mathbb{E}\left\{\sum_{\mathbf{x} \in {\mathcal{C}}}^{} \frac{|{\cal A}^\mathbf{x}|^2}{{|{\cal A}|}^2}\right\}$. Additionally, we need to determine the upper-bounds for the intra- and inter-cluster uplink and downlink error terms, which include $\mathbb{E}\left\{\|\boldsymbol{\epsilon}_\text{u}^{\text{b}\mathbf{o}}\|^2\right\}$, $\mathbb{E}\left\{\|\boldsymbol\epsilon_{\text{d}_{\mathbf{y}_0}}^{\text{b}\mathbf{o}}\|^2\right\}$, $\mathbb{E}\left\{\|\boldsymbol\epsilon_{\text{u}}^{\text{b}}\|^2\right\}$, and $\mathbb{E}\left\{\|\boldsymbol\epsilon_{\text{d}_{\mathbf{y}_0}}^\text{b}\|^2\right\}$. Due to the facts that $|{\cal A}^\mathbf{x}| \sim \text{Binomial}(M,e^{-\text{th}_1}), \forall \mathbf{x}$, $|{\cal A}| \sim \text{Binomial}(CM,e^{-\text{th}_1})$, and there is at least one active device in each cluster, the expected terms are computed as
\begin{align}
\label{intraactive}
&\mathbb{E}\left\{\left. \frac{1}{|{\cal A}^\mathbf{o}|}\right||{\cal A}^\mathbf{o}|>0\right\} = \frac{\mathbb{E}\left\{ \frac{1}{|{\cal A}^\mathbf{o}|}\ \text{and} \ |{\cal A}^\mathbf{o}|>0\right\}}{\mathbb{P}\left\{|{\cal A}^\mathbf{o}|>0\right\}}=\nonumber\\
& \sum_{m=1}^{M} \frac{1}{m} \frac{{M \choose m} e^{-\text{th}_1  m} \left(1-e^{-\text{th}_1}\right)^{M-m}}{1-\left(1-e^{-\text{th}_1}\right)^{M}} \nonumber\\
&= \frac{ \left(1-e^{-\text{th}_1}\right)^{M}}{1-\left(1-e^{-\text{th}_1}\right)^{M}}\sum_{m=1}^{M} {M \choose m}\frac{1}{m} \left(\frac{e^{-\text{th}_1}}{1-e^{-\text{th}_1}}\right)^m,
\end{align}
\vspace{-15pt}
\begin{align}
\label{intraactive2}
&\mathbb{E}\left\{\left. \frac{1}{|{\cal A}^\mathbf{o}|^2}\right||{\cal A}^\mathbf{o}|>0\right\} = \frac{ \left(1-e^{-\text{th}_1}\right)^{M}}{1-\left(1-e^{-\text{th}_1}\right)^{M}}\sum_{m=1}^{M} {M \choose m}\nonumber\\
&\frac{1}{m^2} \left(\frac{e^{-\text{th}_1}}{1-e^{-\text{th}_1}}\right)^m,
\end{align}
\vspace{-15pt}
\begin{align}
\label{interactive}
&\mathbb{E}\left\{\left. \frac{1}{|{\cal A}|}\right||{\cal A}^\mathbf{x}|>0,\forall \mathbf{x}\right\} =\frac{\mathbb{E}\left\{ \frac{1}{|{\cal A}|}\ \text{and} \ |{\cal A}|>C\right\}}{\prod_{\mathbf{x} \in {\mathcal{C}}}^{}\mathbb{P}\left\{|{\cal A}^\mathbf{x}|>0\right\}}=\nonumber\\& \sum_{m=C}^{CM} \frac{1}{m} \frac{{CM \choose m} e^{-\text{th}_1 m} \left(1-e^{-\text{th}_1}\right)^{CM-m}}{\left(1-\left(1-e^{-\text{th}_1}\right)^{M}\right)^C}= \nonumber\\
& \left(\frac{ \left(1-e^{-\text{th}_1}\right)^{M}}{1-\left(1-e^{-\text{th}_1}\right)^{M}}\right)^C\sum_{m=C}^{CM} {CM \choose m}\frac{1}{m} \left(\frac{e^{-\text{th}_1}}{1-e^{-\text{th}_1}}\right)^m,
\end{align}
\vspace{-10pt}
\begin{align}
\label{interactive2}
&\mathbb{E}\left\{\left. \frac{1}{|{\cal A}|^2}\right||{\cal A}^\mathbf{x}|>0,\forall \mathbf{x}\right\} = \left(\frac{ \left(1-e^{-\text{th}_1}\right)^{M}}{1-\left(1-e^{-\text{th}_1}\right)^{M}}\right)^C\times \nonumber\\&\sum_{m=C}^{CM} {CM \choose m}\frac{1}{m^2} \left(\frac{e^{-\text{th}_1}}{1-e^{-\text{th}_1}}\right)^m,
\end{align}
\vspace{-15pt}
\begin{align}
\label{active_both}
&\mathbb{E}\left\{\left.\sum_{\mathbf{x} \in {\mathcal{C}}}^{} \frac{|{\cal A}^\mathbf{x}|^2}{{|{\cal A}|}^2}\right||{\cal A}^\mathbf{x}|>0,\forall \mathbf{x}\right\} = \left(\frac{ \left(1-e^{-\text{th}_1}\right)^{M}}{1-\left(1-e^{-\text{th}_1}\right)^{M}}\right)^C\nonumber\\
&\sum_{m=C}^{MC}{CM \choose m}\frac{\sum_{\left\{(m_1,\cdots,m_C)|\sum_{c=1}^{C}m_c = m\right\}}^{}\sum_{c=1}^{C}m_c^2}{m^2}\times \nonumber\\& \left(\frac{e^{-\text{th}_1}}{1-e^{-\text{th}_1}}\right)^m.
\end{align}
From the facts $\sum_{\mathbf{y} \in {\cal A}^\mathbf{o}}\left(\vartheta_{\text{d}_{\mathbf{y}_0}}^{\mathbf{o}}-\sigma_{\text{g},\mathbf{y}}^\mathbf{o}\right)^2 \leq |{\cal A}^\mathbf{o}|\vartheta_{\text{d}_{\mathbf{y}_0}}^{\text{b}\mathbf{o}\ \hspace{-2pt}2}$ and $\sum_{\mathbf{x}\in {\cal C}}^{} \sum_{\mathbf{y} \in {\cal A}^\mathbf{x}}^{}\left(\vartheta_{\text{d}_{\mathbf{y}_0}}-\sigma_{\text{w},\mathbf{y}}^\mathbf{x}\right)^2 \leq |{\cal A}|\vartheta_{\text{d}_{\mathbf{y}_0}}^{\text{b}\ \hspace{-2pt}2}$, $\mathbb{E}\left\{\|\boldsymbol{\epsilon}_\text{u}^{\mathbf{o}}\|^2\right\}$, $\mathbb{E}\left\{\|\boldsymbol\epsilon_{\text{d}_{\mathbf{y}_0}}^{\mathbf{o}}\|^2\right\}$, $\mathbb{E}\left\{\|\boldsymbol\epsilon_{\text{u}}\|^2\right\}$, and $\mathbb{E}\left\{\|\boldsymbol\epsilon_{\text{d}_{\mathbf{y}_0}}\|^2\right\}$ provided in Subsections IV. B and C can be upper-bounded as
\begin{align}
\label{ubo}
&\mathbb{E}\left\{\|\boldsymbol{\epsilon}_\text{u}^{\text{b}\mathbf{o}}\|^2\right\} =  \mathbb{E}\left\{|{\cal A}^\mathbf{o}|\right\}\vartheta_{\text{d}_{\mathbf{y}_0}}^{\text{b}\mathbf{o}\ \hspace{-2pt}2}+\frac{{\vartheta_{\text{d}_{\mathbf{y}_0}}^{\text{b}\mathbf{o}\ \hspace{-2pt}2}}}{\rho} \Psi \nonumber\\&= \vartheta_{\text{d}_{\mathbf{y}_0}}^{\text{b}\mathbf{o}\ \hspace{-2pt}2}\left(Me^{-\text{th}_1}+\frac{\Psi}{\rho}\right),
\end{align}
\vspace{-15pt}
\begin{align}
\label{dbo}
&\mathbb{E}\left\{\|\boldsymbol{\epsilon}_{\text{d}_{\mathbf{y}_0}}^{\text{b}\mathbf{o}}\|^2\right\}  =\frac{{\vartheta_{\text{d}_{\mathbf{y}_0}}^{\text{b}\mathbf{o}\ \hspace{-2pt}2}}\beta}{\rho} \mathbb{E}\left\{\frac{1}{|f^\mathbf{o}|^2\|\mathbf{y}_0\|^{-\alpha}}\Big| |f^{\mathbf{o}}| > \text{th}_0\right\} \nonumber\\&(\rho \mathbb{E}\left\{|{\cal A}^\mathbf{o}|\right\}+\Psi) = {{\vartheta_{\text{d}_{\mathbf{y}_0}}^{\text{b}\mathbf{o}\ \hspace{-2pt}2}}}\text{Ei}\left(\frac{\text{th}_0}{ \sigma_\text{d}^2}\right)\frac{2\beta}{(2+\alpha)\sigma_\text{d}^2}\frac{R^{\alpha+2}-r_0^{\alpha+2}}{R^2-r_0^2}\nonumber\\&\left(Me^{-\text{th}_1}+\frac{\Psi}{\rho}\right),
\end{align}
which is due to $\mathbb{E}\left\{\|\mathbf{y}_0\|^\alpha\right\} = \int_{r_0}^{R} \frac{2y}{R^2-r_0^2}y^{\alpha} \mathrm{d}y = \frac{2}{2+\alpha}\frac{R^{\alpha+2}-r_0^{\alpha+2}}{R^2-r_0^2}$ and $\mathbb{E}\left\{\frac{1}{|f^\mathbf{o}|^2}\right\} = \frac{1}{\sigma_\text{d}^2}\text{Ei}\left(\frac{\text{th}_0}{ \sigma_\text{d}^2}\right)$, and
\vspace{-5pt}
\begin{align}
\label{udb}
&\mathbb{E}\left\{\|\boldsymbol{\epsilon}_\text{u}^{\text{b}}\|^2\right\} = C \frac{\vartheta_{\text{d}_{\mathbf{y}_0}}^{\text{b}\ \hspace{-2pt}2}}{\vartheta_{\text{d}_{\mathbf{y}_0}}^{\text{b}\mathbf{o}\ \hspace{-2pt}2}}\mathbb{E}\left\{\|\boldsymbol{\epsilon}_\text{u}^{\text{b}\mathbf{o}}\|^2\right\},\nonumber\\&\ \mathbb{E}\left\{\|\boldsymbol{\epsilon}_{\text{d}_{\mathbf{y}_0}}^{\text{b}}\|^2\right\}  = C \frac{\vartheta_{\text{d}_{\mathbf{y}_0}}^{\text{b}\ \hspace{-2pt}2}}{\vartheta_{\text{d}_{\mathbf{y}_0}}^{\text{b}\mathbf{o}\ \hspace{-2pt}2}}\mathbb{E}\left\{\|\boldsymbol{\epsilon}_{\text{d}_{\mathbf{y}_0}}^{\text{b}\mathbf{o}}\|^2\right\}.
\end{align}
The following remarks and design insights can be concluded from Theorem 1. 
\begin{remark}
The first term of the optimality gap decreases with the number of inter-cluster iterations $T$, while the second, error term is increasing, approaching a bound.
\end{remark}
\begin{remark}
	There is a tradeoff for the optimality gap when $\text{th}_1$ increases. The term $\text{Ei}(\text{th}_1)e^{-\text{th}_1}$ in $\Psi$ and then the intra-cluster uplink error term $\mathbb{E}\left\{\|\boldsymbol{\epsilon}_\text{u}^{\text{b}\mathbf{o}}\|^2\right\}$ decrease, however the term $\mathbb{E}\left\{\frac{1}{|{\cal A}^\mathbf{o}|}\right\}$ in \eqref{intraactive} increases. Hence, the impact of $\text{th}_1$ on the convergence in the general case is not evident.
\end{remark}
\begin{remark}
The scaling factors of the intra-cluster uplink and downlink error terms $\mathbb{E}\left\{\|\boldsymbol{\epsilon}_\text{u}^{\text{b}\mathbf{o}}\|^2\right\}$ and $\mathbb{E}\left\{\|\boldsymbol{\epsilon}_{\text{d}_{\mathbf{y}_0}}^{\text{b}\mathbf{o}}\|^2\right\}$ in the optimality gap increase by $\tau$ and $\mu$ with the rate $\mathcal{O}(\mu^3 \tau^2)$. Also, while the scaling factor of the inter-cluster uplink error term $\mathbb{E}\left\{\|\boldsymbol{\epsilon}_\text{u}^{\text{b}}\|^2\right\}$ does not change with them, the scaling factor of the inter-cluster downlink error term $\mathbb{E}\left\{\|\boldsymbol{\epsilon}_{\text{d}_{\mathbf{y}_0}}^{\text{b}}\|^2\right\}$ increases with the rate $\mathcal{O}(\mu \tau)$. Hence, the intra-cluster error terms grow the optimality gap with $\tau$ and $\mu$ much faster compared to the inter-cluster error terms.   
\end{remark}
\begin{remark}
When $\mu \propto \frac{1}{\tau+\gamma}$, the optimality gap has the rate $\mathcal{O}(\text{const}+\frac{1}{\tau+\gamma})$ with $\tau$ and $\gamma$. Hence, increasing $\tau$ and $\gamma$ decrease the gap.
\end{remark}

\begin{remark}
In the optimality gap, the intra-cluster error terms $\mathbb{E}\left\{\|\boldsymbol{\epsilon}_\text{u}^{\text{b}\mathbf{o}}\|^2\right\}$ and $\mathbb{E}\left\{\|\boldsymbol{\epsilon}_{\text{d}_{\mathbf{y}_0}}^{\text{b}\mathbf{o}}\|^2\right\}$ are directly scaled by the inverse squared of the number of active devices in a single cluster $\mathbb{E}\left\{\frac{1}{|{\cal A}^\mathbf{o}|^2}\right\}$. Also, the inter-cluster error terms $\mathbb{E}\left\{\|\boldsymbol{\epsilon}_\text{u}^{\text{b}}\|^2\right\}$ and $\mathbb{E}\left\{\|\boldsymbol{\epsilon}_{\text{d}_{\mathbf{y}_0}}^\text{b}\|^2\right\}$ are scaled by the inverse squared of the total number of active devices in the collaborating clusters $\mathbb{E}\left\{\frac{1}{|{\cal A}|^2}\right\}$. Hence, having higher number of active devices in the learning process can significantly diminish the effect of error terms in the optimality gap.
\end{remark}

\begin{remark}
Increasing the number of collaborating clusters $C$ decreases $\mathbb{E}\left\{\frac{1}{|{\cal A}|}\right\}$ and $\mathbb{E}\left\{\frac{1}{|{\cal A}|^2}\right\}$ with an order
higher than one as in \eqref{interactive} and \eqref{interactive2}. On the other hand, the inter-cluster error terms $\mathbb{E}\left\{\|\boldsymbol{\epsilon}_\text{u}^\text{b}\|^2\right\}$ and $\mathbb{E}\left\{\|\boldsymbol{\epsilon}_{\text{d}_{\mathbf{y}_0}}^\text{b}\|^2\right\}$ linearly increase with $C$. Hence, the optimality gap decreases.  
\end{remark}

\begin{remark}
	From \eqref{upper_error_derive}, the intra- and inter-cluster uplink error terms $\mathbb{E}\left\{\|\boldsymbol{\epsilon}_\text{u}^{\text{b}\mathbf{o}}\|^2\right\}$ and $\mathbb{E}\left\{\|\boldsymbol{\epsilon}_\text{u}^{\text{b}}\|^2\right\}$ linearly increase with the number of devices per cluster, i.e., $M$. However, from \eqref{intraactive2} and \eqref{interactive2}, as $\mathbb{E}\left\{\frac{1}{|{\cal A}^\mathbf{o}|^2}\right\}$ and $\mathbb{E}\left\{\frac{1}{|{\cal A}|^2}\right\}$ decrease with $M$ at orders higher than one and they contribute to all the scaling factors of the error terms, the optimality gap is overally decreased. 
\end{remark}

\begin{remark}
	For a fixed total number of active devices in the collaborating clusters, reducing the number of clusters (or increasing the number of active devices in each cluster) reduces the optimality gap. This results from the constant value of \(\mathbb{E}\left\{\frac{1}{|{\cal A}|^2}\right\}\) and the decreasing value of \(\mathbb{E}\left\{\frac{1}{|{\cal A}^\mathbf{o}|^2}\right\}\), which as the scaling factor directly reduces the terms in \eqref{convergence}.
\end{remark}

\begin{remark}
The scaling factor, or the effect on the optimality gap, of the intra-cluster uplink error term $\mathbb{E}\left\{\|\boldsymbol{\epsilon}_\text{u}^{\text{b}\mathbf{o}}\|^2\right\}$ is more than the one for the intra-cluster downlink error term $\mathbb{E}\left\{\|\boldsymbol{\epsilon}_{\text{d}_{\mathbf{y}_0}}^{\text{b}\mathbf{o}}\|^2\right\}$. That is due to the fact that $\sum_{\mathbf{x} \in {\mathcal{C}}}^{}|{\cal A}^\mathbf{x}|^2 > |{\cal A}|$ and hence $\frac{L^2 \mu^3}{2}\left(\frac{\tau(\tau-1)}{2}+\tau \gamma\right) +\frac{L\mu^2}{2}\tau \mathbb{E}\left\{\sum_{\mathbf{x} \in {\mathcal{C}}}^{} \frac{| {\cal A}^\mathbf{x}|^2}{|{\cal A}|^2} \right\} > \frac{L^2 \mu^3}{2}\left(\frac{\tau(\tau-1)}{2}+\tau \gamma\right) +\frac{L\mu^2}{2} \tau \mathbb{E}\left\{\frac{1}{|{\cal A}|}\right\}$. However, for the inter-cluster error terms, the downlink term $\mathbb{E}\left\{\|\boldsymbol{\epsilon}_{\text{d}_{\mathbf{y}_0}}^\text{b}\|^2\right\}$ has much higher scaling factor than the uplink term $\mathbb{E}\left\{\|\boldsymbol{\epsilon}_\text{u}^\text{b}\|^2\right\}$ since the inequality $L^2 \mu(\tau+{ \gamma}) +\frac{L}{2} \mathbb{E}\left\{\frac{1}{|{\cal A}|}\right\} +L > \frac{L}{2}$ holds. Hence, reducing uplink interference during intra-cluster iterations and reducing downlink interference during inter-cluster iterations can be an efficient way to improve the performance.
\end{remark}

\begin{remark}
From \eqref{upper_error_derive} and  \eqref{ubo}-\eqref{udb}, the uplink error terms $\mathbb{E}\left\{\|\boldsymbol{\epsilon}_\text{u}^{\text{b}\mathbf{o}}\|^2\right\}$ and $\mathbb{E}\left\{\|\boldsymbol{\epsilon}_\text{u}^\text{b}\|^2\right\}$ and the downlink error terms $\mathbb{E}\left\{\|\boldsymbol{\epsilon}_{\text{d}_{\mathbf{y}_0}}^{\text{b}\mathbf{o}}\|^2\right\}$ and $\mathbb{E}\left\{\|\boldsymbol{\epsilon}_{\text{d}_{\mathbf{y}_0}}^\text{b}\|^2\right\}$ linearly and quadratically\footnote{Note that the term $\Psi$ in \eqref{ubo} and \eqref{dbo} has a linear dependency to $\lambda_\text{p}$, as given in \eqref{upper_error_derive}, as well as $\beta$ in \eqref{beta_def}.} increase with the cluster center density $\lambda_\text{p}$, respectively. Hence, the optimality gap increases with the rate $\mathcal{O}(\lambda_\text{p}^2)$.
\end{remark}

In the following corollaries, we present special cases of Theorem 1. For the single-server case when all edge servers are working independently under different learning tasks, i.e., $C = 1$, we have the next simplified convergence result. 
\begin{corollary}
In the case $C = 1$ and the learning rate as in \eqref{learningrate_th1},
the optimality gap for the local learning model of any reference device is
\begin{align}
\label{convergence1}
&\mathbb{E}\left\{F({\mathbf{w}}_{\mathbf{y}_0,0,T}^\mathbf{o})\right\}-F^* \leq {(1-\mu (\tau+\gamma) \delta)}^T \times \nonumber\\& \biggl(F({\mathbf{w}}_{\mathbf{y}_0,0,0}^\mathbf{o})-F^*\biggr)+\frac{1-(1-\mu (\tau+\gamma) \delta)^T}{\mu (\tau+\gamma) \delta}
\biggl[\frac{L^2 \mu^3}{2}\frac{\sigma^2}{B}\times\nonumber\\
&\left(\frac{\tau(\tau-1)}{2}+\tau \gamma \right) \mathbb{E}\left\{ \frac{1}{|{\cal A}^\mathbf{o}|}\right\}+\frac{L\mu^2}{2} \frac{\sigma^2}{B} \tau \mathbb{E}\left\{\frac{1}{|{\cal A}^\mathbf{o}|}\right\} +\nonumber\\
&\frac{ L^2 \mu^3}{4} \frac{\sigma^2}{B} \gamma (\gamma-1) +\frac{L\mu^2}{2} \gamma \frac{\sigma^2}{B}\mathbb{E}\left\{\frac{1}{|{\cal A}^\mathbf{o}|}\right\}+\mathbb{E}\left\{\frac{1}{{|{\cal A}^\mathbf{o}|^2} }\right\}\nonumber\\
&\biggl(\frac{L^2 \mu^3}{2}\left(\frac{\tau(\tau-1)}{2}+\tau \gamma\right) +\frac{L\mu^2}{2}\tau+\frac{L}{2} \biggr)  \mathbb{E}\left\{\|\boldsymbol\epsilon_\text{u}^{\text{b}\mathbf{o}}\|^2\right\}+\nonumber\\
&\mathbb{E}\left\{\frac{1}{{|{\cal A}^\mathbf{o}|^2} }\right\} \biggl(\frac{L^2 \mu^3}{2}\left(\frac{\tau(\tau-1)}{2}+\tau \gamma\right) +\frac{L\mu^2}{2} \tau \mathbb{E}\left\{\frac{1}{|{\cal A}^\mathbf{o}|}\right\} \nonumber\\&+{L^2 \mu(\tau+{ \gamma})}  +\frac{L}{2} \mathbb{E}\left\{\frac{1}{|{\cal A}^\mathbf{o}|}\right\} +L\biggr)\mathbb{E}\left\{\|\boldsymbol\epsilon_{\text{d}_{\mathbf{y}_0}}^{\text{b}\mathbf{o}}\|^2\right\} \biggr].
\end{align}	
\end{corollary}
\begin{IEEEproof}
It comes from Theorem 1 when ${\cal A} = {\cal A}^\mathbf{o}$ and $\mathbb{E}\left\{\sum_{\mathbf{x} \in {\mathcal{C}}}^{} \frac{|{\cal A}^\mathbf{x}|^2}{{|{\cal A}|}^2}\right\} = 1$.
\end{IEEEproof}
When all the edge servers are collaboratively working under the same task, i.e., $C \to \infty$, the convergence result is simplified in the next corollary. 
\begin{corollary}
	In the case $C \to \infty$ and the learning rate as in \eqref{learningrate_th1},
	the optimality gap for the local learning model of any reference device is
	\begin{align}
\label{convergence2}
&\mathbb{E}\left\{F({\mathbf{w}}_{\mathbf{y}_0,0,T}^\mathbf{o})\right\}-F^* \leq {(1-\mu (\tau+\gamma) \delta)}^T\times\nonumber\\& \biggl(F({\mathbf{w}}_{\mathbf{y}_0,0,0}^\mathbf{o})-F^*\biggr)+\frac{1-(1-\mu (\tau+\gamma) \delta)^T}{\mu (\tau+\gamma) \delta}\biggl[\frac{L^2 \mu^3}{2}\frac{\sigma^2}{B}\times\nonumber\\&\left(\frac{\tau(\tau-1)}{2}+\tau \gamma \right) \mathbb{E}\left\{ \frac{1}{|{\cal A}^\mathbf{o}|}\right\}+\frac{ L^2 \mu^3}{4} \frac{\sigma^2}{B} \gamma (\gamma-1) +\nonumber\\&\mathbb{E}\left\{\frac{1}{{|{\cal A}^\mathbf{o}|^2} }\right\}\biggl(\frac{L^2 \mu^3}{2}\left(\frac{\tau(\tau-1)}{2}+\tau \gamma\right) \biggr)  \mathbb{E}\left\{\|\boldsymbol\epsilon_\text{u}^{\text{b}\mathbf{o}}\|^2\right\}+\nonumber\\&\mathbb{E}\left\{\frac{1}{{|{\cal A}^\mathbf{o}|^2} }\right\} \biggl(\frac{L^2 \mu^3}{2}\left(\frac{\tau(\tau-1)}{2}+\tau \gamma\right)\hspace{0pt} \biggr)\mathbb{E}\left\{\|\boldsymbol\epsilon_{\text{d}_{\mathbf{y}_0}}^{\text{b}\mathbf{o}}\|^2\right\}\biggr].
\end{align}
\end{corollary}
\begin{IEEEproof}
	It comes from Theorem 1 when $\mathbb{E}\left\{ \frac{1}{|{\cal A}|}\right\} = 0$ and $\mathbb{E}\left\{ \frac{1}{|{\cal A}|^2}\right\} = 0$. Also, $\mathbb{E}\left\{\sum_{\mathbf{x} \in {\mathcal{C}}}^{} \frac{|{\cal A}^\mathbf{x}|^2}{{|{\cal A}|}^2}\right\} = 0$ since $|{\cal A}|^2 = \sum_{\mathbf{x} \in {\mathcal{C}}}^{}|{\cal A}^\mathbf{x}|^2 + \sum_{\substack{\mathbf{x}, \mathbf{x}^\prime \in {\mathcal{C}}\\ \mathbf{x} \neq \mathbf{x}^\prime}}{}2|{\cal A}^\mathbf{x}| |{\cal A}^{\mathbf{x}^\prime}|$ and $\lim_{C \to \infty}  \sum_{\substack{\mathbf{x}, \mathbf{x}^\prime \in {\mathcal{C}}\\ \mathbf{x} \neq \mathbf{x}^\prime}}{}2|{\cal A}^\mathbf{x}| |{\cal A}^{\mathbf{x}^\prime}|= \infty$.
\end{IEEEproof}
\begin{remark}
Four terms in the optimality gap given in Theorem 1, the two terms of the inter-cluster errors and $\frac{L\mu^2}{2} \frac{\sigma^2}{B} \tau \mathbb{E}\left\{\frac{1}{|{\cal A}^\mathbf{o}|}\right\} \mathbb{E}\left\{\sum_{\mathbf{x} \in {\mathcal{C}}}^{} \frac{|{\cal A}^\mathbf{x}|^2}{{|{\cal A}|}^2}\right\}$ and $\frac{L\mu^2}{2} \gamma \frac{\sigma^2}{B}\mathbb{E}\left\{\frac{1}{|{\cal A}|}\right\}$, vanish when $C \to \infty$.
\end{remark}
\begin{remark}
The scaling factors of the intra-cluster error terms $\mathbb{E}\left\{\|\boldsymbol\epsilon_\text{u}^{\text{b}\mathbf{o}}\|^2\right\}$ and $\mathbb{E}\left\{\|\boldsymbol\epsilon_{\text{d}_{\mathbf{y}_0}}^{\text{b}\mathbf{o}}\|^2\right\}$ aproaches to equal terms when $C \to \infty$, thus their effect on the convergence will be the same. 
\end{remark}
\vspace{-10pt}
\section{Experimental Results}
The learning task over the collaborating clusters is the classification
on the standard MNIST and CIFAR-10 datasets. The classifier model for MNIST (CIFAR-10) is implemented using a CNN, which consists of two (four) $3 \times 3$ convolution layers with
ReLU activation (the (two) first with 32 channels, the (two) second with
64), each (two) followed by a $2 \times 2$ max pooling; a fully connected
layer with 128 units and ReLU activation; and a final softmax
output layer. We consider both i.i.d. and non-i.i.d. distribution of dataset 
samples over the devices. The number of samples at different devices is different and comes from the power law distribution $\sim 110 n^{-2}, 100 \leq n \leq 1000$. For non-i.i.d. case, each device has samples of only two classes, similar to \cite{salehi_sg, ding, downlink}. The performance is measured
as the learning accuracy with reference to the test dataset over global inter-cluster iteration count $t$. Each
performance result is evaluated as the average of
10 realization samples to account for random network distributions.

% (582,026 parameters in total)
\begin{comment}
\begin{table}
	\caption {Parameter Values} 
	\vspace{-18pt}
	\begin{center}
		\resizebox{13.2cm}{!} {
			\begin{tabular}{| l | l | l | l | l | l | l | l | l | l | l | l | l | l | l | 1 |}
				
				\hline
				\hline
				{$\lambda_{\rm p}$}&{$r_0$}&{$R$}& $\mu$& $C$& $M$& $P_\text{u}$&$\left\{P_\text{d},\sigma_\text{d}^2\right\}$& $\left\{\text{th}_0,\text{th}_1\right\}$ & $\alpha$&$\left\{\tau,\gamma\right\}$&$T$&$W$&$N_\text{b}$&$\left\{c,f\right\}$&$B$\\ \hline
				$20$ $\text{Km}^{-2}$& $4 \ \text{m}$ &$30 \ \text{m}$&$0.01$ & 3 & 15 & 1 & $\left\{1,10\right\}$&$\left\{0.01,0.5\right\}$& 4&$\left\{6,2\right\}$&40&$1 \ \text{MHz}$&2.5 $\text{Mbit}$&$\left\{20 \ \text{Cycles/bit},1 \ \text{GHz}\right\}$& 100\\ \hline	
				\hline
		\end{tabular}}
	\end{center}
	\vspace{-32pt}
\end{table}
\end{comment}

\begin{table}
	\caption {Parameter Values}
	\vspace{-6pt}
	\begin{center}
		\resizebox{9cm}{!} {
			\begin{tabular}{| l | l | l | l | l | l | l | l | l |}
				\hline
				\hline
				$\lambda_{\rm p}$ & $r_0$ & $R$ & $\mu$ & $C$ & $M$ & $P_\text{u}$ & $\left\{P_\text{d},\sigma_\text{d}^2\right\}$ & $\left\{\text{th}_0,\text{th}_1\right\}$ \\ 
				\hline
				$20 \ \text{Km}^{-2}$ & $4 \ \text{m}$ & $30 \ \text{m}$ & $0.01$ & $3$ & $15$ & $1$ & $\left\{1,10\right\}$ & $\left\{0.01,0.5\right\}$ \\ 
				\hline	
				\hline
				$\alpha$ & $\left\{\tau,\gamma\right\}$ & $B$ &  $T$ & $W$ & $N_\text{b}$ & $\left\{c,f\right\}$ & & \\ 
				\hline
				$4$ & $\left\{6,2\right\}$ & $60$ & $40$ & $1 \ \text{MHz}$ & $2.5 \ \text{Mbit}$ & $\left\{20 \ \text{Cycles/bit},1 \ \text{GHz}\right\}$ & & \\
				\hline
			\end{tabular}
		}
	\end{center}
	\vspace{-12pt}
\end{table}

\begin{figure}[tb!]
	\vspace{0pt}
	\centering
	\includegraphics[width =2.9in]{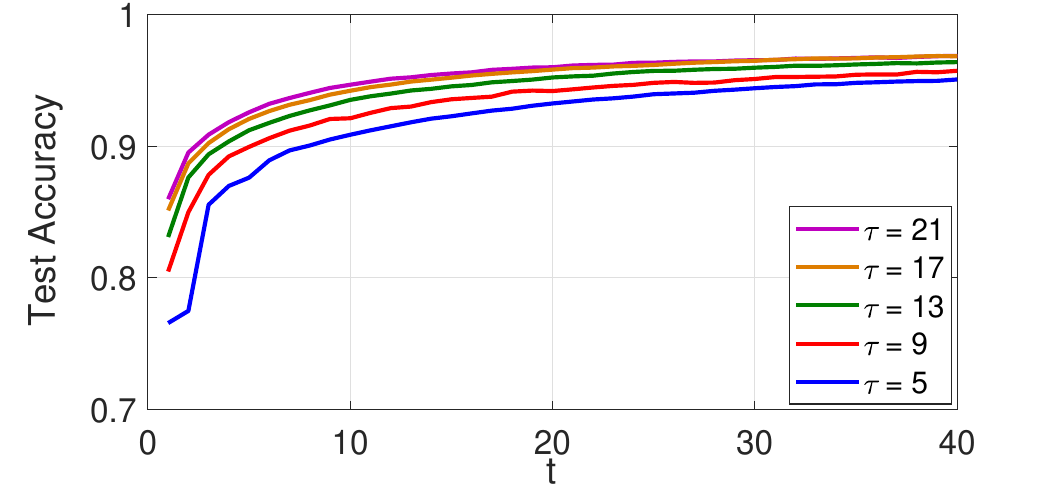} %mvalid.eps
	\vspace{-5pt}
	\caption{Test accuracy as a function of global iterations $t$ (i.i.d.)}
	\vspace{-5pt}
\end{figure}

\begin{figure}[tb!]
	\vspace{0pt}
	\centering
	\includegraphics[width =2.9in]{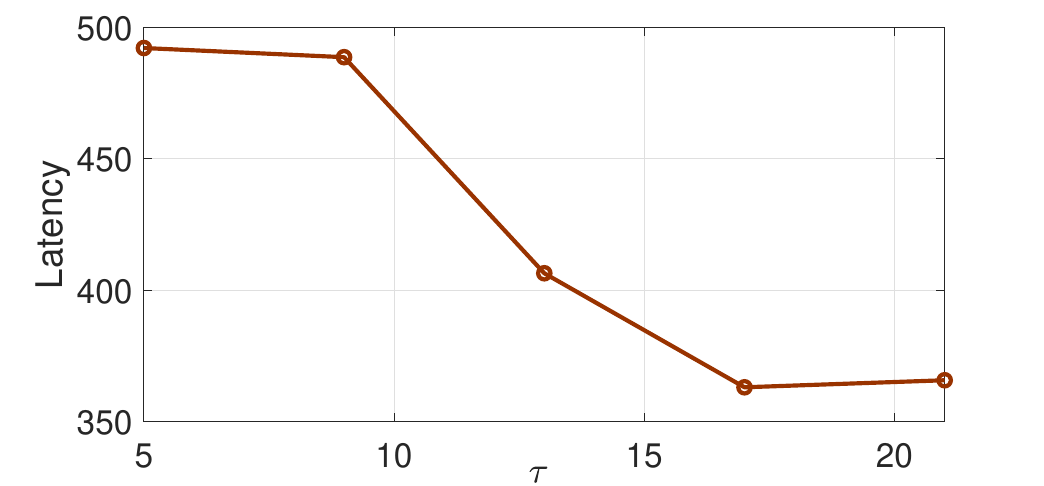} %mvalid.eps
	\vspace{-5pt}
	\caption{Latency in seconds as a function of intra-cluster iterations $\tau$ (i.i.d.)}
	\vspace{-5pt}
\end{figure}

\begin{figure}[tb!]
	\vspace{0pt}
	\centering
	\includegraphics[width =2.9in]{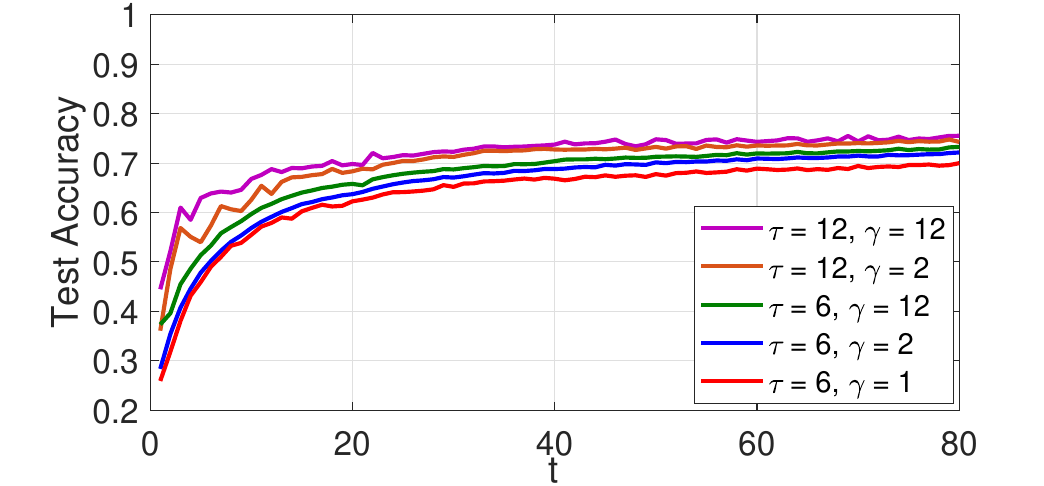} %mvalid.eps
	\vspace{-5pt}
	\caption{Test accuracy as a function of global iterations $t$ (i.i.d.)}
	\vspace{-5pt}
\end{figure}

\begin{figure}[tb!]
	\vspace{0pt}
	\centering
	\includegraphics[width =2.9in]{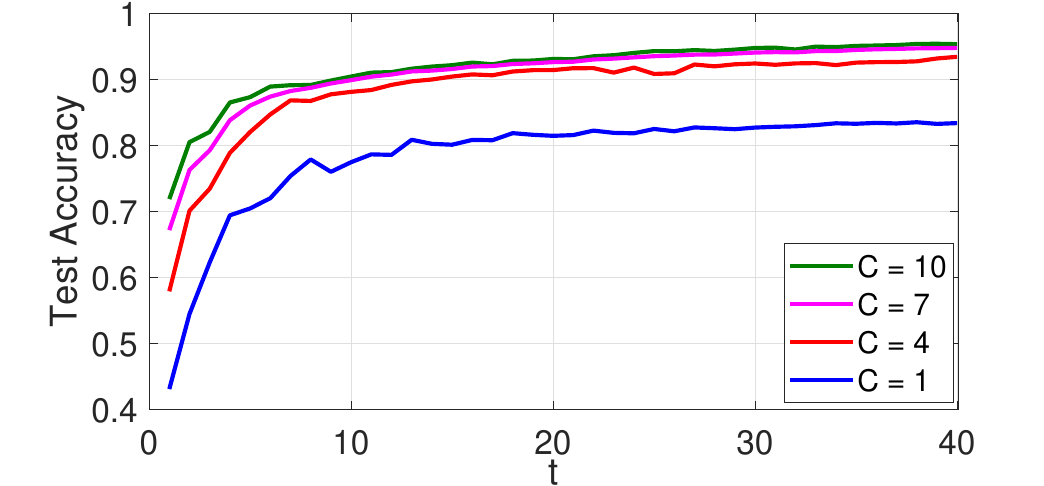} %mvalid.eps
	\vspace{-5pt}
	\caption{Test accuracy as a function of global iterations $t$ (non-i.i.d.)}
	\vspace{-10pt}
\end{figure}

\begin{figure}[tb!]
	\vspace{0pt}
	\centering
	\includegraphics[width =2.9in]{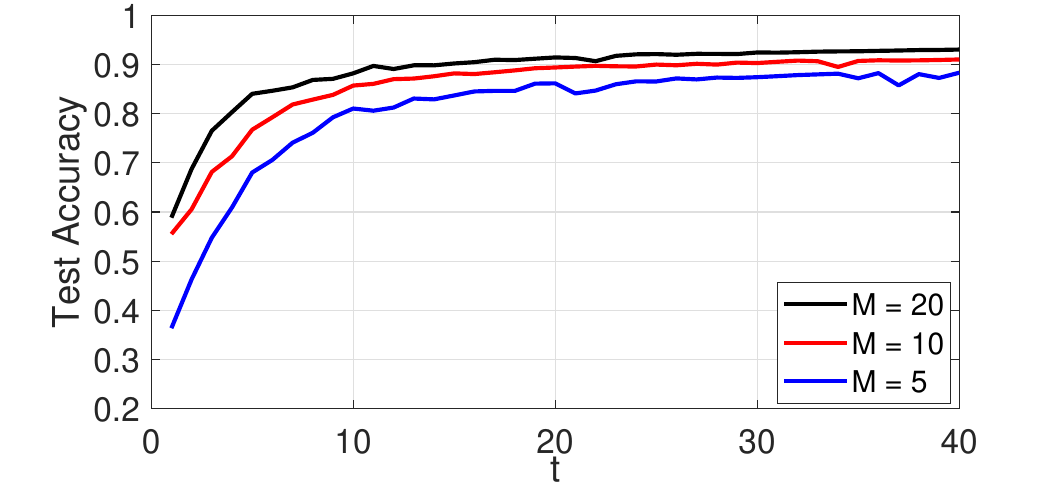} %mvalid.eps
	\vspace{-5pt}
	\caption{Test accuracy as a function of global iterations $t$ (non-i.i.d.)}
	\vspace{-5pt}
\end{figure}

\begin{figure}[tb!]
	\vspace{0pt}
	\centering
	\includegraphics[width =2.9in]{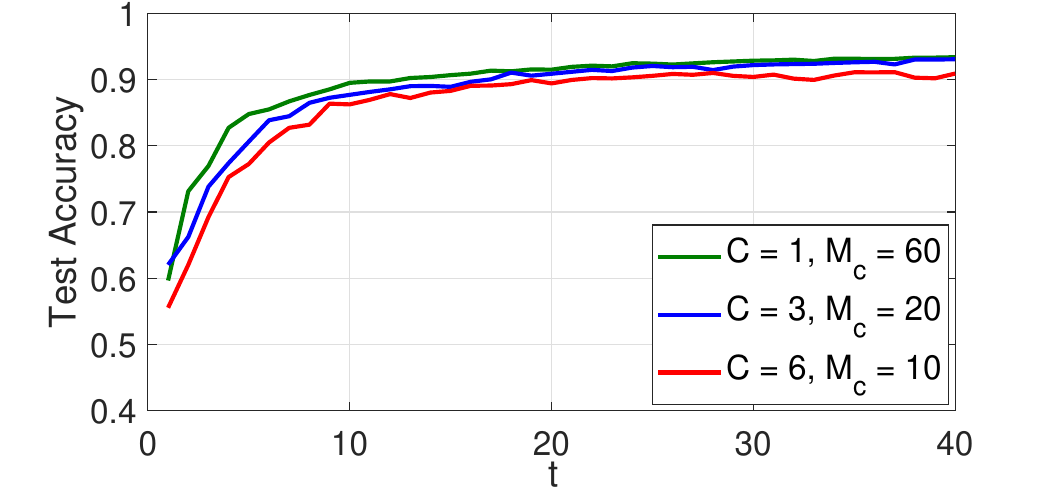} %mvalid.eps
	\vspace{-5pt}
	\caption{Test accuracy as a function of global iterations $t$ (non-i.i.d. and $M_\text{c} C = 60$)}
	\vspace{-5pt}
\end{figure}

\begin{figure}[tb!]
	\vspace{0pt}
	\centering
	\includegraphics[width =2.9in]{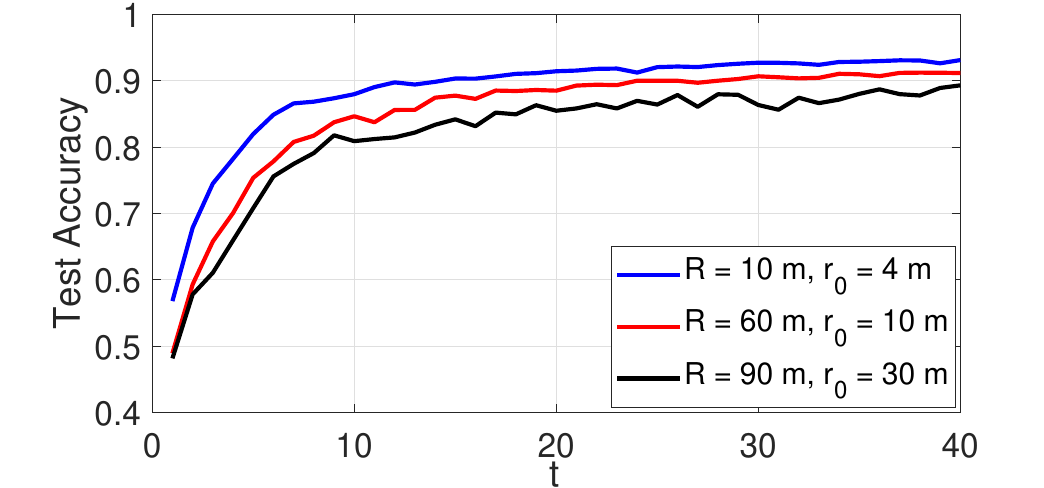} %mvalid.eps
	\vspace{-5pt}
	\caption{Test accuracy as a function of global iterations $t$ (non-i.i.d.)}
	\vspace{-5pt}
\end{figure}

\begin{figure}[tb!]
	\vspace{0pt}
	\centering
	\includegraphics[width =2.9in]{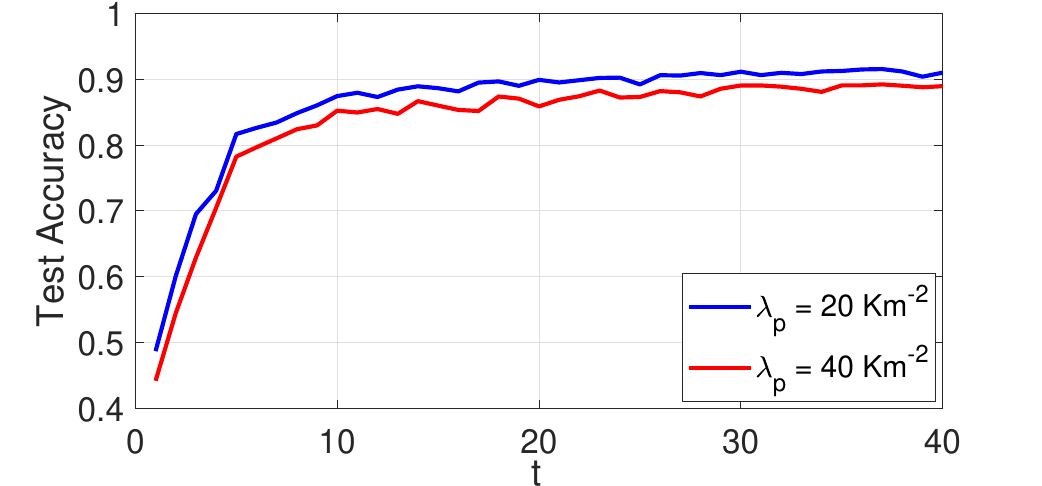} %mvalid.eps
	\vspace{-5pt}
	\caption{Test accuracy as a function of global iterations $t$ (non-i.i.d.)}
	\vspace{-5pt}
\end{figure}

\begin{figure}[tb!]
	\vspace{0pt}
	\centering
	\includegraphics[width =2.9in]{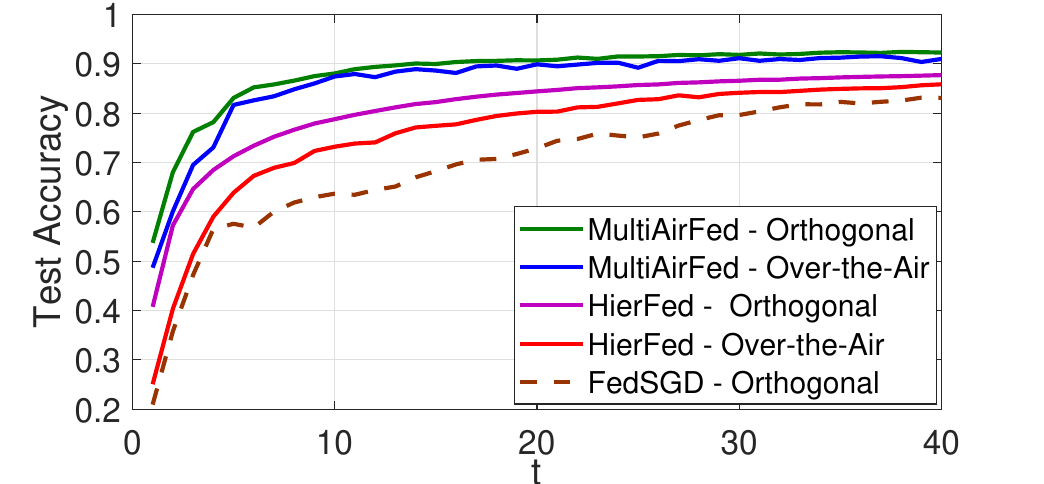} %mvalid.eps
	\vspace{-5pt}
	\caption{Test accuracy as a function of global iterations $t$ (non-i.i.d. and $\lambda_\text{p} = 20 \ \text{Km}^{-2}$)}
	\vspace{-10pt}
\end{figure}

\begin{figure}[tb!]
	\vspace{0pt}
	\centering
	\includegraphics[width =3.5in]{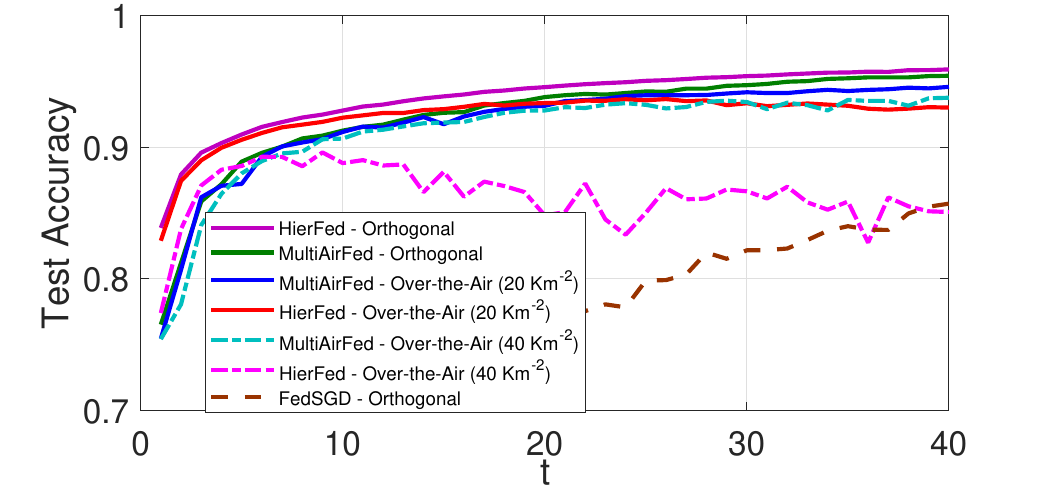} %mvalid.eps
	\vspace{-5pt}
	\caption{Test accuracy as a function of global iterations $t$ (i.i.d.)}
	\vspace{-10pt}
\end{figure}

In Fig. 2, the accuracy is shown for different intra-cluster iterations $\tau$ in the MNIST and i.i.d. scenario. As observed, increasing $\tau$ or $t$ improves the learning performance, justifying \textit{Remarks 1} and \textit{4}. The improvement gap is decreased in higher $\tau$ or $t$. Also, increasing $\tau$ accelerates convergence in terms of $t$. It shows that a minimum number of intra- and inter-cluster iterations can ensure a desirable performance. The latency $T_\text{L}$ given in \eqref{delay} is plotted in Fig. 3 for different $\tau$ in Fig. 2 when the target learning accuracy $95 \%$ is achieved. We assume $t_\text{BH} = 10 t_\text{BC}$. As observed, the latency is minimized at $\tau = 17$. For higher values, the convergence rate does not increase sufficiently to compensate for the longer time for the intra-cluster iterations.  

In Fig. 4, the accuracy is shown for different $\tau$ and local iterations $\gamma$ in the CIFAR-10 and i.i.d. scenario. The performance is improved with the increase in $\gamma$, which justifies \textit{Remark 4}. Also, by comparing the cases $(\tau = 6, \gamma = 12)$ and $(\tau = 12, \gamma = 2)$, the greater impact of $\tau$ compared to $\gamma$ on the performance is demonstrated. This is mainly because of the detrimental effect of the term $\frac{L^2 \mu^3}{4} \frac{\sigma^2\gamma (\gamma-1)}{B}$ on the optimality gap in \eqref{convergence}.

Figs 5-7 evaluate the effect of $C$ and $M$, based on the MNIST and non-i.i.d. scenario. On Fig. 5, we can observe that multi-server collaboration can significantly improve the accuracy. 
%compared to the single-server case. 
It justifies \textit{Remark 6}. That is because of accessing a diverse set of intra-cluster learning models and increasing the total active devices in the learning process over different clusters. Furthermore, as the inter-cluster uplink and downlink error increases, the degree of improvement diminishes at higher values of $C$. Fig. 6 demonstrates that the performance is improved as $M$ increases since the number of active devices that can participate in the learning process increases. It justifies \textit{Remark 7}.
In Fig. 7, the number of collaborating devices is kept constant $M_\text{c}\times C=60$, for $C = 1, 3, 6$, while in the non-collaborating clusters $M_\text{n} = 15$. The results suggest that consolidating a greater portion of collaborating devices within fewer clusters enhances the performance. This observation aligns with \textit{Remark 8} and can be attributed to the engagement of a larger number of devices in intra-cluster iterations.

Figs 8 and 9 studies the effects of the cluster size, as a function of $R$ and $r_0$, and the cluster density $\lambda_\text{p}$ in the MNIST and non-i.i.d. scenario. In Fig. 8, as the cluster size grows, the performance diminishes. This decline can be attributed to the devices in a cluster becoming closer to other clusters, leading to amplified interference. Similarly, Fig. 9 illustrates a reduction in accuracy with an increment in \(\lambda_\text{p}\), aligning with \textit{Remark 10}. This behavior is the consequence of the increasing interference.

In Figs 10 and 11, the learning performance of {\fontfamily{lmtt}\selectfont
	MultiAirFed} is compared with the conventional hierarchical FL ({\fontfamily{lmtt}\selectfont HierFed}) in \cite{letaief, bennis, tony} and {\fontfamily{lmtt}\selectfont
	FedSGD} in \cite{mcmahan} in the MNIST scenario. In {\fontfamily{lmtt}\selectfont FedSGD}, the edge servers act as simple relay nodes. In each iteration, gradients are aggregated at the edge server, and then are directly forwarded to the core server. Thus, the gradients from all the devices in the collaborating clusters are aggregated in each iteration.
{\fontfamily{lmtt}\selectfont
HierFed} differs from {\fontfamily{lmtt}\selectfont
MultiAirFed} in the intra-cluster iteration, as model parameters are uploaded to and aggregated at the edge servers, and the initial state at each local device is synchronized. This allows executing $\gamma >1$ local decent steps at each intra-cluster iteration. In the numerical results, $\gamma = 2$.
We consider the "over-the-air" transmission scheme proposed in Section IV both for {\fontfamily{lmtt}\selectfont
MultiAirFed} and for {\fontfamily{lmtt}\selectfont HierFed}. Additionally, we implement all the methods with "orthogonal" transmissions in both uplink and downlink, which eliminates interference by assuming unlimited communication resources. 
Fig. 10 considers non-i.i.d. data distribution.
%To better evaluate the impact of interference, we also consider a denser network with $\lambda_\text{p} = 40 \ \text{Km}^{-2}$ for the i.i.d. case. 
The results indicate that {\fontfamily{lmtt}\selectfont MultiAirFed} outperforms {\fontfamily{lmtt}\selectfont HierFed} by a substantial margin, for both transmission schemes. This highlights the robustness of {\fontfamily{lmtt}\selectfont MultiAirFed} against both data heterogeneity and interference. The less demanding i.i.d. scenario is shown on Fig. 11, however, now considering also a denser network with $\lambda_\text{p} = 40 \ \text{Km}^{-2}$. {\fontfamily{lmtt}\selectfont HierFed} outperforms {\fontfamily{lmtt}\selectfont MultiAirFed} when interference is absent, due to the multiple local steps, its performance is significantly impacted under the higher interference. While the accuracy is increased fast in the first iterations, the learning does not converge. This supports our reasoning in Section III, and motivates the use of {\fontfamily{lmtt}\selectfont MultiAirFed}. On both Figs 10 and 11, the performance of {\fontfamily{lmtt}\selectfont FedSGD} is weak even under orthogonal transmission, as this scheme does not take advantage of the hierarchical structure. Therefore, we do not evaluate the effect of interference.

\vspace{-4pt}
\section{Conclusions}
In this paper, we proposed a new two-level federated learning algorithm that leverages the hierarchical network architecture and intra- and inter-cluster collaborations for a higher communication efficiency and learning accuracy. To implement the proposed algorithm over wireless distributed systems independent of their scale and with minimum resource requirements, we presented an over-the-air aggregation scheme for the uplink and a bandwidth-limited broadcast scheme for the downlink, and determined how uplink and downlink interference impacts gradient and model aggregations in the algorithm. To minimize the interference-induced distortion on the estimations, we incorporated and optimized normalizing factors. We utilized PCP to characterize the spatial distribution of the devices and edge servers, derived a convergence bound of the learning process, and presented design remarks. Our results show that the PCP based modeling leads to useful insights on how the network parameters affect the interference and consequently the learning performance. The presented experimental results confirm the analytic findings and  demonstrate that the proposed gradient based hierarchical FL outperforms existing solutions, and achieves high accuracy, despite interference and data heterogeneity.

\vspace{-4pt}
\appendices
\section{Proof of Theorem 1}
In the proof, we show the index for the intra- and inter-cluster iterations. Then, the update of the learning model at global inter-cluster iteration $t+1$ is represented as
\begin{align}
\label{update}
&{\mathbf{w}}_{\mathbf{y}_0,0,t+1}^{\mathbf{o}} = \frac{1}{|{\cal A}_{\tau,t}|} \sum_{\mathbf{x} \in {\mathcal{C}}}^{} \sum_{\mathbf{y} \in {\cal A}_{\tau,t}^{\mathbf{x}}} \mathbf{w}_{\mathbf{y},\tau,\gamma,t}^{\mathbf{x}} +\frac{\boldsymbol{\epsilon_\text{u}}_{{t}}}{|{\cal A}_{\tau,t}|}+\frac{\boldsymbol{\epsilon_\text{d}}_{\mathbf{y}_0,{t}}}{|{\cal A}_{\tau,t}|}=\nonumber\\&\frac{1}{|{\cal A}_{\tau,t}|} \sum_{\mathbf{x} \in {\mathcal{C}}}^{} \sum_{\mathbf{y} \in {\cal A}_{\tau,t}^{\mathbf{x}}} \biggl(\mathbf{w}_{\mathbf{y},0,t}^{\mathbf{x}} - \mu_t \sum_{i=0}^{\tau-1} {\mathbf{g}}_{\mathbf{y},i,t}^\mathbf{x}-\mu_t \sum_{i=0}^{\gamma-1}\nonumber\\&\nabla F_{\mathbf{y}}^{\mathbf{x}}(\mathbf{w}_{\mathbf{y},\tau,i,t}^{\mathbf{x}})\biggr) +\frac{\boldsymbol{\epsilon_\text{u}}_{{t}}}{|{\cal A}_{\tau,t}|}+\frac{\boldsymbol{\epsilon_\text{d}}_{\mathbf{y}_0,{t}}}{|{\cal A}_{\tau,t}|}=-\frac{\mu_t}{|{\cal A}_{\tau,t}|} \sum_{\mathbf{x} \in {\mathcal{C}}}^{} \sum_{\mathbf{y} \in {\cal A}_{\tau,t}^{\mathbf{x}}}  \sum_{i=0}^{\tau-1}\nonumber\\& {\mathbf{g}}_{\mathbf{y},i,t}^\mathbf{x}-\frac{\mu_t}{|{\cal A}_{\tau,t}|} \sum_{\mathbf{x} \in {\mathcal{C}}}^{} \sum_{\mathbf{y} \in {\cal A}_{\tau,t}^{\mathbf{x}}}  \sum_{i=0}^{\gamma-1} \nabla F_{\mathbf{y}}^{\mathbf{x}}(\mathbf{w}_{\mathbf{y},\tau,i,t}^{\mathbf{x}}) +\frac{1}{|{\cal A}_{\tau,t}|} \sum_{\mathbf{x} \in {\mathcal{C}}}^{} \nonumber\\&\sum_{\mathbf{y} \in {\cal A}_{\tau,t}^{\mathbf{x}}} \mathbf{w}_{\mathbf{y},0,t}^{\mathbf{x}} +\frac{\boldsymbol{\epsilon_\text{u}}_{{t}}}{|{\cal A}_{\tau,t}|}+\frac{\boldsymbol{\epsilon_\text{d}}_{\mathbf{y}_0,{t}}}{|{\cal A}_{\tau,t}|} =-\frac{\mu_t}{|{\cal A}_{\tau,t}|} \sum_{\mathbf{x} \in {\mathcal{C}}}^{} \sum_{\mathbf{y} \in {\cal A}_{\tau,t}^\mathbf{x}}^{} \sum_{i=0}^{\tau-1} \nonumber\\
&\biggl({\mathbf{g}}_{i,t}^\mathbf{x}+\frac{\boldsymbol{\epsilon_\text{u}}_{i,t}^\mathbf{x}}{|{\cal A}_{i,t}^\mathbf{x}|}+\frac{\boldsymbol{\epsilon_\text{d}}_{\mathbf{y},i,t}^\mathbf{x}}{|{\cal A}_{i,t}^\mathbf{x}|}\biggr)- \frac{\mu_t}{|{\cal A}_{\tau,t}|} \sum_{\mathbf{x} \in {\mathcal{C}}}^{} \sum_{\mathbf{y} \in {\cal A}_{\tau,t}^{\mathbf{x}}}  \sum_{i=0}^{\gamma-1} \nonumber\\&\nabla F_{\mathbf{y}}^{\mathbf{x}}(\mathbf{w}_{\mathbf{y},\tau,i,t}^{\mathbf{x}})+\frac{1}{|{\cal A}_{\tau,t}|}\sum_{\mathbf{x} \in {\mathcal{C}}}^{} \sum_{\mathbf{y} \in {\cal A}_{\tau,t}^{\mathbf{x}}} \biggl(\mathbf{w}_{\mathbf{y}_0,0,t}^{\mathbf{o}}+\frac{\boldsymbol{\epsilon_\text{d}}_{\mathbf{y},t-1}}{|{\cal A}_{\tau,t-1}|}-\nonumber\\
&\frac{\boldsymbol{\epsilon_\text{d}}_{\mathbf{y}_0,{t-1}}}{|{\cal A}_{\tau,t-1}|}\biggr) +\frac{\boldsymbol{\epsilon_\text{u}}_{{t}}}{|{\cal A}_{\tau,t}|}+\frac{\boldsymbol{\epsilon_\text{d}}_{\mathbf{y}_0,{t}}}{|{\cal A}_{\tau,t}|}=-\frac{\mu_t}{|{\cal A}_{\tau,t}|} \sum_{\mathbf{x} \in {\mathcal{C}}}^{} |{\cal A}_{\tau,t}^\mathbf{x}| \sum_{i=0}^{\tau-1} {\mathbf{g}}_{i,t}^\mathbf{x}\nonumber\\&- \frac{\mu_t}{|{\cal A}_{\tau,t}|} \sum_{\mathbf{x} \in {\mathcal{C}}}^{} \sum_{\mathbf{y} \in {\cal A}_{\tau,t}^{\mathbf{x}}}  \sum_{i=0}^{\gamma-1} \nabla F_{\mathbf{y}}^{\mathbf{x}}(\mathbf{w}_{\mathbf{y},\tau,i,t}^{\mathbf{x}})-\frac{\mu_t}{|{\cal A}_{\tau,t}|} \sum_{\mathbf{x} \in {\mathcal{C}}}^{} \sum_{\mathbf{y} \in {\cal A}_{\tau,t}^\mathbf{x}}^{} \nonumber\\&\sum_{i=0}^{\tau-1} \left(\frac{\boldsymbol{\epsilon_\text{u}}_{i,t}^\mathbf{x}}{|{\cal A}_{i,t}^\mathbf{x}|}+\frac{\boldsymbol{\epsilon_\text{d}}_{\mathbf{y},i,t}^\mathbf{x}}{|{\cal A}_{i,t}^\mathbf{x}|}\right)+\frac{1}{|{\cal A}_{\tau,t}|} \sum_{\mathbf{x} \in {\mathcal{C}}}^{} \sum_{\mathbf{y} \in {\cal A}_{\tau,t}^{\mathbf{x}}} \biggl(\frac{\boldsymbol{\epsilon_\text{d}}_{\mathbf{y},t-1}}{|{\cal A}_{\tau,t-1}|}-\nonumber\\&\frac{\boldsymbol{\epsilon_\text{d}}_{\mathbf{y}_0,{t-1}}}{|{\cal A}_{\tau,t-1}|}\biggr) +\frac{\boldsymbol{\epsilon_\text{u}}_{{t}}}{|{\cal A}_{\tau,t}|}+\frac{\boldsymbol{\epsilon_\text{d}}_{\mathbf{y}_0,{t}}}{|{\cal A}_{\tau,t}|}+\mathbf{w}_{\mathbf{y}_0,0,t}^\mathbf{o} =-\frac{\mu_t}{|{\cal A}_{\tau,t}|} \sum_{\mathbf{x} \in {\mathcal{C}}}^{} |{\cal A}_{\tau,t}^\mathbf{x}|\nonumber\\
& \sum_{i=0}^{\tau-1} {\mathbf{g}}_{i,t}^\mathbf{x}- \frac{\mu_t}{|{\cal A}_{\tau,t}|} \sum_{\mathbf{x} \in {\mathcal{C}}}^{} \sum_{\mathbf{y} \in {\cal A}_{\tau,t}^{\mathbf{x}}}  \sum_{i=0}^{\gamma-1} \nabla F_{\mathbf{y}}^{\mathbf{x}}(\mathbf{w}_{\mathbf{y},\tau,i,t}^{\mathbf{x}})-\frac{\mu_t}{|{\cal A}_{\tau,t}|} \sum_{\mathbf{x} \in {\mathcal{C}}}^{}\nonumber\\
& | {\cal A}_{\tau,t}^\mathbf{x}| \sum_{i=0}^{\tau-1} \frac{\boldsymbol{\epsilon_\text{u}}_{i,t}^\mathbf{x}}{|{\cal A}_{i,t}^\mathbf{x}|}-\frac{\mu_t}{|{\cal A}_{\tau,t}|} \sum_{\mathbf{x} \in {\mathcal{C}}}^{} \sum_{\mathbf{y} \in {\cal A}_{\tau,t}^\mathbf{x}}^{} \sum_{i=0}^{\tau-1} \frac{\boldsymbol{\epsilon_\text{d}}_{\mathbf{y},i,t}^\mathbf{x}}{|{\cal A}_{i,t}^\mathbf{x}|}+\frac{1}{|{\cal A}_{\tau,t}|} \sum_{\mathbf{x} \in {\mathcal{C}}}^{} \nonumber\\
&\sum_{\mathbf{y} \in {\cal A}_{\tau,t}^{\mathbf{x}}} \left(\frac{\boldsymbol{\epsilon_\text{d}}_{\mathbf{y},t-1}}{|{\cal A}_{\tau,t-1}|}-\frac{\boldsymbol{\epsilon_\text{d}}_{\mathbf{y}_0,{t-1}}}{|{\cal A}_{\tau,t-1}|}\right) +\frac{\boldsymbol{\epsilon_\text{u}}_{{t}}}{|{\cal A}_{\tau,t}|}+\frac{\boldsymbol{\epsilon_\text{d}}_{\mathbf{y}_0,{t}}}{|{\cal A}_{\tau,t}|}+\mathbf{w}_{\mathbf{y}_0,0,t}^\mathbf{o}.
\end{align}
According to \eqref{update} and the $L$-Lipschitz continuous property in Assumption 1, we have
\begin{align}
\label{lip}
&F(\mathbf{w}_{\mathbf{y}_0,0,t+1}^\mathbf{o}) - F(\mathbf{w}_{\mathbf{y}_0,0,t}^\mathbf{o}) \leq \nabla F( \mathbf{w}_{\mathbf{y}_0,0,t}^\mathbf{o})^\top \bigl({\mathbf{w}}_{\mathbf{y}_0,0,t+1}^\mathbf{o} -\nonumber\\
& {\mathbf{w}}_{\mathbf{y}_0,0,t}^\mathbf{o}\bigr)+\frac{L}{2} \|{\mathbf{w}}_{\mathbf{y}_0,0,t+1}^\mathbf{o} - {\mathbf{w}}_{\mathbf{y}_0,0,t}^\mathbf{o}\|^2 =\nabla F( {\mathbf{w}}_{\mathbf{y}_0,0,t}^\mathbf{o})^\top \nonumber\\
&\biggl(-\frac{\mu_t}{|{\cal A}_{\tau,t}|} \sum_{\mathbf{x} \in {\mathcal{C}}}^{} |{\cal A}_{\tau,t}^\mathbf{x}| \sum_{i=0}^{\tau-1} {\mathbf{g}}_{i,t}^\mathbf{x}- \frac{\mu_t}{|{\cal A}_{\tau,t}|} \sum_{\mathbf{x} \in {\mathcal{C}}}^{} \sum_{\mathbf{y} \in {\cal A}_{\tau,t}^{\mathbf{x}}}  \sum_{i=0}^{\gamma-1} \nonumber\\
&\nabla F_{\mathbf{y}}^{\mathbf{x}}(\mathbf{w}_{\mathbf{y},\tau,i,t}^{\mathbf{x}})-\frac{\mu_t}{|{\cal A}_{\tau,t}|} \sum_{\mathbf{x} \in {\mathcal{C}}}^{} | {\cal A}_{\tau,t}^\mathbf{x}| \sum_{i=0}^{\tau-1} \frac{\boldsymbol{\epsilon_\text{u}}_{i,t}^\mathbf{x}}{|{\cal A}_{i,t}^\mathbf{x}|}-\frac{\mu_t}{|{\cal A}_{\tau,t}|} \sum_{\mathbf{x} \in {\mathcal{C}}}^{} \nonumber
\end{align}
\begin{align}
&\sum_{\mathbf{y} \in {\cal A}_{\tau,t}^\mathbf{x}}^{} \sum_{i=0}^{\tau-1} \frac{\boldsymbol{\epsilon_\text{d}}_{\mathbf{y},i,t}^\mathbf{x}}{|{\cal A}_{i,t}^\mathbf{x}|}+\frac{1}{|{\cal A}_{\tau,t}|} \sum_{\mathbf{x} \in {\mathcal{C}}}^{} \sum_{\mathbf{y} \in {\cal A}_{\tau,t}^{\mathbf{x}}} \biggl(\frac{\boldsymbol{\epsilon_\text{d}}_{\mathbf{y},t-1}}{|{\cal A}_{\tau,t-1}|}-\frac{\boldsymbol{\epsilon_\text{d}}_{\mathbf{y}_0,{t-1}}}{|{\cal A}_{\tau,t-1}|}\nonumber\\
&\biggr) +\frac{\boldsymbol{\epsilon_\text{u}}_{{t}}}{|{\cal A}_{\tau,t}|}+\frac{\boldsymbol{\epsilon_\text{d}}_{\mathbf{y}_0,{t}}}{|{\cal A}_{\tau,t}|}\biggr)+\frac{L}{2} \biggl\Vert -\frac{\mu_t}{|{\cal A}_{\tau,t}|} \sum_{\mathbf{x} \in {\mathcal{C}}}^{} |{\cal A}_{\tau,t}^\mathbf{x}| \sum_{i=0}^{\tau-1} {\mathbf{g}}_{i,t}^\mathbf{x}-\nonumber\\
& \frac{\mu_t}{|{\cal A}_{\tau,t}|} \sum_{\mathbf{x} \in {\mathcal{C}}}^{} \sum_{\mathbf{y} \in {\cal A}_{\tau,t}^{\mathbf{x}}}  \sum_{i=0}^{\gamma-1} \nabla F_{\mathbf{y}}^{\mathbf{x}}(\mathbf{w}_{\mathbf{y},\tau,i,t}^{\mathbf{x}})-\frac{\mu_t}{|{\cal A}_{\tau,t}|} \sum_{\mathbf{x} \in {\mathcal{C}}}^{} | {\cal A}_{\tau,t}^\mathbf{x}| \sum_{i=0}^{\tau-1}\nonumber\\
& \frac{\boldsymbol{\epsilon_\text{u}}_{i,t}^\mathbf{x}}{|{\cal A}_{i,t}^\mathbf{x}|}-\frac{\mu_t}{|{\cal A}_{\tau,t}|} \sum_{\mathbf{x} \in {\mathcal{C}}}^{} \sum_{\mathbf{y} \in {\cal A}_{\tau,t}^\mathbf{x}}^{} \sum_{i=0}^{\tau-1} \frac{\boldsymbol{\epsilon_\text{d}}_{\mathbf{y},i,t}^\mathbf{x}}{|{\cal A}_{i,t}^\mathbf{x}|}+\frac{1}{|{\cal A}_{\tau,t}|} \sum_{\mathbf{x} \in {\mathcal{C}}}^{} \sum_{\mathbf{y} \in {\cal A}_{\tau,t}^{\mathbf{x}}}\nonumber\\
& \left(\frac{\boldsymbol{\epsilon_\text{d}}_{\mathbf{y},t-1}}{|{\cal A}_{\tau,t-1}|}-\frac{\boldsymbol{\epsilon_\text{d}}_{\mathbf{y}_0,{t-1}}}{|{\cal A}_{\tau,t-1}|}\right) +\frac{\boldsymbol{\epsilon_\text{u}}_{{t}}}{|{\cal A}_{\tau,t}|}+\frac{\boldsymbol{\epsilon_\text{d}}_{\mathbf{y}_0,{t}}}{|{\cal A}_{\tau,t}|}\biggr\Vert^2.
\end{align}
By taking expectation on both sides of \eqref{lip} and considering the independency of error terms, we continue as
\begin{align}
\label{lipexp}
&\mathbb{E}\left\{F({\mathbf{w}}_{\mathbf{y}_0,0,t+1}^\mathbf{o})-F({\mathbf{w}}_{\mathbf{y}_0,0,t}^\mathbf{o})\right\} \leq -\frac{\mu_t}{|{\cal A}_{\tau,t}|} \sum_{\mathbf{x} \in {\mathcal{C}}}^{} |{\cal A}_{\tau,t}^\mathbf{x}| \sum_{i=0}^{\tau-1} \nonumber\\&\mathbb{E}\left\{\nabla F( {\mathbf{w}}_{\mathbf{y}_0,0,t}^\mathbf{o})^\top{\mathbf{g}}_{i,t}^\mathbf{x}\right\} - \frac{\mu_t}{|{\cal A}_{\tau,t}|} \sum_{\mathbf{x} \in {\mathcal{C}}}^{} \sum_{\mathbf{y} \in {\cal A}_{\tau,t}^{\mathbf{x}}}  \sum_{i=0}^{\gamma-1} \nonumber\\
&\mathbb{E}\left\{\nabla F( {\mathbf{w}}_{\mathbf{y}_0,0,t}^\mathbf{o})^\top\nabla F_{\mathbf{y}}^{\mathbf{x}}(\mathbf{w}_{\mathbf{y},\tau,i,t}^{\mathbf{x}})\right\}
+\frac{L\mu_t^2}{2} \mathbb{E}\biggl\{\biggl\Vert \sum_{\mathbf{x} \in {\mathcal{C}}}^{} \frac{|{\cal A}_{\tau,t}^\mathbf{x}|}{|{\cal A}_{\tau,t}|} \sum_{i=0}^{\tau-1}\nonumber\\
& {\mathbf{g}}_{i,t}^\mathbf{x}+\frac{1}{|{\cal A}_{\tau,t}|} \sum_{\mathbf{x} \in {\mathcal{C}}}^{} \sum_{\mathbf{y} \in {\cal A}_{\tau,t}^{\mathbf{x}}}  \sum_{i=0}^{\gamma-1} \nabla F_{\mathbf{y}}^{\mathbf{x}}(\mathbf{w}_{\mathbf{y},\tau,i,t}^{\mathbf{x}})\biggr\Vert^2\biggr\}+\frac{L\mu_t^2}{2|{\cal A}_{\tau,t}|^2}\nonumber\\
&\sum_{\mathbf{x} \in {\mathcal{C}}}^{} | {\cal A}_{\tau,t}^\mathbf{x}|^2 \sum_{i=0}^{\tau-1} \frac{\mathbb{E}\left\{\|\boldsymbol{\epsilon_\text{u}}_{i,t}^\mathbf{x}\|^2\right\}}{|{\cal A}_{i,t}^\mathbf{x}|^2}+\frac{L\mu_t^2}{2|{\cal A}_{\tau,t}|^2} \sum_{\mathbf{x} \in {\mathcal{C}}}^{} \sum_{\mathbf{y} \in {\cal A}_{\tau,t}^\mathbf{x}}^{} \sum_{i=0}^{\tau-1}\nonumber\\
& \frac{\mathbb{E}\left\{\|\boldsymbol{\epsilon_\text{d}}_{\mathbf{y},i,t}^\mathbf{x}\|^2\right\}}{|{\cal A}_{i,t}^\mathbf{x}|^2}+\frac{L}{2|{\cal A}_{\tau,t}|^2} \sum_{\mathbf{x} \in {\mathcal{C}}}^{} \sum_{\mathbf{y} \in {\cal A}_{\tau,t}^{\mathbf{x}}} \frac{\mathbb{E}\left\{\|\boldsymbol{\epsilon_\text{d}}_{\mathbf{y},t-1}\|^2\right\}}{|{\cal A}_{\tau,t-1}|^2}+\frac{L}{2}\nonumber\\&\frac{\mathbb{E}\left\{\|\boldsymbol{\epsilon_\text{d}}_{\mathbf{y}_0,{t-1}}\|^2\right\}}{|{\cal A}_{\tau,t-1}|^2} +\frac{L}{2}\frac{\mathbb{E}\left\{\|\boldsymbol{\epsilon_\text{u}}_{{t}}\|^2\right\}}{|{\cal A}_{\tau,t}|^2}+\frac{L}{2}\frac{\mathbb{E}\left\{\|\boldsymbol{\epsilon_\text{d}}_{\mathbf{y}_0,{t}}\|^2\right\}}{|{\cal A}_{\tau,t}|^2}.
\end{align}
Next, we bound the first term of the right-hand side (RHS) in \eqref{lipexp}. We can write its inner-sum term as
\begin{align}
\label{curlexp}
&\mathbb{E}\left\{\nabla F( {\mathbf{w}}_{\mathbf{y}_0,0,t}^\mathbf{o})^\top {\mathbf{g}}_{i,t}^{\mathbf{x}}\right\} = \mathbb{E}\biggl\{\nabla F( {\mathbf{w}}_{\mathbf{y}_0,0,t}^\mathbf{o})^\top \frac{1}{|{\cal A}_{i,t}^\mathbf{x}|} \sum_{\mathbf{y}\in {\cal A}_{i,t}^\mathbf{x}}^{} \nonumber\\
&\nabla F_{\mathbf{y}}^{\mathbf{x}}(\mathbf{w}_{{\mathbf{y}},i,t}^\mathbf{x})\biggr\} =\mathbb{E}\left\{ \frac{\sum_{\mathbf{y}\in {\cal A}_{i,t}^\mathbf{x}}^{} \nabla F( {\mathbf{w}}_{\mathbf{y}_0,0,t}^\mathbf{o})^\top \nabla F_{\mathbf{y}}^{\mathbf{x}}(\mathbf{w}_{{\mathbf{y}},i,t}^\mathbf{x})}{|{\cal A}_{i,t}^\mathbf{x}|} \right\} \nonumber\\
&= \frac{\sum_{{\mathbf{y}}\in {\cal A}_{i,t}^\mathbf{x}}^{} \mathbb{E}\left\{\nabla F( {\mathbf{w}}_{\mathbf{y}_0,0,t}^\mathbf{o})^\top \nabla F(\mathbf{w}_{{\mathbf{y}},i,t}^\mathbf{x})\right\}}{|{\cal A}_{i,t}^\mathbf{x}|}.
\end{align}
Using the equality $\|\mathbf{a}_1-\mathbf{a}_2\|^2 = \|\mathbf{a}_1\|^2+\|\mathbf{a}_2\|^2- 2\mathbf{a}_1^\top \mathbf{a}_2$ for any vectors $\mathbf{a}_1$ and $\mathbf{a}_2$, the term in the sum in \eqref{curlexp} can be written as 
\begin{align}
\label{diffexp}
&\mathbb{E}\left\{\nabla F( {\mathbf{w}}_{\mathbf{y}_0,0,t}^\mathbf{o})^\top \nabla F(\mathbf{w}_{{\mathbf{y}},i,t}^\mathbf{x})\right\} =\nonumber\\
&  \frac{1}{2} \mathbb{E}\left\{\|\nabla F( {\mathbf{w}}_{\mathbf{y}_0,0,t}^\mathbf{o})\|^2\right\} +\frac{1}{2} \mathbb{E}\left\{\|\nabla F(\mathbf{w}_{{\mathbf{y}},i,t}^\mathbf{x})\|^2\right\} \nonumber\\
& - \frac{1}{2} \mathbb{E}\left\{\|\nabla F( {\mathbf{w}}_{\mathbf{y}_0,0,t}^\mathbf{o})- \nabla F(\mathbf{w}_{{\mathbf{y}},i,t}^\mathbf{x})\|^2\right\}.
\end{align}
From Assumption 1, the last term in \eqref{diffexp} is bounded as
\begin{align}
\label{diffnorm}
&\mathbb{E}\left\{\|\nabla F( {\mathbf{w}}_{\mathbf{y}_0,0,t}^\mathbf{o})- \nabla F(\mathbf{w}_{{\mathbf{y}},i,t}^\mathbf{x})\|^2\right\} \leq L^2 \mathbb{E}\bigl\{\|\mathbf{w}_{\mathbf{y}_0,0,t}^\mathbf{o}-\nonumber\\&\mathbf{w}_{{\mathbf{y}},i,t}^\mathbf{x}\|^2\bigr\}= L^2  \mathbb{E}\biggl\{\biggl\Vert\mathbf{w}_{\mathbf{y},0,t}^\mathbf{x}-\frac{\boldsymbol{\epsilon_\text{d}}_{\mathbf{y},t-1}}{|{\cal A}_{\tau,t-1}|}+\frac{\boldsymbol{\epsilon_\text{d}}_{\mathbf{y}_0,{t-1}}}{|{\cal A}_{\tau,t-1}|}-\mathbf{w}_{{\mathbf{y}},i,t}^\mathbf{x}\biggr\Vert^2\nonumber\\&\biggr\} = L^2 \mathbb{E}\biggl\{\biggl\Vert-\mu_t \sum_{j=0}^{i-1} \mathbf{g}_{\mathbf{y},j,t}^\mathbf{x}-\frac{\boldsymbol{\epsilon_\text{d}}_{\mathbf{y},t-1}}{|{\cal A}_{\tau,t-1}|}+\frac{\boldsymbol{\epsilon_\text{d}}_{\mathbf{y}_0,{t-1}}}{|{\cal A}_{\tau,t-1}|}\biggr \Vert^2\biggr\}=L^2\nonumber\\
&\mu_t^2 \mathbb{E}\biggl\{\biggl\Vert\sum_{j=0}^{i-1}{\mathbf{g}}_{\mathbf{y},j,t}^\mathbf{x} \biggr\Vert^2\biggr\} +L^2\frac{\mathbb{E}\left\{\|\boldsymbol{\epsilon_\text{d}}_{\mathbf{y},t-1}\|^2\right\}}{|{\cal A}_{\tau,t-1}|^2} +L^2\nonumber\\
&\frac{{\mathbb{E}\left\{\|\boldsymbol{\epsilon_\text{d}}_{\mathbf{y}_0,{t-1}}\|^2\right\}}}{{|{\cal A}_{\tau,t-1}|^2} }= L^2 \mu_t^2 \mathbb{E}\biggl\{\biggl\Vert\sum_{j=0}^{i-1}{\mathbf{g}}_{j,t}^\mathbf{x}+\frac{\boldsymbol {\epsilon_\text{u}}_{j,t}^{\mathbf{x}}  }{{|{\cal A}_{j,t}^\mathbf{x}|} }+\frac{\boldsymbol {\epsilon_\text{d}}_{\mathbf{y},j,t}^{\mathbf{x}}  }{{|{\cal A}_{j,t}^\mathbf{x}|} } \biggr\Vert^2\biggr\}\nonumber\\&+ L^2\frac{\mathbb{E}\left\{\|\boldsymbol{\epsilon_\text{d}}_{\mathbf{y},t-1}\|^2\right\}}{{|{\cal A}_{\tau,t-1}|^2} }+L^2\frac{\mathbb{E}\left\{\|\boldsymbol{\epsilon_\text{d}}_{\mathbf{y}_0,{t-1}}\|^2\right\}}{{|{\cal A}_{\tau,t-1}|^2} }= L^2 \mu_t^2\nonumber\\& \mathbb{E}\biggl\{\biggl\Vert\sum_{j=0}^{i-1}{\mathbf{g}}_{j,t}^\mathbf{x} \biggr\Vert^2\biggr\}+L^2 \mu_t^2 \sum_{j=0}^{i-1} \frac{\mathbb{E}\left\{\|\boldsymbol {\epsilon_\text{u}}_{j,t}^{\mathbf{x}} \|^2 \right\}}{{|{\cal A}_{j,t}^\mathbf{x}|^2} } +L^2 \mu_t^2 \sum_{j=0}^{i-1}\nonumber\\
& \frac{\mathbb{E}\left\{\|\boldsymbol {\epsilon_\text{d}}_{\mathbf{y},j,t}^{\mathbf{x}} \|^2 \right\}}{{|{\cal A}_{j,t}^\mathbf{x}|^2} }+ L^2\frac{\mathbb{E}\left\{\|\boldsymbol{\epsilon_\text{d}}_{\mathbf{y},t-1}\|^2\right\}}{{|{\cal A}_{\tau,t-1}|^2} }+L^2\frac{\mathbb{E}\left\{\|\boldsymbol{\epsilon_\text{d}}_{\mathbf{y}_0,{t-1}}\|^2\right\}}{{|{\cal A}_{\tau,t-1}|^2} },
\end{align}
where using the equality $\mathbb{E}\left\{\|\mathbf{a}\|^2\right\} = \|\mathbb{E}\left\{\mathbf{a}\right\}\|^2 + \mathbb{E}\left\{\|\mathbf{a}-\mathbb{E}\left\{\mathbf{a}\right\}\|^2\right\}$, we have
\begin{align}
\label{normexp}
&\mathbb{E}\biggl\{\biggl\Vert\sum_{j=0}^{i-1}{\mathbf{g}}_{j,t}^\mathbf{x} \biggr\Vert^2\biggr\} = \mathbb{E}\biggl\{\biggl\Vert  \sum_{j=0}^{i-1}\frac{1}{|{\cal A}_{j,t}^\mathbf{x}|}\sum_{\mathbf{y} \in {\cal A}_{j,t}^\mathbf{x} }^{} \nabla F_\mathbf{y}^\mathbf{x}(\mathbf{w}_{\mathbf{y},j,t}^\mathbf{x})\biggr\Vert^2\biggr\} \nonumber\\&= \mathbb{E}\biggl\{\biggl\Vert  \sum_{j=0}^{i-1}\frac{1}{|{\cal A}_{j,t}^\mathbf{x}|}\sum_{\mathbf{y} \in {\cal A}_{j,t}^\mathbf{x} }^{} \nabla F(\mathbf{w}_{\mathbf{y},j,t}^\mathbf{x})\biggr\Vert^2\biggr\}+
\nonumber\\
&\mathbb{E}\biggl\{\biggl\Vert \sum_{j=0}^{i-1}\frac{1}{|{\cal A}_{j,t}^\mathbf{x}|}\sum_{\mathbf{y} \in {\cal A}_{j,t}^\mathbf{x} }^{} \left(\nabla F_\mathbf{y}^\mathbf{x}(\mathbf{w}_{\mathbf{y},j,t}^\mathbf{x})- \nabla F(\mathbf{w}_{\mathbf{y},j,t}^\mathbf{x})\right)\biggr\Vert^2\biggr\},
\end{align}
where the first term of RHS can be upper-bounded as
\begin{align}
\label{RHS1}
&\mathbb{E}\biggl\{\biggl\Vert  \sum_{j=0}^{i-1}\frac{1}{|{\cal A}_{j,t}^\mathbf{x}|}\sum_{\mathbf{y} \in {\cal A}_{j,t}^\mathbf{x} }^{} \nabla F(\mathbf{w}_{\mathbf{y},j,t}^\mathbf{x})\biggr\Vert^2\biggr\} \nonumber\\
&\stackrel{(a)}{\leq}  i\sum_{j=0}^{i-1}\mathbb{E}\biggl\{\biggl\Vert \frac{1}{|{\cal A}_{j,t}^\mathbf{x}|}\sum_{\mathbf{y} \in {\cal A}_{j,t}^\mathbf{x} }^{} \nabla F(\mathbf{w}_{\mathbf{y},j,t}^\mathbf{x})\biggr\Vert^2\biggr\}\nonumber\\&\stackrel{(b)}{\leq}  i \sum_{j=0}^{i-1} \frac{1}{|{\cal A}_{j,t}^\mathbf{x}|}\sum_{\mathbf{y} \in {\cal A}_{j,t}^\mathbf{x}}^{} \mathbb{E}\left\{\|\nabla F(\mathbf{w}_{\mathbf{y},j,t}^\mathbf{x})\|^2\right\},
\end{align}
where $(a)$ comes from the inequality
of arithmetic and geometric means, i.e., $(\sum_{i=1}^{I}a_i)^2\leq I \sum_{i=1}^{I}a_i^2$, and $(b)$ is from the convexity of the function $\|.\|^2$. The second term of RHS in \eqref{normexp} can be upper-bounded as
\begin{align}
\label{RHS2}
&\mathbb{E}\biggl\{\biggl\Vert \sum_{j=0}^{i-1}\frac{1}{|{\cal A}_{j,t}^\mathbf{x}|}\sum_{\mathbf{y} \in {\cal A}_{j,t}^\mathbf{x} }^{} \left(\nabla F_\mathbf{y}^\mathbf{x}(\mathbf{w}_{\mathbf{y},j,t}^\mathbf{x})- \nabla F(\mathbf{w}_{\mathbf{y},j,t}^\mathbf{x})\right)\biggr\Vert^2\biggr\} \nonumber\\
&\stackrel{(c)}{=} \sum_{j=0}^{i-1}\mathbb{E}\biggl\{\biggl\Vert\frac{1}{|{\cal A}_{j,t}^\mathbf{x}|}\sum_{\mathbf{y} \in {\cal A}_{j,t}^\mathbf{x} }^{} \left(\nabla F_\mathbf{y}^\mathbf{x}(\mathbf{w}_{\mathbf{y},j,t}^\mathbf{x})- \nabla F(\mathbf{w}_{\mathbf{y},j,t}^\mathbf{x})\right)\biggr\Vert^2\biggr\}\nonumber
\end{align}
\begin{align}
&\stackrel{(d)}{=}\sum_{j=0}^{i-1}\frac{1}{|{\cal A}_{j,t}^\mathbf{x}|^2}\sum_{\mathbf{y} \in {\cal A}_{j,t}^\mathbf{x} }^{} \mathbb{E}\biggl\{\biggl\Vert\nabla F_\mathbf{y}^\mathbf{x}(\mathbf{w}_{\mathbf{y},j,t}^\mathbf{x})- \nabla F(\mathbf{w}_{\mathbf{y},j,t}^\mathbf{x})\biggr\Vert^2\biggr\}\nonumber\\
&\stackrel{(e)}{\leq} \sum_{j=0}^{i-1}\frac{1}{|{\cal A}_{j,t}^\mathbf{x}|^2}\sum_{\mathbf{y} \in {\cal A}_{j,t}^\mathbf{x} }^{} \frac{\sigma^2}{B} = \frac{\sigma^2}{B} \sum_{j=0}^{i-1} \frac{1}{|{\cal A}_{j,t}^\mathbf{x}|},
\end{align}
where $(c)$ and $(d)$ are due to the fact that for any $\mathbf{y}_1\not = \mathbf{y}_2, j_1 \not = j_2$
\begin{align}
&\mathbb{E}\biggl\{\left(\nabla F_{\mathbf{y}_1}^\mathbf{x}(\mathbf{w}_{\mathbf{y}_1,j_1,t}^\mathbf{x})- \nabla F(\mathbf{w}_{\mathbf{y}_1,j_1,t}^\mathbf{x})\right)^\top\times \nonumber\\& \left(\nabla F_{\mathbf{y}_2}^\mathbf{x}(\mathbf{w}_{\mathbf{y}_2,j_2,t}^\mathbf{x})- \nabla F(\mathbf{w}_{\mathbf{y}_2,j_2,t}^\mathbf{x})\right)\biggr\} = \mathbb{E}\biggl\{\mathbb{E}_{{\xi}_{\mathbf{y}_1,j_1}^\mathbf{x}}\biggl\{\nonumber\\&\left(\nabla F_{\mathbf{y}_1}^\mathbf{x}(\mathbf{w}_{\mathbf{y}_1,j_1,t}^\mathbf{x})- \nabla F(\mathbf{w}_{\mathbf{y}_1,j_1,t}^\mathbf{x})\right)^\top\biggr\} \times \nonumber\\&\left(\nabla F_{\mathbf{y}_2}^\mathbf{x}(\mathbf{w}_{\mathbf{y}_2,j_2,t}^\mathbf{x})- \nabla F(\mathbf{w}_{\mathbf{y}_2,j_2,t}^\mathbf{x})\right)|{\xi}_{\mathbf{y}_1,j_1}^\mathbf{x}\biggr\} = 0,
\end{align}
where $\mathbb{E}_{{\xi}_{\mathbf{y}_1,j_1}^\mathbf{x}}\left\{\nabla F_{\mathbf{y}_1}^\mathbf{x}(\mathbf{w}_{\mathbf{y}_1,j_1,t}^\mathbf{x})- \nabla F(\mathbf{w}_{\mathbf{y}_1,j_1,t}^\mathbf{x})\right\} = \mathbf{0}$. Then, $(e)$ comes from the Assumption 2. Replacing \eqref{RHS1} and \eqref{RHS2} in \eqref{normexp} and then replacing the result in \eqref{diffnorm}, we have
\begin{align}
\label{normdiff}
&\mathbb{E}\left\{\|\nabla F( {\mathbf{w}}_{\mathbf{y}_0,0,t}^\mathbf{o})- \nabla F(\mathbf{w}_{\mathbf{y},i,t}^\mathbf{x})\|^2\right\} \leq  L^2 \mu_t^2 i \sum_{j=0}^{i-1} \frac{1}{|{\cal A}_{j,t}^\mathbf{x}|}\nonumber\\&\sum_{\mathbf{y} \in {\cal A}_{j,t}^\mathbf{x}}^{} {\mathbb{E}\left\{\|\nabla F(\mathbf{w}_{\mathbf{y},j,t}^\mathbf{x})\|^2\right\}}+L^2 \mu_t^2\frac{\sigma^2}{B} \sum_{j=0}^{i-1} \frac{1}{|{\cal A}_{j,t}^\mathbf{x}|} +L^2 \mu_t^2 \sum_{j=0}^{i-1}\nonumber\\& \frac{\mathbb{E}\left\{\|\boldsymbol {\epsilon_\text{u}}_{j,t}^{\mathbf{x}} \|^2 \right\}}{{|{\cal A}_{j,t}^\mathbf{x}|^2} } +L^2 \mu_t^2 \sum_{j=0}^{i-1} \frac{\mathbb{E}\left\{\|\boldsymbol {\epsilon_\text{d}}_{\mathbf{y},j,t}^{\mathbf{x}} \|^2 \right\}}{{|{\cal A}_{j,t}^\mathbf{x}|^2} }\nonumber\\&+ L^2\frac{\mathbb{E}\left\{\|\boldsymbol{\epsilon_\text{d}}_{\mathbf{y},t-1}\|^2\right\}}{{|{\cal A}_{\tau,t-1}|^2} }+L^2\frac{\mathbb{E}\left\{\|\boldsymbol{\epsilon_\text{d}}_{\mathbf{y}_0,{t-1}}\|^2\right\}}{{|{\cal A}_{\tau,t-1}|^2} },
\end{align}
and then replacing \eqref{normdiff} in \eqref{diffexp} and replacing the result in \eqref{curlexp}, we obtain the following bound  
\begin{align}
\label{1RHS1}
&-\frac{\mu_t}{|{\cal A}_{\tau,t}|} \sum_{\mathbf{x} \in {\mathcal{C}}}^{} |{\cal A}_{\tau,t}^\mathbf{x}| \sum_{i=0}^{\tau-1} \mathbb{E}\left\{\nabla F( {\mathbf{w}}_{\mathbf{y}_0,0,t}^\mathbf{o})^\top{\mathbf{g}}_{i,t}^\mathbf{x}\right\} \leq -\frac{\mu_t \tau}{2}\times\nonumber\\
& \mathbb{E}\left\{\|\nabla F( {\mathbf{w}}_{\mathbf{y}_0,0,t}^\mathbf{o})\|^2\right\}-\frac{\mu_t}{2|{\cal A}_{\tau,t}|} \sum_{\mathbf{x} \in {\mathcal{C}}}^{} |{\cal A}_{\tau,t}^\mathbf{x}| \sum_{i=0}^{\tau-1}\frac{1}{|{\cal A}_{i,t}^\mathbf{x}|}\sum_{{\mathbf{y}}\in {\cal A}_{i,t}^\mathbf{x}}^{}\nonumber\\&\mathbb{E}\left\{\|\nabla F( {\mathbf{w}}_{\mathbf{y},i,t}^\mathbf{x})\|^2\right\}+\frac{L^2 \mu_t^3}{2{|{\cal A}_{\tau,t}|}} \sum_{\mathbf{x} \in {\mathcal{C}}}^{} |{\cal A}_{\tau,t}^\mathbf{x}| \sum_{i=0}^{\tau-1} i \sum_{j=0}^{i-1} \frac{1}{|{\cal A}_{j,t}^\mathbf{x}|}\nonumber\\&\sum_{\mathbf{y} \in {\cal A}_{j,t}^\mathbf{x}}^{} \mathbb{E}\left\{\|\nabla F(\mathbf{w}_{\mathbf{y},j,t}^\mathbf{x})\|^2\right\}+\frac{L^2 \mu_t^3}{2{|{\cal A}_{\tau,t}|}}\frac{\sigma^2}{B} \sum_{\mathbf{x} \in {\mathcal{C}}}^{} |{\cal A}_{\tau,t}^\mathbf{x}| \sum_{i=0}^{\tau-1} \sum_{j=0}^{i-1} \nonumber\\
&\frac{1}{|{\cal A}_{j,t}^\mathbf{x}|}+\frac{L^2 \mu_t^3}{2{|{\cal A}_{\tau,t}|}}\sum_{\mathbf{x} \in {\mathcal{C}}}^{} |{\cal A}_{\tau,t}^\mathbf{x}| \sum_{i=0}^{\tau-1} \sum_{j=0}^{i-1} \frac{\mathbb{E}\left\{\|\boldsymbol {\epsilon_\text{u}}_{j,t}^{\mathbf{x}} \|^2 \right\}}{{|{\cal A}_{j,t}^\mathbf{x}|^2} } +\frac{L^2 \mu_t^3}{2{|{\cal A}_{\tau,t}|}}\nonumber\\
&\sum_{\mathbf{x} \in {\mathcal{C}}}^{} |{\cal A}_{\tau,t}^\mathbf{x}| \sum_{i=0}^{\tau-1} \frac{1}{|{\cal A}_{i,t}^\mathbf{x}|}\sum_{\mathbf{y} \in {\cal A}_{i,t}^\mathbf{x}}^{}\sum_{j=0}^{i-1} \frac{\mathbb{E}\left\{\|\boldsymbol {\epsilon_\text{d}}_{\mathbf{y},j,t}^{\mathbf{x}} \|^2 \right\}}{{|{\cal A}_{j,t}^\mathbf{x}|^2} }+\frac{L^2 \mu_t}{2{|{\cal A}_{\tau,t}|}}\nonumber\\
&\sum_{\mathbf{x} \in {\mathcal{C}}}^{} |{\cal A}_{\tau,t}^\mathbf{x}| \sum_{i=0}^{\tau-1} \frac{1}{|{\cal A}_{i,t}^\mathbf{x}|}\sum_{\mathbf{y} \in {\cal A}_{i,t}^\mathbf{x}}^{} \frac{\mathbb{E}\left\{\|\boldsymbol{\epsilon_\text{d}}_{\mathbf{y},t-1}\|^2\right\}}{{|{\cal A}_{\tau,t-1}|^2} } \nonumber
\end{align}
\begin{align}
&+\frac{L^2 \mu_t \tau}{2} \frac{\mathbb{E}\left\{\|\boldsymbol{\epsilon_\text{d}}_{\mathbf{y}_0,{t-1}}\|^2\right\}}{{|{\cal A}_{\tau,t-1}|^2} }.
\end{align}
Next, following the same approach as in \eqref{curlexp}-\eqref{normdiff}, we can bound the second term of RHS in \eqref{lipexp} as
\begin{align}
\label{new_prod}
&-\frac{\mu_t}{|{\cal A}_{\tau,t}|} \sum_{\mathbf{x} \in {\mathcal{C}}}^{} \sum_{\mathbf{y} \in {\cal A}_{\tau,t}^{\mathbf{x}}}  \sum_{i=0}^{\gamma-1} \mathbb{E}\left\{\nabla F( {\mathbf{w}}_{\mathbf{y}_0,0,t}^\mathbf{o})^\top\nabla F_\mathbf{y}^\mathbf{x}(\mathbf{w}_{\mathbf{y},\tau,i,t}^{\mathbf{x}})\right\} \leq \nonumber\\&-\frac{\mu_t \gamma}{2}\mathbb{E}\left\{\|\nabla F( {\mathbf{w}}_{\mathbf{y}_0,0,t}^\mathbf{o})\|^2\right\} -\frac{\mu_t}{2|{\cal A}_{\tau,t}|}\sum_{\mathbf{x} \in {\mathcal{C}}}^{} \sum_{\mathbf{y} \in {\cal A}_{\tau,t}^{\mathbf{x}}}  \sum_{i=0}^{\gamma-1} \nonumber\\&\mathbb{E}\left\{\|\nabla F(\mathbf{w}_{\mathbf{y},\tau,i,t}^{\mathbf{x}})\|^2\right\} +\frac{ L^2 \mu_t^3}{4} \frac{\sigma^2}{B} \gamma (\gamma-1)+\frac{ L^2 \mu_t^3}{2|{\cal A}_{\tau,t}|}\sum_{\mathbf{x} \in {\mathcal{C}}}^{}\nonumber\\& \sum_{\mathbf{y} \in {\cal A}_{\tau,t}^{\mathbf{x}}}  \sum_{i=0}^{\gamma-1} i\sum_{j=0}^{i-1}\mathbb{E}\left\{\|\nabla F(\mathbf{w}_{\mathbf{y},\tau,i,t}^{\mathbf{x}})\|^2\right\}+\frac{L^2 \mu_t^3 \tau \gamma}{2|{\cal A}_{\tau,t}|} \sum_{\mathbf{x} \in {\mathcal{C}}}^{}  |{\cal A}_{\tau,t}^{\mathbf{x}}|\nonumber\\&  \sum_{j=0}^{\tau-1} \frac{1}{|{\cal A}_{j,t}^\mathbf{x}|}\sum_{\mathbf{y} \in {\cal A}_{j,t}^\mathbf{x}}^{} {\mathbb{E}\left\{\|\nabla F(\mathbf{w}_{\mathbf{y},j,t}^\mathbf{x})\|^2\right\}}+ \frac{L^2 \mu_t^3 \gamma}{2|{\cal A}_{\tau,t}|}\frac{\sigma^2}{B}\sum_{\mathbf{x} \in {\mathcal{C}}}^{}  |{\cal A}_{\tau,t}^{\mathbf{x}}| \nonumber\\
& \sum_{j=0}^{\tau-1} \frac{1}{|{\cal A}_{j,t}^\mathbf{x}|} +\frac{L^2 \mu_t^3 \gamma}{2|{\cal A}_{\tau,t}|}\sum_{\mathbf{x} \in {\mathcal{C}}}^{}  |{\cal A}_{\tau,t}^{\mathbf{x}}| \sum_{j=0}^{\tau-1} \frac{\mathbb{E}\left\{\|\boldsymbol {\epsilon_\text{u}}_{j,t}^{\mathbf{x}} \|^2 \right\}}{{|{\cal A}_{j,t}^\mathbf{x}|^2} } +\frac{L^2 \mu_t^3 \gamma}{2|{\cal A}_{\tau,t}|}\nonumber\\
&\sum_{\mathbf{x} \in {\mathcal{C}}}^{} \sum_{\mathbf{y} \in {\cal A}_{\tau,t}^{\mathbf{x}}} \sum_{j=0}^{\tau-1} \frac{\mathbb{E}\left\{\|\boldsymbol {\epsilon_\text{d}}_{\mathbf{y},j,t}^{\mathbf{x}} \|^2 \right\}}{{|{\cal A}_{j,t}^\mathbf{x}|^2} }+ \frac{L^2 \mu_t \gamma}{2|{\cal A}_{\tau,t}|}\sum_{\mathbf{x} \in {\mathcal{C}}}^{} \sum_{\mathbf{y} \in {\cal A}_{\tau,t}^{\mathbf{x}}}\nonumber\\&\frac{\mathbb{E}\left\{\|\boldsymbol{\epsilon_\text{d}}_{\mathbf{y},t-1}\|^2\right\}}{{|{\cal A}_{\tau,t-1}|^2} }+\frac{L^2 \mu_t \gamma}{2|{\cal A}_{\tau,t-1}|^2}{\mathbb{E}\left\{\|\boldsymbol{\epsilon_\text{d}}_{\mathbf{y}_0,{t-1}}\|^2\right\}}.
\end{align}
Next, we bound the third term of the RHS in \eqref{lipexp} as 
\begin{align}
\label{2RHS2}
&\mathbb{E}\biggl\{\biggl\Vert \sum_{\mathbf{x} \in {\cal A}}^{} \frac{|{\cal A}_{\tau,t}^\mathbf{x}|}{|{\cal A}_{\tau,t}|} \sum_{i=0}^{\tau-1} {\mathbf{g}}_{i,t}^\mathbf{x}+\frac{1}{|{\cal A}_{\tau,t}|}\sum_{\mathbf{x} \in {\mathcal{C}}}^{} \sum_{\mathbf{y} \in {\cal A}_{\tau,t}^{\mathbf{x}}}  \sum_{i=0}^{\gamma-1} \nonumber\\
&\nabla F_{\mathbf{y}}^{\mathbf{x}}(\mathbf{w}_{\mathbf{y},\tau,i,t}^{\mathbf{x}})\biggr\Vert^2\biggr\} =\mathbb{E}\biggl\{\biggl\Vert \sum_{\mathbf{x} \in {\mathcal{C}}}^{} \frac{|{\cal A}_{\tau,t}^\mathbf{x}|}{{|{\cal A}_{\tau,t}|}} \sum_{i=0}^{\tau-1} \frac{1}{|{\cal A}_{i,t}^\mathbf{x}|}\sum_{\mathbf{y} \in {\cal A}_{i,t}^\mathbf{x} }^{}\nonumber\\
& \nabla F(\mathbf{w}_{\mathbf{y},i,t}^\mathbf{x})+\frac{1}{|{\cal A}_{\tau,t}|}\sum_{\mathbf{x} \in {\mathcal{C}}}^{} \sum_{\mathbf{y} \in {\cal A}_{\tau,t}^{\mathbf{x}}}  \sum_{i=0}^{\gamma-1} \nabla F(\mathbf{w}_{\mathbf{y},\tau,i,t}^{\mathbf{x}})\biggr\Vert^2\biggr\}+\nonumber\\
&\mathbb{E}\biggl\{\biggl\Vert \sum_{\mathbf{x} \in {\mathcal{C}}}^{} \frac{|{\cal A}_{\tau,t}^\mathbf{x}|}{{|{\cal A}_{\tau,t}|}} \sum_{i=0}^{\tau-1} \frac{1}{|{\cal A}_{i,t}^\mathbf{x}|}\sum_{\mathbf{y} \in {\cal A}_{i,t}^\mathbf{x} }^{} \bigl(\nabla F_\mathbf{y}^\mathbf{x}(\mathbf{w}_{\mathbf{y},i,t}^\mathbf{x}) - \nonumber\\
&\nabla F(\mathbf{w}_{\mathbf{y},i,t}^\mathbf{x})\bigr)\biggr\Vert^2\biggr\}+\mathbb{E}\biggl\{\biggl\Vert \frac{1}{|{\cal A}_{\tau,t}|}\sum_{\mathbf{x} \in {\mathcal{C}}}^{} \sum_{\mathbf{y} \in {\cal A}_{\tau,t}^{\mathbf{x}}}  \sum_{i=0}^{\gamma-1}\nonumber\\& \left(\nabla F_{\mathbf{y}}^{\mathbf{x}}(\mathbf{w}_{\mathbf{y},\tau,i,t}^{\mathbf{x}})- \nabla F(\mathbf{w}_{\mathbf{y},\tau,i,t}^{\mathbf{x}})\right)\biggr\Vert^2\biggr\},
\end{align}
where
\begin{align}
\label{2RHS21}
&\mathbb{E}\biggl\{\biggl\Vert \sum_{\mathbf{x} \in {\mathcal{C}}}^{} \frac{|{\cal A}_{\tau,t}^\mathbf{x}|}{{|{\cal A}_{\tau,t}|}} \sum_{i=0}^{\tau-1} \frac{1}{|{\cal A}_{i,t}^\mathbf{x}|}\sum_{\mathbf{y} \in {\cal A}_{i,t}^\mathbf{x} }^{} \nabla F(\mathbf{w}_{\mathbf{y},i,t}^\mathbf{x})+\frac{1}{|{\cal A}_{\tau,t}|}\sum_{\mathbf{x} \in {\mathcal{C}}}^{}\nonumber\\
& \sum_{\mathbf{y} \in {\cal A}_{\tau,t}^{\mathbf{x}}}  \sum_{i=0}^{\gamma-1} \nabla F(\mathbf{w}_{\mathbf{y},\tau,i,t}^{\mathbf{x}})\biggr\Vert^2\biggr\} \stackrel{(f)}{\leq}\hspace{-3pt}  \sum_{\mathbf{x} \in {\mathcal{C}}}^{}\frac{|{\cal A}_{\tau,t}^\mathbf{x}|}{{|{\cal A}_{\tau,t}|}}  \mathbb{E}\biggl\{\biggl\Vert\sum_{i=0}^{\tau-1} \frac{1}{|{\cal A}_{i,t}^\mathbf{x}|}\nonumber\\
&\sum_{\mathbf{y} \in {\cal A}_{i,t}^\mathbf{x} }^{} \nabla F(\mathbf{w}_{\mathbf{y},i,t}^\mathbf{x})\biggr\Vert^2\biggr\}+\frac{1}{|{\cal A}_{\tau,t}|}\sum_{\mathbf{x} \in {\mathcal{C}}}^{} \sum_{\mathbf{y} \in {\cal A}_{\tau,t}^{\mathbf{x}}}  \mathbb{E}\biggl\{\biggl\Vert\sum_{i=0}^{\gamma-1} \nonumber
\end{align}
\begin{align}
&\nabla F(\mathbf{w}_{\mathbf{y},\tau,i,t}^{\mathbf{x}})\biggr\Vert^2\biggr\} \stackrel{(g)}{\leq}   \tau \sum_{\mathbf{x} \in {\mathcal{C}}}^{} \frac{|{\cal A}_{\tau,t}^\mathbf{x}|}{{|{\cal A}_{\tau,t}|}} \sum_{i=0}^{\tau-1} \mathbb{E}\biggl\{\biggl\Vert\frac{1}{|{\cal A}_{i,t}^\mathbf{x}|}\sum_{\mathbf{y} \in {\cal A}_{i,t}^\mathbf{x} }^{} \nonumber\\&\nabla F(\mathbf{w}_{\mathbf{y},i,t}^\mathbf{x})\biggr\Vert^2\biggr\} +\frac{\gamma}{|{\cal A}_{\tau,t}|}\sum_{\mathbf{x} \in {\mathcal{C}}}^{} \sum_{\mathbf{y} \in {\cal A}_{\tau,t}^{\mathbf{x}}}  \sum_{i=0}^{\gamma-1} \mathbb{E}\biggl\{\biggl\Vert\nabla F(\mathbf{w}_{\mathbf{y},\tau,i,t}^{\mathbf{x}})\biggr\Vert^2\nonumber\\
&\biggr\}\stackrel{(h)}{\leq}   \tau  \sum_{\mathbf{x} \in {\mathcal{C}}}^{}\frac{|{\cal A}_{\tau,t}^\mathbf{x}|}{{|{\cal A}_{\tau,t}|}} \sum_{i=0}^{\tau-1} \frac{1}{|{\cal A}_{i,t}^\mathbf{x}|}\sum_{\mathbf{y} \in {\cal A}_{i,t}^\mathbf{x} }^{} \mathbb{E}\left\{\left\Vert\nabla F(\mathbf{w}_{\mathbf{y},i,t}^\mathbf{x})\right\Vert^2\right\}+\nonumber\\&\frac{\gamma}{|{\cal A}_{\tau,t}|}\sum_{\mathbf{x} \in {\mathcal{C}}}^{} \sum_{\mathbf{y} \in {\cal A}_{\tau,t}^{\mathbf{x}}}  \sum_{i=0}^{\gamma-1} \mathbb{E}\left\{\left\Vert\nabla F(\mathbf{w}_{\mathbf{y},\tau,i,t}^{\mathbf{x}})\right\Vert^2\right\},
\end{align}
where $(f)$ and $(h)$ follows from the convexity of $\|.\|^2$, and $(g)$ is from the inequality of arithmetic and geometric means. Also, the second term of RHS in \eqref{2RHS2} can be upper-bounded as
\begin{align}
\label{2RHS22}
&\mathbb{E}\biggl\{\biggl\Vert \sum_{\mathbf{x} \in {\mathcal{C}}}^{} \frac{|{\cal A}_{\tau,t}^\mathbf{x}|}{{|{\cal A}_{\tau,t}|}} \sum_{i=0}^{\tau-1} \frac{1}{|{\cal A}_{i,t}^\mathbf{x}|}\sum_{\mathbf{y} \in {\cal A}_{i,t}^\mathbf{x} }^{} \bigl(\nabla F_\mathbf{y}^\mathbf{x}(\mathbf{w}_{\mathbf{y},i,t}^\mathbf{x}) - \nonumber\\&\nabla F(\mathbf{w}_{\mathbf{y},i,t}^\mathbf{x})\bigr)\biggr\Vert^2\biggr\}\stackrel{(i)}{=} \sum_{\mathbf{x} \in {\mathcal{C}}}^{} \frac{|{\cal A}_{\tau,t}^\mathbf{x}|^2}{{|{\cal A}_{\tau,t}|}^2} \sum_{i=0}^{\tau-1} \frac{1}{|{\cal A}_{i,t}^\mathbf{x}|^2}\sum_{\mathbf{y} \in {\cal A}_{i,t}^\mathbf{x} }^{} \nonumber\\&\mathbb{E}\left\{\left\Vert\nabla F_\mathbf{y}^\mathbf{x}(\mathbf{w}_{\mathbf{y},i,t}^\mathbf{x}) - \nabla F(\mathbf{w}_{\mathbf{y},i,t}^\mathbf{x})\right\Vert^2\right\} \stackrel{(j)}{\leq} \sum_{\mathbf{x} \in {\mathcal{C}}}^{} \frac{|{\cal A}_{\tau,t}^\mathbf{x}|^2}{{|{\cal A}_{\tau,t}|}^2} \sum_{i=0}^{\tau-1} \nonumber\\&\frac{1}{|{\cal A}_{i,t}^\mathbf{x}|^2}\sum_{\mathbf{y} \in {\cal A}_{i,t}^\mathbf{x} }^{} \frac{\sigma^2}{B} = \frac{\sigma^2}{B} \sum_{\mathbf{x} \in {\mathcal{C}}}^{} \frac{|{\cal A}_{\tau,t}^\mathbf{x}|^2}{{|{\cal A}_{\tau,t}|}^2} \sum_{i=0}^{\tau-1} \frac{1}{|{\cal A}_{i,t}^\mathbf{x}|},
\end{align}
where $(i)$ is due to the independency and $(j)$ is from the Assumption 2. Following a similar approach as in \eqref{2RHS22}, for the third term of RHS in \eqref{2RHS2}, we have
\begin{align}
\label{new_var}
&\mathbb{E}\biggl\{\biggl\Vert \frac{1}{|{\cal A}_{\tau,t}|}\sum_{\mathbf{x} \in {\mathcal{C}}}^{} \sum_{\mathbf{y} \in {\cal A}_{\tau,t}^{\mathbf{x}}}  \sum_{i=0}^{\gamma-1} \bigl(\nabla F_{\mathbf{y}}^{\mathbf{x}}(\mathbf{w}_{\mathbf{y},\tau,i,t}^{\mathbf{x}})-\nonumber\\& \nabla F(\mathbf{w}_{\mathbf{y},\tau,i,t}^{\mathbf{x}})\bigr)\biggr\Vert^2\biggr\} \leq  \frac{1}{|{\cal A}_{\tau,t}|^2}|{\cal A}_{\tau,t}| \gamma \frac{\sigma^2}{B} = \frac{\gamma \frac{\sigma^2}{B}}{|{\cal A}_{\tau,t}|}.
\end{align}
Now, replacing \eqref{2RHS21}-\eqref{new_var} in \eqref{2RHS2} and replacing the result with \eqref{1RHS1}-\eqref{new_prod} in \eqref{lipexp} and then using the bound $\sum_{i=0}^{\tau-1} i \sum_{j=0}^{i-1} \frac{1}{|{\cal A}_{j,t}^\mathbf{x}|}\sum_{\mathbf{y} \in {\cal A}_{j,t}^\mathbf{x}}^{} \mathbb{E}\left\{\|\nabla F(\mathbf{w}_{\mathbf{y},j,t}^\mathbf{x})\|^2\right\} \leq \sum_{i=0}^{\tau-1} i \times \sum_{i=0}^{\tau-1} \frac{1}{|{\cal A}_{i,t}^\mathbf{x}|}\sum_{\mathbf{y} \in {\cal A}_{i,t}^\mathbf{x}}^{} \mathbb{E}\bigl\{\|\nabla F(\mathbf{w}_{\mathbf{y},i,t}^\mathbf{x})\|^2\bigr\} = \frac{\tau(\tau-1)}{2}\sum_{i=0}^{\tau-1} \frac{1}{|{\cal A}_{i,t}^\mathbf{x}|}\sum_{\mathbf{y} \in {\cal A}_{i,t}^\mathbf{x}}^{} \mathbb{E}\left\{\|\nabla F(\mathbf{w}_{\mathbf{y},i,t}^\mathbf{x})\|^2\right\}$, and due to the symmetry and independency of the network distribution for different intra- and inter-cluster iterations, which lead to \eqref{intraactive}-\eqref{active_both} having the same value and $\mathbb{E}\left\{\|\boldsymbol {\epsilon_\text{u}}_{j,t}^{\mathbf{x}} \|^2 \right\} \leq \mathbb{E}\left\{\|\boldsymbol {\epsilon_\text{u}^{\text{b}\mathbf{o}}} \|^2 \right\}$, $\mathbb{E}\left\{\|\boldsymbol {\epsilon_\text{d}}_{\mathbf{y},j,t}^{\mathbf{x}} \|^2 \right\}\leq \mathbb{E}\left\{\|\boldsymbol {\epsilon_{\text{d}_{\mathbf{y}_0}}^{\text{b}\mathbf{o}}} \|^2 \right\}$, $\mathbb{E}\left\{\|\boldsymbol {\epsilon_\text{u}}_{j,t} \|^2 \right\} \leq \mathbb{E}\left\{\|\boldsymbol {\epsilon_\text{u}^{\text{b}}} \|^2 \right\}$, and $\mathbb{E}\left\{\|\boldsymbol {\epsilon_\text{d}}_{\mathbf{y},j,t}^{\mathbf{x}} \|^2 \right\}\leq \mathbb{E}\left\{\|\boldsymbol {\epsilon_{\text{d}_{\mathbf{y}_0}}^{\text{b}}} \|^2 \right\}$ given in \eqref{ubo}-\eqref{udb} for all $\mathbf{x}$, $\mathbf{y}$, $i$ and $t$, we obtain the following bound on \eqref{lipexp}
\begin{align}
&\mathbb{E}\left\{F({\mathbf{w}}_{\mathbf{y}_0,0,t+1}^\mathbf{o})-F({\mathbf{w}}_{\mathbf{y}_0,0,t}^\mathbf{o})\right\} \leq -\frac{\mu_t (\tau+\gamma)}{2}\times\nonumber\\
& \mathbb{E}\left\{\|\nabla F( {\mathbf{w}}_{\mathbf{y}_0,0,t}^\mathbf{o})\|^2\right\}-\frac{\mu_t}{2}
\biggl(1- \frac{L^2\mu_t^2 \tau (\tau-1)}{2}-{L\mu_t \tau}-\nonumber
\end{align}
\begin{align}
&{L^2 \mu_t^2 \tau \gamma}\biggr)\mathbb{E}\biggl\{\sum_{\mathbf{x} \in {\mathcal{C}}}^{} \frac{|{\cal A}_{\tau,t}^\mathbf{x}|}{{|{\cal A}_{\tau,t}|}} \sum_{i=0}^{\tau-1} \frac{1}{|{\cal A}_{i,t}^\mathbf{x}|}\sum_{\mathbf{y} \in {\cal A}_{i,t}^\mathbf{x} }^{} \left\Vert\nabla F(\mathbf{w}_{\mathbf{y},i,t}^\mathbf{x})\right\Vert^2\biggr\}\nonumber\\
&-\frac{\mu_t}{2}
\left(1- \frac{L^2\mu_t^2 \gamma (\gamma-1)}{2}-{L\mu_t \gamma}\right)\mathbb{E}\biggl\{\frac{1}{|{\cal A}_{\tau,t}|}\sum_{\mathbf{x} \in {\mathcal{C}}}^{} \sum_{\mathbf{y} \in {\cal A}_{\tau,t}^{\mathbf{x}}}  \nonumber\\&\sum_{i=0}^{\gamma-1} \|\nabla F(\mathbf{w}_{\mathbf{y},\tau,i,t}^{\mathbf{x}})\|^2\biggr\}+\frac{L^2 \mu_t^3}{2}\frac{\sigma^2}{B}\frac{\tau(\tau-1)}{2} \mathbb{E}\left\{ \frac{1}{|{\cal A}^\mathbf{o}|}\right\}+\nonumber\\&\frac{ L^2 \mu_t^3}{4} \frac{\sigma^2}{B} \gamma (\gamma-1)+\frac{L^2 \mu_t^3}{2}\frac{\tau(\tau-1)}{2} \mathbb{E}\left\{\frac{1}{{|{\cal A}^\mathbf{o}|^2} }\right\}\mathbb{E}\left\{\|\boldsymbol \epsilon_\text{u}^{\text{b}\mathbf{o}} \|^2 \right\} \nonumber\\
&+\frac{L^2 \mu_t^3}{2}\frac{\tau(\tau-1)}{2} \mathbb{E}\left\{\frac{1}{{|{\cal A}^\mathbf{o}|^2} }\right\}\mathbb{E}\left\{\|\boldsymbol \epsilon_{\text{d}_{\mathbf{y}_0}}^{\text{b}\mathbf{o}} \|^2 \right\}+\frac{L^2 \mu_t\tau}{2 }\nonumber\\&\mathbb{E}\left\{\frac{1}{{|{\cal A}|^2} }\right\} \mathbb{E}\left\{\|\boldsymbol\epsilon_{\text{d}_{\mathbf{y}_0}}^\text{b}\|^2\right\} +\frac{L^2 \mu_t \tau}{2} \mathbb{E}\left\{\frac{1}{{|{\cal A}|^2} }\right\} \mathbb{E}\left\{\|\boldsymbol\epsilon_{\text{d}_{\mathbf{y}_0}}^\text{b}\|^2\right\}\nonumber\\&
+\frac{L\mu_t^2}{2} \frac{\sigma^2}{B} \tau \mathbb{E}\left\{\frac{1}{|{\cal A}^\mathbf{o}|}\right\} \mathbb{E}\left\{\sum_{\mathbf{x} \in {\mathcal{C}}}^{} \frac{|{\cal A}^\mathbf{x}|^2}{{|{\cal A}|}^2}\right\}  +
\frac{L\mu_t^2}{2} \gamma \frac{\sigma^2}{B}\mathbb{E}\left\{\frac{1}{|{\cal A}|}\right\}\nonumber\\
&+\frac{L\mu_t^2}{2}\tau \mathbb{E}\left\{\frac{1}{{|{\cal A}^\mathbf{o}|^2} }\right\} \mathbb{E}\left\{\|\boldsymbol\epsilon_\text{u}^{\text{b}\mathbf{o}}\|^2\right\}\mathbb{E}\left\{\sum_{\mathbf{x} \in {\mathcal{C}}}^{} \frac{| {\cal A}^\mathbf{x}|^2}{|{\cal A}|^2} \right\} +\frac{L\mu_t^2}{2} \tau \nonumber\\
&\mathbb{E}\left\{\frac{1}{|{\cal A}|}\right\} \mathbb{E}\left\{\frac{1}{{|{\cal A}^\mathbf{o}|^2} }\right\}\mathbb{E}\left\{\|\boldsymbol\epsilon_{\text{d}_{\mathbf{y}_0}}^{\text{b}\mathbf{o}}\|^2\right\}+\frac{L}{2} \mathbb{E}\left\{\frac{1}{|{\cal A}|}\right\} \mathbb{E}\left\{\frac{1}{{|{\cal A}|^2}} \right\}\nonumber\\
&\mathbb{E}\left\{\|\boldsymbol\epsilon_{\text{d}_{\mathbf{y}_0}}^{\text{b}}\|^2\right\}+\frac{L}{2}\mathbb{E}\left\{\frac{1}{{|{\cal A}|^2} }\right\}\mathbb{E}\left\{\|\boldsymbol\epsilon_{\text{d}_{\mathbf{y}_0}}^{\text{b}}\|^2\right\} +\frac{L}{2}\mathbb{E}\left\{\frac{1}{{|{\cal A}|^2} }\right\}\nonumber\\
&\mathbb{E}\left\{\|\boldsymbol\epsilon_\text{u}^\text{b}\|^2\right\}+\frac{L}{2}\mathbb{E}\left\{\frac{1}{{|{\cal A}|^2} }\right\}\mathbb{E}\left\{\|\boldsymbol\epsilon_{\text{d}_{\mathbf{y}_0}}^\text{b}\|^2\right\}+ \frac{L^2 \mu_t^3 \tau\gamma}{2}\frac{\sigma^2}{B} \nonumber\\
&\mathbb{E}\left\{ \frac{1}{|{\cal A}^\mathbf{o}|}\right\} +\frac{L^2 \mu_t^3 \tau \gamma}{2} \mathbb{E}\left\{\frac{1}{{{|{\cal A}^\mathbf{o}|^2} }}\right\}\mathbb{E}\left\{\|\boldsymbol {\epsilon}_\text{u}^{\text{b}\mathbf{o}} \|^2 \right\} +\frac{L^2 \mu_t^3 \tau \gamma}{2}\nonumber\\
& \mathbb{E}\left\{\frac{1}{{{|{\cal A}^\mathbf{o}|^2} }}\right\}{\mathbb{E}\left\{\|\boldsymbol {\epsilon_{\text{d}_{\mathbf{y}_0}}^{\text{b}\mathbf{o}}} \|^2 \right\}}+ \frac{L^2 \mu_t \gamma}{2}\mathbb{E}\left\{\frac{1}{{{|{\cal A}|^2} }}\right\}\mathbb{E}\left\{\|\boldsymbol {\epsilon_{\text{d}_{\mathbf{y}_0}}^{\text{b}}}\|^2\right\}\nonumber\\
&+\frac{L^2 \mu_t \gamma}{2}\mathbb{E}\left\{\frac{1}{|{\cal A}|^2}\right\}{\mathbb{E}\left\{\|\boldsymbol {\epsilon_{\text{d}_{\mathbf{y}_0}}^{\text{b}}}\|^2\right\}}.
\end{align}
Thus, under small enough $\mu_t$ and the following conditions
\begin{align}
&1- \frac{L^2\mu_t^2 \tau (\tau-1)}{2}-{L\mu_t \tau} -{L^2 \mu_t^2 \tau \gamma}\geq 0, \nonumber\\& 1- \frac{L^2\mu_t^2 \gamma (\gamma-1)}{2}-{L\mu_t \gamma} \geq 0,
\end{align}
and applying Assumption 3, we have for any $t \in \left\{0,\cdots,T-1\right\}$
\begin{align}
\label{finalstep}
&\mathbb{E}\left\{F({\mathbf{w}}_{\mathbf{y}_0,0,t+1}^\mathbf{o})\right\}-F^* \leq  {\left(1-\mu_t (\tau+\gamma) \delta\right)}\times \nonumber\\&\Bigl(\mathbb{E}\left\{F({\mathbf{w}}_{\mathbf{y}_0,0,t}^\mathbf{o})\right\}-F^*\Bigr)+\frac{L^2 \mu_t^3}{2}\frac{\sigma^2}{B}\left(\frac{\tau(\tau-1)}{2}+\tau \gamma \right) \nonumber\\&\mathbb{E}\left\{ \frac{1}{|{\cal A}^\mathbf{o}|}\right\}+\frac{ L^2 \mu_t^3}{4} \frac{\sigma^2}{B} \gamma (\gamma-1)+\frac{L\mu_t^2}{2} \frac{\sigma^2}{B} \tau \mathbb{E}\left\{\frac{1}{|{\cal A}^\mathbf{o}|}\right\} \nonumber\\
&\mathbb{E}\left\{\sum_{\mathbf{x} \in {\mathcal{C}}}^{} \frac{|{\cal A}^\mathbf{x}|^2}{{|{\cal A}|}^2}\right\} +
\frac{L\mu_t^2}{2} \gamma \frac{\sigma^2}{B}\mathbb{E}\left\{\frac{1}{|{\cal A}|}\right\}+\mathbb{E}\left\{\frac{1}{{|{\cal A}^\mathbf{o}|^2} }\right\}\nonumber\\
&\biggl(\frac{L^2 \mu_t^3}{2}\left(\frac{\tau(\tau-1)}{2}+\tau \gamma\right) +\frac{L\mu_t^2}{2}\tau \mathbb{E}\left\{\sum_{\mathbf{x} \in {\mathcal{C}}}^{} \frac{| {\cal A}^\mathbf{x}|^2}{|{\cal A}|^2} \right\}\biggr)  \nonumber\\
&\mathbb{E}\left\{\|\boldsymbol\epsilon_\text{u}^{\text{b}\mathbf{o}}\|^2\right\}+\mathbb{E}\left\{\frac{1}{{|{\cal A}^\mathbf{o}|^2} }\right\} \biggl(\frac{L^2 \mu_t^3}{2}\left(\frac{\tau(\tau-1)}{2}+\tau \gamma\right) +\nonumber
\end{align}
\begin{align}
&\frac{L\mu_t^2}{2} \tau \mathbb{E}\left\{\frac{1}{|{\cal A}|}\right\} \biggr)\mathbb{E}\left\{\|\boldsymbol\epsilon_{\text{d}_{\mathbf{y}_0}}^{\text{b}\mathbf{o}}\|^2\right\}+\mathbb{E}\left\{\frac{1}{{|{\cal A}|^2}}\right\}\biggl({L^2 \mu_t(\tau+{ \gamma})}+  \nonumber\\&\frac{L}{2} \mathbb{E}\left\{\frac{1}{|{\cal A}|}\right\} +L\biggr)\mathbb{E}\left\{\|\boldsymbol {\epsilon_{\text{d}_{\mathbf{y}_0}}^{\text{b}}}\|^2\right\} +\frac{L}{2}\mathbb{E}\left\{\frac{1}{{|{\cal A}|^2} }\right\}\mathbb{E}\left\{\|\boldsymbol\epsilon_\text{u}^\text{b}\|^2\right\}.
\end{align}
This bound connects the inter-cluster iterative steps $t+1$ and $t$. 
To get the bound of Theorem 1, we can replace $\mathbb{E}\left\{F({\mathbf{w}}_{\mathbf{y}_0,0,t}^\mathbf{o})\right\}-F^*$ on RHS with the same one step bound for $t$ and $t-1$. Repeating the procedure over $\left\{t-1,\cdots,0\right\}$, and from the equality $\sum_{i=0}^{t-1} c^i = \frac{1-c^t}{1-c}$ for any $c < 1$, the proof is complete.

\begin{IEEEbiography}[{\includegraphics[width=1in,height=1.25in,clip,keepaspectratio]{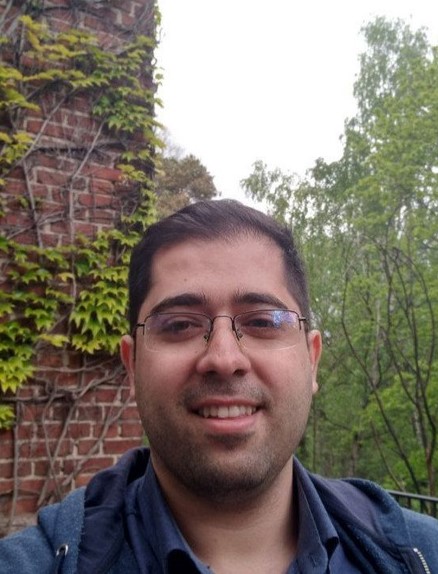}}]{Seyed Mohammad Azimi-Abarghouyi} received the B.Sc. degree (with honors) from Isfahan University of Technology, Isfahan, Iran, in 2012, and the M.Sc. and Ph.D. degrees from Sharif University of Technology, Tehran, Iran, in 2014 and 2020 respectively, all in electrical engineering. In 2017, he was a Visiting Research Scholar at Chalmers University of Technology, Gothenburg, Sweden. He is currently a Postdoctoral Fellow with KTH Royal Institute of Technology, Stockholm, Sweden. His research interests include communication theory, distributed optimization, machine learning, and point processes. He was selected as an Outstanding Researcher at Sharif University of Technology and received the best PhD Thesis Award in electrical engineering in Iran from the 2022 30\textsuperscript{th} International Conference on Electrical Engineering (ICEE). He was the recipient of several fellowship awards from Iran's National Elites Foundation and the Knut and Alice Wallenberg Foundation. He received the 2018 IEEE GLOBECOM Best Paper Award. He is a winner of Marie Sk\l{}odowska-Curie Fellowship awards in both 2020 and 2022. He was also selected as an Exemplary Reviewer for {\sc IEEE Wireless Communications Letters} in 2021 and 2022, and for {\sc IEEE Communications Letters} in 2023.
\end{IEEEbiography}

\begin{IEEEbiography}[{\includegraphics[width=1in,height=1.25in,clip,keepaspectratio]{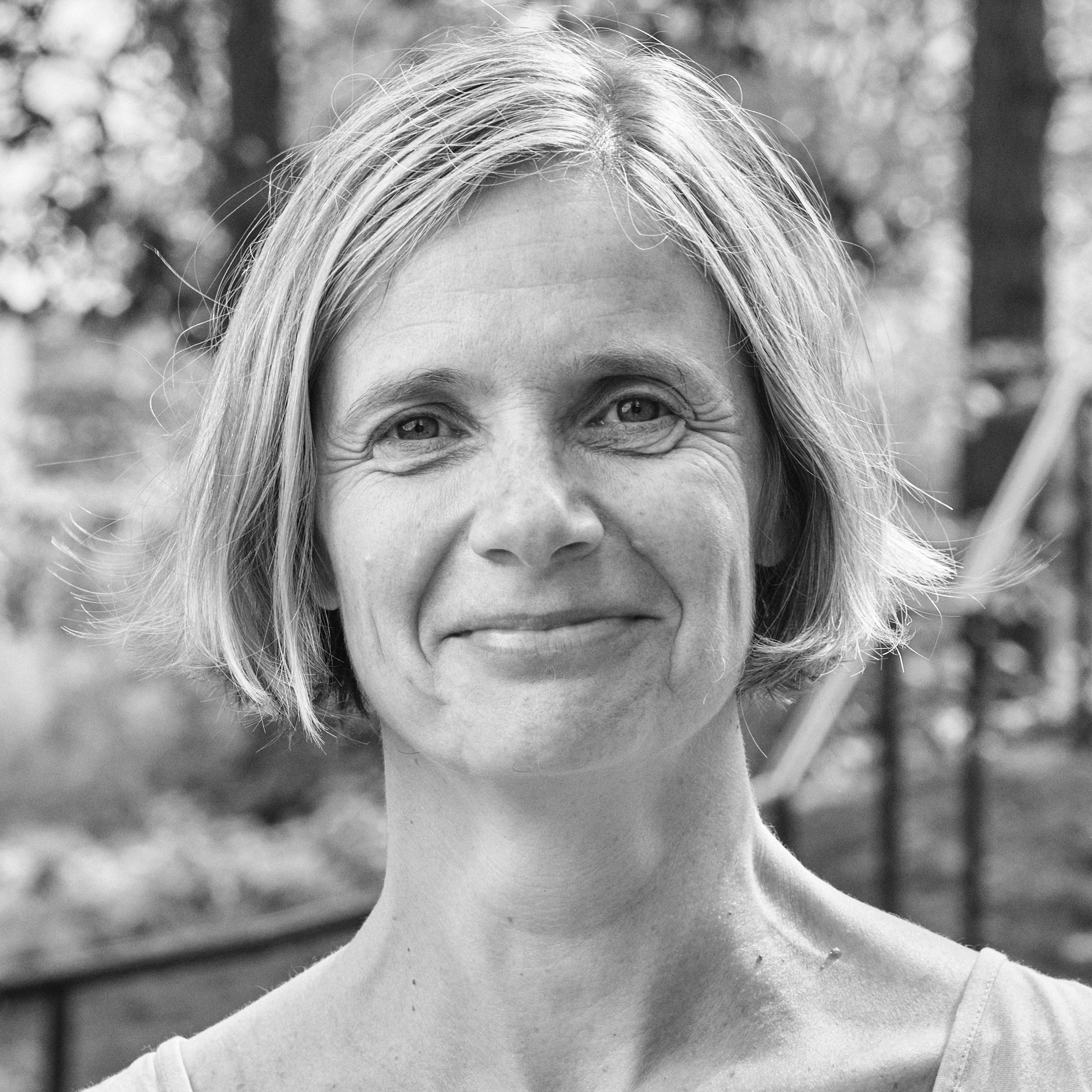}}]{Viktoria Fodor} (Member, IEEE) received the M.Sc. and Ph.D. degrees in computer engineering from the Budapest University of Technology and Economics, Budapest, Hungary, in 1992 and 1999, respectively, and the Habilitation degree (docent) from the KTH Royal Institute of Technology, Sweden, in 2011. She is a Professor of communication networks with the KTH Royal Institute of Technology. In 1998, she was a Senior Researcher with Hungarian Telecommunication Company. Since 1999, she has been with KTH Royal Institute of Technology. She has published more than 100 scientific publications. Her current research interests include performance evaluation of networks and distributed systems, stochastic modeling, and protocol design, with focus on edge computing and machine learning over networks. She is an Associate Editor of IEEE Transactions on Network and Service Management.
\end{IEEEbiography}

\end{document}